\documentclass[aps,pre,onecolumn,groupedaddress,longbibliography,nofootinbib,showkeys,showpacs]{revtex4-1}

\usepackage{latexsym}
\usepackage{amsmath}
\usepackage{amsfonts}
\usepackage{graphicx}
\usepackage{mathptmx}      
\usepackage{bm}
\usepackage{enumerate}
\usepackage{subfigure}
\usepackage{color}
\usepackage{hyperref}
\definecolor{DarkGreen}{RGB}{0,100,0}

\begin{document}


\title{Random State Technology\footnote{To appear in J. Phys. Soc. Jpn.}}

\author{Fengpin Jin}
\affiliation{Institute for Advanced Simulation, J\"ulich Supercomputing Centre,\\
Forschungszentrum J\"ulich, D-52425 J\"ulich, Germany}
\author{Dennis Willsch}
\affiliation{Institute for Advanced Simulation, J\"ulich Supercomputing Centre,\\
Forschungszentrum J\"ulich, D-52425 J\"ulich, Germany}
\author{Madita Willsch}
\affiliation{Institute for Advanced Simulation, J\"ulich Supercomputing Centre,\\
Forschungszentrum J\"ulich, D-52425 J\"ulich, Germany}
\author{Hannes Lagemann}
\affiliation{Institute for Advanced Simulation, J\"ulich Supercomputing Centre,\\
Forschungszentrum J\"ulich, D-52425 J\"ulich, Germany}
\affiliation{RWTH Aachen University, D-52056 Aachen, Germany}
\author{Kristel Michielsen}
\affiliation{Institute for Advanced Simulation, J\"ulich Supercomputing Centre,\\
Forschungszentrum J\"ulich, D-52425 J\"ulich, Germany}
\affiliation{RWTH Aachen University, D-52056 Aachen, Germany}
\author{Hans De Raedt}
\email{deraedthans@gmail.com}
\thanks{Corresponding author}
\affiliation{Institute for Advanced Simulation, J\"ulich Supercomputing Centre,\\
Forschungszentrum J\"ulich, D-52425 J\"ulich, Germany}
\affiliation{Zernike Institute for Advanced Materials,\\
University of Groningen, Nijenborgh 4, NL-9747 AG Groningen, The Netherlands}

\date{\today}

\begin{abstract}
We review and extend, in a self-contained way, the mathematical foundations of
numerical simulation methods that are based on the use of random states. The
power and versatility of this simulation technology is illustrated by
calculations of physically relevant properties such as the density of states of
large single particle systems, the specific heat, current-current correlations,
density-density correlations, and electron spin resonance spectra of many-body
systems. We explore a new field of applications of the random state technology
by showing that it can be used to analyze numerical simulations and experiments
that aim to realize quantum supremacy on a noisy intermediate-scale quantum
processor. Additionally, we show that concepts of the random state technology
prove useful in quantum information theory.
\end{abstract}


\maketitle

\section{Introduction}\label{INT}
Random numbers are ubiquitous in computational and theoretical physics.
A prime example is random matrix theory, whose application in physics emerged when Eugene Wigner introduced the notion of random matrices to model the nuclei of heavy atoms,
explaining the level spacing in the spectra of the nuclei~\cite{WIGN55}.
In the meantime, random matrix theory has found applications in many other fields such as
quantum chaos~\cite{BOHI84}, quantum optics~\cite{RUSS17}, and quantum computing~\cite{BOIX18}, as well as in other disciplines of engineering and finance \cite{edelman2005randomMatrixTheory}.
In quantum statistical mechanics, a way to study
eigenstate thermalization is to introduce a small perturbation in the form of a random matrix~\cite{DEUT91}.
The resulting model is assumed to be applicable to physical situations, in particular to the study of equilibration of local observables~\cite{SRED94,REGO08}.

In the past two decades, the concept of the so-called canonical typicality
has been developed~\cite{GOLD06,GOLD10,GOLD15,POPE06,REIM07,REIM10,REIM15}.
In essence, the idea is that the reduced density matrix of a system
is a thermal state if the state of the whole system is
a pure state of the overwhelming majority of wave functions,
that is a random state on the subspace defined by a small energy interval
containing the energy of the microcanonical ensemble.
Similar ideas appear in the works from Bocchieri and Loinger~\cite{BOCC59}, and Tasaki~\cite{TASA98},
who did not emphasize the role of the random state though.
Gemmer et al. studied the process of equilibration of a two-level system coupled to a special many-level bath and
derived a rate equation for the two levels by Hilbert space averaging~\cite{GEMM04}.
The derivation implicitly uses the notion of the bath being in a pure random state.
The idea of the thermal pure quantum state formulation of statistical mechanics was given in Ref.~\cite{SUGI12,SUGI13},
some elements of which were already contained in S. Lloyd’s Phd thesis in 1988~\cite{LLOY13}.
This thesis was barely noticed in the community of typicality until the relevant chapter was made available on arXiv in 2013~\cite{LLOY13}.

In the numerical-simulation arena, (pseudo-)random numbers are key to all Monte Carlo methods~\cite{HAMM64,LAND00,PRES03}.
The focus of this paper is on the theory and application of a numerical method
for computing properties of quantum systems, based on the use of a random state, which in essence is
just a collection of random amplitudes defining the state vector of the quantum system.
We coin this numerical tool to calculate various static and dynamic properties of quantum systems
as {\bf random state technology}.
In the literature, this technology is also referred to as quantum dynamical
typicality~\cite{GEMM03a,BART09,STEIN14,STEIN16h}.

Calculating traces of square matrices is one of the basic
computational problems in quantum statistical physics.
If calculating each matrix element $\langle i|X|i\rangle$ for $i=1,\ldots,D$
takes ${\cal O}(D)$ arithmetic operations, computing $\mathbf{Tr\;} X=\sum_{i=1}^D \langle i|X|i\rangle$
requires ${\cal O}(D^2)$ arithmetic operations.
In general, $D$ is the dimension of the Hilbert space containing the state vectors
describing the quantum system. $D$ grows exponentially with the system size;
for instance, for a system of $n$ qubits, $D=2^n$.
Random state technology
makes calculations feasible that would otherwise be out of reach with present-day supercomputers
because it changes the operation count from
${\cal O}(D^2)$ to ${\cal O}(D)$, a significant reduction if $D$ is large.

To the best of our knowledge, the basic idea of the random state technology is due to Alben \emph{et al.},
who used a random phase vector to reduce the numerical work of calculating the density of states (DOS) for
a particle moving on a lattice~\cite{ALBE75}.
Similar ideas have been used in the quantum transfer matrix Monte Carlo Method~\cite{IMAD86}, the estimation of the
eigenvalue spectrum (by calculating the moments of the distribution of eigenvalues)~\cite{SKIL88},
the calculation of the DOS and other properties of the Anderson model and
of the spin-$1/2$ Heisenberg model~\cite{RAED89,RAED93,FUKA94},
and in the calculation of the DOS~\cite{DRAB93,SILV94,SILV97} and linear response functions~\cite{IITA97,NOMU97b,IITA99,NISH20}.
A rigorous justification of the working principle of the random state technology
and applications to the calculation of the eigenvalues of very large matrices, the density of states (DOS), and static properties of spin models
were given by Hams and De~Raedt~\cite{HAMS00}.

One area of applications for which the random state technology proves very useful
is the study of decoherence, thermalization, and dissipation of a quantum system interacting with a bath,
that is an open quantum system~\cite{BREU02}.
A simple but approximate method to treat the dynamics of an open quantum system
is to use a master equation for the reduced density matrix of the quantum system~\cite{BREU02}.
A direct, approximation-free simulation of a relatively large system (quantum system + bath)
becomes possible~\cite{YUAN06,YUAN08,YUAN09,YUAN11,JIN10x,RAED12z,NOVO16,ZHAO16,RAED17b,GELM03,GELM04,KATZ08}
if we use random state technology to represent the bath state by a projected random state,
the so-called thermal pure quantum state~\cite{SUGI12,SUGI13}. 
Another important class of applications is
calculation of various static and dynamic properties, such as
the DOS, the specific heat, the diffusion constant, the optical conductivity, etc.\
for systems such as the spin-$1/2$ Heisenberg model~\cite{MONN14,STEIN14,STEI16,STEI17a,YAMA18,RICH18,RICH19},
the Hubbard model~\cite{JIN15,STEI17b},
tight-binding models of 2D materials~\cite{YUAN10d,YUAN10e,YUAN14g,YUAN15}, etc.

For the efficient calculation of dynamic properties,
the random state technology has to be supplemented by an efficient method
for computing the result of applying the time-evolution operator to the state vector.
Methods most often used are Suzuki-Trotter product formula algorithms~\cite{RAED87,RAED06},
the Chebyshev polynomial algorithm~\cite{TALE84,DOBR03,WEIS06},
a Lanczos-iteration based method~\cite{PREL17,SCHN20}, and the Runge-Kutta method~\cite{TARE13,STEIN14}.

The paper is structured as follows.
In Sec.~\ref{PRE} we present an extensive, rigorous analysis of the random state technology
using basic tools of calculus only.
We derive general results for the mean and variance for different random states,
and use Markov's lemma to derive bounds.
Then, we discuss various applications of the random state technology such as
the calculation of the DOS (Sec.~\ref{DOS}), the specific heat (Sec.~\ref{QSP}),
current-current and density-density correlations, and electron spin resonance (ESR) spectra
(Sec.~\ref{QDYN}).
Sections~\ref{AQSUP} and~\ref{QINF} apply
the ideas of the random state technology to analyze numerical simulations and experiments that aim to establish quantum supremacy on a noisy intermediate-scale quantum processor~\cite{GOOG19} and generalize a statement in quantum information theory~\cite{nielsen2002gatefidelity,Gilchrist2005fidelities}.
Sec.~\ref{frm} summarizes the material presented in this paper.

\section{Theoretical background}\label{PRE}

The underlying idea of the random state approach is that accurate approximations
to $\mathbf{Tr\;} X$ may be obtained by computing $\langle\Phi|X|\Phi\rangle$
with the pure state $|\Phi\rangle$ defined by
\begin{eqnarray}
|\Phi\rangle=\sum_{j=1}^D c_j |j\rangle
,
\label{PRE0}
\end{eqnarray}
where
$\{|j\rangle\; | \; j=1,\ldots,D\}$ denotes (any) complete set of orthonormal basis states
spanning the $D$-dimensional Hilbert space
and where
the complex-valued amplitudes $\{c_j=c_j(\bm\xi);|\;j=1,\ldots,D\}$ are
functions of complex-valued random variables $\bm\xi=(\xi_1,\ldots,\xi_D)$,
i.e. $\xi_j=a_j+i b_j$ with real-valued $a_j$ and $b_j$,
distributed according to a probability density $p(\bm\xi)$,
We denote the expectation and variance of a random variable $z=z(\bm\xi)$ with respect to $p(\bm\xi)$
by $\mathrm{E}[z]$ and $\mathrm{Var}[z]=\mathrm{E}[z^2]-\mathrm{E}^2[z]$, respectively.
By construction, we have $\mathrm{E}[1]=1$ and $\mathrm{Var}[1]=0$.

We focus on the case where
$p(\bm\xi)=p(a_1,b_1,\ldots,a_D,b_D)$ is a symmetric function of the $a$'s and $b$'s, meaning that its value does not change if we interchange its arguments.
Furthermore, we impose the constraint that
$p(\bm\xi)=p(a_1,b_1,\ldots,a_D,b_D)$
is an even function of each of the $a$'s and $b$'s
and that both the real and imaginary part of $c_j(\bm\xi)=c_j(a_1,b_1,\ldots,a_D,b_D)$
are odd functions of each of the $a$'s and $b$'s.
Exploiting these symmetries, it is easy to see that
\begin{eqnarray}
\mathrm{E}\left[ c^\ast_i \right]&=&\mathrm{E}\left[ c^{}_j \right]=
\mathrm{E}\left[ c^{}_i c^{}_j \right]=\mathrm{E}\left[ c^\ast_i c^\ast_j \right]=0,
\label{PRE1a}
\\
\mathrm{E}\left[ c^\ast_i c^{}_j \right]&=&
\delta_{i,j} \mathrm{E}\left[| c^{}_i|^2 \right] = \delta_{i,j} \mathrm{E}\left[|c|^2 \right],
\label{PRE1b}
\\
\mathrm{E}\left[ c^\ast_i c^{}_j  c^{}_k c^\ast_l
\right]&=&
(1-\delta_{i,l})[\delta_{i,j}\delta_{k,l}+\delta_{i,k}\delta_{j,l}]
\mathrm{E}\left[| c^{}_i|^2 | c^{}_l|^2 \right]
+\delta_{i,l}\delta_{i,j}\delta_{k,l}
\mathrm{E}\left[| c^{}_i|^4 \right]
\nonumber \\
&=& (1-\delta_{i,l})[\delta_{i,j}\delta_{k,l}+\delta_{i,k}\delta_{j,l}] \mathrm{E}\left[|c|^2 |\widehat c|^2 \right]
 +\delta_{i,l}\delta_{i,j}\delta_{k,l} \mathrm{E}\left[|c|^4 \right]
,
\label{PRE1c}
\end{eqnarray}
where we used the symmetries of $p(\bm\xi)$ to drop subscripts in the argument of $\mathrm E\left[.\right]$
and introduced $\widehat{c}$ to keep track
of the fact that $|c|^2$ and $|\widehat c|^2$ correspond to two different random variables $\xi_j$ and $\xi_k$ with $j\not=k$.

In this paper, we limit the discussion to the random states listed in Table~\ref{tab1}.
We refer to states generated by the methods implied by Case A and B as Gaussian random states and those generated
by the method implied by Case C as random phase states~\cite{IITA04}.
Salient features of these random states are that in practice they are easy to generate
and that exact results for their statistical properties can be derived.
In Appendix~\ref{PICK}, we present two different algorithms for Case A and show that the Gaussian random state is drawn from a
uniform distribution on the $D$-dimensional sphere (Haar measure).
The Gaussian random state generated by Case B does not have this property by itself;
however, as we normalize the state vector in most applications of quantum theory,
Case B is very similar to Case A.
Generating the random phase state is close to trivial.

\begin{table*}[ht]
\caption{%
Overview of the combination of probability densities $p(\bm\xi)$ and amplitudes $c_j(\bm\xi)$ that we consider in this paper.
Columns four to six give the  moments that appear in Eqs.~(\ref{PRE1b}) and~(\ref{PRE1c}).
Columns seven and eight give the exact expressions (see Appendix~\ref{SOM} for the proof)
of the average and the variance of ${D\langle\Phi|X|\Phi\rangle}/{\langle\Phi|\Phi\rangle}$, respectively.
}
\begin{ruledtabular}
\begin{tabular}{cccccccc}
Case &$p(\bm\xi)$ &  $c_j(\bm\xi)$ & $\mathrm E\left[|c|^2\right]$ & $\mathrm E\left[|c|^2 |\widehat c|^2 \right]$ & $\mathrm E\left[|c|^4\right]$
&$\mathrm{E}\left[\frac{D\langle\Phi|X|\Phi\rangle}{\langle\Phi|\Phi\rangle}\right]$
& $\mathrm{Var}\left[\frac{D\langle\Phi|X|\Phi\rangle}{\langle\Phi|\Phi\rangle}\right]$\\
\noalign{\medskip}\hline\noalign{\medskip}
A&$\prod_{j=1}^D \frac{1}{\pi}e^{-|\xi_{j}|^2}$ &  $\frac{\xi_j}{\sqrt{|\xi_{1}|^2+\ldots+|\xi_{D}|^2}}$ &$\frac{1}{D}$&$\frac{1}{D(D+1)}$&$\frac{2}{D(D+1)}$
& $\mathbf{Tr\;}X$ & $\frac{D\mathbf{Tr\;}XX^\dagger-\left|\mathbf{Tr\;}X\right|^2}{D+1}$\\
B&$\prod_{j=1}^D \frac{1}{\pi}e^{-|\xi_{j}|^2}$ &  $\frac{\xi_j}{\sqrt{D}} $&$\frac{1}{D}$&$\frac{1}{D^2}$&$\frac{2}{D^2}$
& $\mathbf{Tr\;}X$ & $\frac{D\mathbf{Tr\;}XX^\dagger-\left|\mathbf{Tr\;}X\right|^2}{D+1}$\\
C&$\left(\frac{1}{2\pi}\right)^D$&  $\frac{1}{\sqrt{D}}e^{ia_j}$&$\frac{1}{D}$&$\frac{1}{D^2}$&$\frac{1}{D^2}$
& $\mathbf{Tr\;}X$ & $\mathbf{Tr\;}XX^\dagger-\sum_{i=1}^D |\langle i|X|i\rangle|^2$\\
\end{tabular}
\end{ruledtabular}
\label{tab1}
\end{table*}

In the following, we first prove general results using the symmetries of the probability density only.
Then, if no further progress can be made, we use Table~\ref{tab1} to obtain explicit expressions
for the Cases A, B, and C.
Using Eq.~(\ref{PRE1b}), we find
\begin{eqnarray}
\mathrm{E}\left[\langle \Phi|X|\Phi\rangle\right]&=&
\sum_{i,j=1}^D \mathrm{E}\left[c^\ast_i c^{}_j\right] \langle i|X|j\rangle = \mathrm{E}\left[|c|^2 \right] \mathbf{Tr\;}X.
\label{PRE2a}
\end{eqnarray}
Evaluating Eq.~(\ref{PRE2a}) for the special case $X=\openone$ yields $\mathrm{E}\left[|c|^2 \right]=\mathrm{E}\left[\langle \Phi|\Phi\rangle\right]/D$. Using this to eliminate $\mathrm{E}\left[|c|^2 \right]$, we have
\begin{eqnarray}
 \mathbf{Tr\;}X &=& \frac{D\mathrm{E}\left[\langle \Phi|X|\Phi\rangle\right]}{\mathrm{E}\left[\langle \Phi|\Phi\rangle\right]}
,
\label{PRE3}
\end{eqnarray}
which shows that the ${\cal O}(D^2)$
computational effort of evaluating the trace (see above) is replaced by the ${\cal O}(D)$
effort of evaluating the matrix elements in Eq.~(\ref{PRE3}) and the estimation of
the expectation values $\mathrm E\left[.\right]$.

A further, significant reduction of work is possible if
we can obtain accurate estimates of the averages in Eq.~(\ref{PRE3})
by using only one ``realization'' of the random vector $\bm\xi$.
Therefore, in practice, we generate a random vector $\bm\xi$, construct the pure state $|\Phi\rangle$,
compute $D\langle\Phi|X|\Phi\rangle/\langle\Phi|\Phi\rangle$
and expect that
\begin{center}
\framebox{
\parbox[t]{0.25\hsize}{%
\begin{eqnarray}
 \mathbf{Tr\;}X &\approx& \frac{D\langle\Phi|X|\Phi\rangle}{\langle \Phi|\Phi\rangle}
.
\label{PRE3z}
\end{eqnarray}
}}%
\end{center}
Markov's lemma~\cite{GRIM01} (see below) tells us that if the variance of ${D\langle\Phi|X|\Phi\rangle}/{\langle \Phi|\Phi\rangle}$
vanishes, the probability that ${D\langle\Phi|X|\Phi\rangle}/{\langle \Phi|\Phi\rangle}$
is equal to $\mathbf{Tr\;}X$ approaches one.
Therefore, in order to assess the usefulness of the ``one random state'' approximation,
we have to compute the variance of  ${D\langle\Phi|X|\Phi\rangle}/{\langle \Phi|\Phi\rangle}$.

In Appendix~\ref{SOM}, we derive the exact expressions for the average and the variance
of the right-hand side of Eq.~(\ref{PRE3z}).
These expressions are listed in the seventh and eight column of Table~\ref{tab1}, respectively.
From column seven, it follows directly that on average
${D\langle\Phi|X|\Phi\rangle}/{\langle \Phi|\Phi\rangle}$ is equal to $ \mathbf{Tr\;}X$.
Note that in Case A and B, the variance does not depend on the choice of the
basis, whereas in Case C, it does. Therefore, in Case C,  a proper choice (depending on $X$)
of the basis may help to reduce the variance~\cite{IITA04}.

At this stage of the discussion, $X$ can be any matrix with entries of any size.
In order to have a numerically meaningful measure of the statistical fluctuations
we define the relative variance $\mathrm{rVar}$ by
\begin{eqnarray}
\mathrm{rVar}\left[\frac{D\langle\Phi|X|\Phi\rangle}{\langle \Phi|\Phi\rangle}\right]&=&
\frac{\mathrm{Var}\left[ \frac{D\langle\Phi|X|\Phi\rangle}{\langle \Phi|\Phi\rangle}\right]}{\left(\mathrm{E}\left[ \frac{D\langle\Phi|X|\Phi\rangle}{\langle \Phi|\Phi\rangle}\right]\right)^2}
=
\frac{\mathrm{Var}\left[ \frac{D\langle\Phi|X|\Phi\rangle}{\langle \Phi|\Phi\rangle}\right]}{\left|\mathbf{Tr\;}X\right|^2}
\;.
\label{PRE4a}
\end{eqnarray}
Clearly, Eq.~(\ref{PRE4a}) only makes sense if $\mathrm{E}\left[ X\right]$ is not a small number.
If it is, we should take $\mathrm{Var}\left[ X\right]$ as a measure for the absolute
statistical error.
Let us consider the physically relevant case where $X=X^\dagger$ with real eigenvalues $\lambda_j$, $j=1,\ldots,D$.
Then, for Cases A and B we have
\begin{eqnarray}
\mathrm{rVar}\left[\frac{D\langle\Phi|X|\Phi\rangle}{\langle \Phi|\Phi\rangle}\right]&=&
\frac{D\mathbf{Tr\;}XX^\dagger-\left|\mathbf{Tr\;}X\right|^2}{(D+1)\left|\mathbf{Tr\;}X\right|^2}
= \frac{D^{-1}\sum_{j=1}^D \lambda_j^2 - \left(D^{-1}\sum_{j=1}^D \lambda_j\right)^2}{
(D+1)\left(D^{-1}\sum_{j=1}^D \lambda_j\right)^2}
=\frac{1}{(D+1)}
\frac{\overline{\lambda^2}-\overline{\lambda}^2}{\overline{\lambda}^2}
\;,
\nonumber \\
\label{VAR2}
\end{eqnarray}
and for Case C we have
\begin{eqnarray}
\mathrm{rVar}\left[\frac{D\langle\Phi|X|\Phi\rangle}{\langle \Phi|\Phi\rangle}\right]&\le&
\frac{\mathbf{Tr\;}XX^\dagger}{\left|\mathbf{Tr\;}X\right|^2}
= \frac{D^{-1}\sum_{j=1}^D \lambda_j^2}{D\left(D^{-1}\sum_{j=1}^D \lambda_j\right)^2}
=
\frac{1}{D}
\frac{\overline{\lambda^2}}{\overline{\lambda}^2}
,
\label{VAR2a}
\end{eqnarray}
where $\overline{\lambda^k}\equiv D^{-1}\sum_{j=1}^D \lambda_j^k$ denotes the average with respect to the index $j$.
The last factor in Eq.~(\ref{VAR2}) is the relative mean square deviation of the eigenvalues of $X$.
If we assume that this number does not increase faster than $D+1$, a reasonable assumption for
physically relevant problems, then
$\mathrm{rVar}\left[{D\langle\Phi|X|\Phi\rangle}/{\langle \Phi|\Phi\rangle}\right]$
vanishes as $D\to\infty$.
Similarly, if ${\overline{\lambda^2}}/{\overline{\lambda}^2}$  does not increase faster than $D$,
then the left-hand side of Eq.~(\ref{VAR2a}) vanishes as $D\to\infty$.

The vanishing of the relative variances as $D\to\infty$ is the key idea behind
any successful, practical application of the random state technology.
Indeed, once it has been shown that relative variances vanish as $D\to\infty$,
for large $D$ (which is a very large number in most cases of interest)
it suffices to use only one realization of a random state $|\Phi\rangle$
to obtain an accurate estimate of $\mathbf{Tr\;}X$, implying that the computational burden has been reduced by a factor $D$.
This can be shown by the following argument.
Appealing to a variant of Markov's inequality (Chebyshev's inequality)~\cite{GRIM01}
\begin{eqnarray}
P(|A-\langle A\rangle |\ge \epsilon) \le\frac{\mathrm{Var}\left[A\right]}{\epsilon^2}
\;,\quad \epsilon>0
,
\label{VAR0}
\end{eqnarray}
we see that if the variance of $A$ is sufficiently small,
the probability that the value of this random variable deviates from its average value will also be small.
We use this basic result in the form
\begin{eqnarray}
P\left(\left|\frac{D\langle\Phi|X|\Phi\rangle}{\langle \Phi|\Phi\rangle}-\mathbf{Tr\;}X
\right|\ge \epsilon\left|\mathbf{Tr\;}X\right| \right)&\le&\frac{
\mathrm{Var}\left[{D\langle\Phi|X|\Phi\rangle}/{\langle \Phi|\Phi\rangle}\right]}{\epsilon^2\left|\mathbf{Tr\;}X\right|^2}
\nonumber \\
&=&\epsilon^{-2}\;\mathrm{rVar}\left[\frac{D\langle\Phi|X|\Phi\rangle}{\langle \Phi|\Phi\rangle}\right]
\;,\quad \epsilon>0
\label{VAR1}
\end{eqnarray}
This shows that if
$(D\mathbf{Tr\;}XX^\dagger-\left|\mathbf{Tr\;}X\right|)/\left|\mathbf{Tr\;}X\right|^2$ (Case A and B)
or
$\mathbf{Tr\;}XX^\dagger/\left|\mathbf{Tr\;}X\right|^2$ (Case C)
do not increase faster than $D$,
the probability that the right-hand side of Eq.~(\ref{PRE3z}), computed from one realization
of a random state, deviates from its expectation value $\mathbf{Tr\;}X$, vanishes as $D$ increases.

It is instructive to estimate the expected error of using Eq.~(\ref{PRE3z}) in a different, somewhat more general manner.
Let us multiply both sides of Eq.~(\ref{PRE3z}) by $\langle\Phi|\Phi\rangle$
and define
\begin{eqnarray}
{\widetilde X}&=&D \langle \Phi|X|\Phi\rangle - \langle\Phi|\Phi\rangle\mathbf{Tr\;}X
.
\label{PRE8}
\end{eqnarray}
Obviously, Eq.~(\ref{PRE3}) implies $\mathrm{E}\left[{\widetilde X}\right]=0$.
The mean square deviation is given by
\begin{eqnarray}
\mathrm{Var}\left[{\widetilde X}\right]=\mathrm{E}\left[\left|D \langle \Phi|X|\Phi\rangle - \langle\Phi|\Phi\rangle\mathbf{Tr\;}X\right|^2\right].
\label{PRE9}
\end{eqnarray}
By straightforward application of Eqs.~(\ref{PRE1a})--~(\ref{PRE1c})
we find
\begin{eqnarray}
\mathrm{E}\left[|\langle \Phi|X|\Phi\rangle\right|^2]&=&
\sum_{i,j=1}^D \sum_{k,l=1}^D \mathrm{E}\left[c^\ast_i c^{}_j  c^{}_k c^\ast_l\right]
\langle i|X|j\rangle\langle k|X|l\rangle^\ast
\nonumber \\
&=&
 \mathrm{E}\left[|c|^2 |\widehat c|^2 \right]
 \sum_{i,j=1}^D\sum_{k,l=1}^D[\delta_{i,j}\delta_{k,l}+\delta_{i,k}\delta_{j,l}]\langle i|X|j\rangle\langle k|X|l\rangle^\ast
\nonumber \\
&& +
 \left(\mathrm{E}\left[|c|^4 \right] - 2\mathrm{E}\left[|c|^2 |\widehat c|^2 \right]\right)
 \sum_{i=1}^D  \langle i|X|i\rangle\langle i|X|i\rangle^\ast
\label{PRE2b}
\nonumber \\
&=&
 \mathrm{E}\left[|c|^2 |\widehat c|^2 \right] \left( \mathbf{Tr\;} XX^\dagger +\left|\mathbf{Tr\;}X\right|^2\right)
\nonumber \\
&& +
 \left(\mathrm{E}\left[|c|^4 \right] - 2\mathrm{E}\left[|c|^2 |\widehat c|^2 \right]\right)
 \sum_{i=1}^D  |\langle i|X|i\rangle|^2
\\
\mathrm{E}\left[\langle \Phi|X|\Phi\rangle\langle \Phi|\Phi\rangle^\ast\right]&=&
\sum_{i,j=1}^D\sum_{k,l=1}^D
\mathrm{E}\left[c^\ast_i c^{}_j  c^{}_l c^\ast_k\right]
\langle i|X|j\rangle
\langle k|l\rangle^\ast
\nonumber \\
&=& \left\{ (D-1)\mathrm{E}\left[|c|^2 |\widehat c|^2 \right] +  \mathrm{E}\left[|c|^4 \right] \right\} \mathbf{Tr\;} X
,
\label{PRE2c}
\end{eqnarray}
and therefore
\begin{eqnarray}
\mathrm{Var}\left[{\widetilde X}\right]
&=&
D^2 \mathrm{E}\left[|\langle \Phi|X|\Phi\rangle\right|^2]+ \mathrm{E}\left[|\langle \Phi|\Phi\rangle\right|^2] \left|\mathbf{Tr\;}X \right|^2
-2D\Re\left( \mathrm{E}\left[
\langle \Phi|X|\Phi\rangle \langle \Phi|\Phi\rangle^\ast\right]  \mathbf{Tr\;}X \right)
\nonumber \\
&=&
 D^2\mathrm{E}\left[|c|^2 |\widehat c|^2 \right] \left( \mathbf{Tr\;} XX^\dagger +\left|\mathbf{Tr\;}X\right|^2\right)
 +
 D^2\left(\mathrm{E}\left[|c|^4 \right] - 2\mathrm{E}\left[|c|^2 |\widehat c|^2 \right]\right) \sum_{i=1}^D  |\langle i|X|i\rangle|^2
\nonumber \\
&&+
 D(D-1) \mathrm{E}\left[|c|^2 |\widehat c|^2 \right]\left|\mathbf{Tr\;}X\right|^2
 +
 D\mathrm{E}\left[|c|^4 \right] \left|\mathbf{Tr\;}X\right|^2
\nonumber \\&&
-
2 D(D-1)\mathrm{E}\left[|c|^2 |\widehat c|^2 \right]\left|\mathbf{Tr\;}X\right|^2 - 2D  \mathrm{E}\left[|c|^4 \right] \left|\mathbf{Tr\;}X\right|^2
\nonumber \\
&=&
 D^2 \mathrm{E}\left[|c|^2 |\widehat c|^2 \right] \mathbf{Tr\;} XX^\dagger +
 D\left(\mathrm{E}\left[|c|^2 |\widehat c|^2 \right] - \mathrm{E}\left[|c|^4 \right]\right)
 \left|\mathbf{Tr\;}X\right|^2
\nonumber \\&&
 +
 D^2\left(\mathrm{E}\left[|c|^4 \right] - 2\mathrm{E}\left[|c|^2 |\widehat c|^2 \right]\right) \sum_{i=1}^D  |\langle i|X|i\rangle|^2
\nonumber \\
&=&\left\{
\begin{array}{cc}
({D\mathbf{Tr\;}XX^\dagger-\left|\mathbf{Tr\;}X\right|^2})/{(D+1)}& \quad,\quad \hbox{Case A}\\
({D\mathbf{Tr\;}XX^\dagger-\left|\mathbf{Tr\;}X\right|^2})/{D}& \quad,\quad \hbox{Case B}\\
\mathbf{Tr\;}XX^\dagger-\sum_{i=1}^D |\langle i|X|i\rangle|^2& \quad,\quad \hbox{Case C}\\
\end{array}
\right.
,
\label{PRE4}
\end{eqnarray}
where the last line follows by substituting the corresponding  values of
$\mathrm{E}\left[|c|^2 \right]$,
$\mathrm{E}\left[|c|^2 |\widehat c|^2 \right]$, and
$\mathrm{E}\left[|c|^4 \right]$,
given in Table~\ref{tab1}.
Except for Case B where there is an irrelevant difference ($D+1$ instead of $D$) in the denominator,
the final expressions given by Eq.~(\ref{PRE4}) agree with the exact expressions for the variance
given in Table~\ref{tab1}.
Therefore, the expressions in Eq.~(\ref{PRE4}) for the mean square deviation $\mathrm{Var}\left[{\widetilde X}\right]$
lead to the same conclusion as the one obtained for the exact expressions listed in Table~\ref{tab1}.
\section{Application: density of states}\label{DOS}

To our knowledge, the first application of the random state technology was
for the calculation of the density of states (DOS) of an alloy~\cite{ALBE75}.
In this section, we present a discussion of this kind of application from a
more general perspective. An example implementation that can be used to reproduce the figures presented in this section and compute the DOS of many other lattices is available online at~\cite{JugitRST}.
In the following, we set $\hbar=1$ for convenience.

Given a state $|\Psi\rangle$, the local density of states $\mathrm{LDOS}_\Psi(\omega)$
of a system described by a Hamiltonian $H$ is defined by 
\begin{eqnarray}
\mathrm{LDOS}_\Psi(\omega) &=& \langle\Psi|\delta(\omega - H)|\Psi\rangle
=\frac{1}{2\pi}\int_{-\infty}^{+\infty} e^{i\omega t} \langle\Psi|e^{-itH} |\Psi\rangle \, dt
\nonumber \\
&\approx& \frac{1}{2\pi}\int_{-T}^{+T}  e^{i\omega t} \langle\Psi|e^{-itH} |\Psi\rangle \, dt
,
\label{DOS0}
\end{eqnarray}
where $[-T,T]$ denotes the relevant time interval (see below for a criterion on how to choose $T$ given $H$). In practice, $T$ is limited by
the amount of computational work it takes to calculate the matrix element in Eq.~(\ref{DOS0}).
If $|\Psi\rangle$ is taken to be a random state $|\Phi\rangle$, applying Eq.~(\ref{PRE2a}) yields,
\begin{eqnarray}
\mathrm{E}\left[
\mathrm{LDOS}_\Psi(\omega)
\right]
&=& \frac{1}{2\pi} \mathrm{E}\left[|c|^2 \right]
\int_{-\infty}^{+\infty} e^{i\omega t} \mathbf{Tr\;} e^{-itH} \, dt
= \mathrm{E}\left[|c|^2 \right] \sum_{n=1}^D \delta(\omega-E_n)
\nonumber \\
&=&
D\,\mathrm{E}\!\left[|c|^2 \right]\mathrm{DOS}(\omega)
,
\label{DOS0a}
\end{eqnarray}
where $\mathrm{DOS}(\omega)=\sum_{n=1}^D\delta(\omega-E_n)/D$ is, by definition, the density of states.
For numerical work, problems with large $D$ are demanding, but by
using the random state technology we can reduce the computational burden by writing
\begin{eqnarray}
\mathrm{DOS}(\omega)
\approx \frac{1}{2\pi D\,\mathrm{E}\!\left[|c|^2 \right]}\int_{-T}^{+T}  e^{i\omega t} \langle\Phi|e^{-itH} |\Phi\rangle \, dt
.
\label{DOS0b}
\end{eqnarray}
In practice, we perform the integration over time in Eq.~(\ref{DOS0}) by means
of the Discrete Fourier Transform (DFT).
To this end we rewrite Eq.~(\ref{DOS0}) as
\begin{eqnarray}
\mathrm{LDOS}_\Psi(k\pi/T)
&=&\frac{T}{2 \pi N}\sum_{j=-N}^{N-1}  e^{\pi ijk/N } \langle\Psi|e^{-ij\tau H} |\Psi\rangle
= \frac{T}{2 \pi N} \mathrm{DFT}\!\left[  \langle\Psi|e^{-ij\tau H} |\Psi\rangle \right],
\label{DOS1a}
\\
\mathrm{DOS}(k\pi/T)
&=&\frac{T}{2\pi N D\mathrm{E}\left[|c|^2 \right]}\sum_{j=-N}^{N-1}  e^{\pi ijk/N } \langle\Phi|e^{-ij\tau H} |\Phi\rangle
\nonumber \\
&=& \frac{T}{2 \pi N D\mathrm{E}\left[|c|^2 \right]} \mathrm{DFT}\!\left[ \langle\Phi|e^{-ij\tau H} |\Phi\rangle \right]
,
\label{DOS1b}
\end{eqnarray}
where $\tau=T/N$ is the time step at which we sample the function $\langle\Psi|e^{-it H} |\Psi\rangle$.
Accordingly, we obtain $\mathrm{DOS}(\omega)$ at frequencies $\omega=k\pi/T$ for $k=-N,\ldots,N-1$.
As before, we see that the random state technology reduces the amount
of computational work from ${\cal O}(D^2)$ to ${\cal O}(D)$.

\begin{figure}[t]
    \centering
    \includegraphics{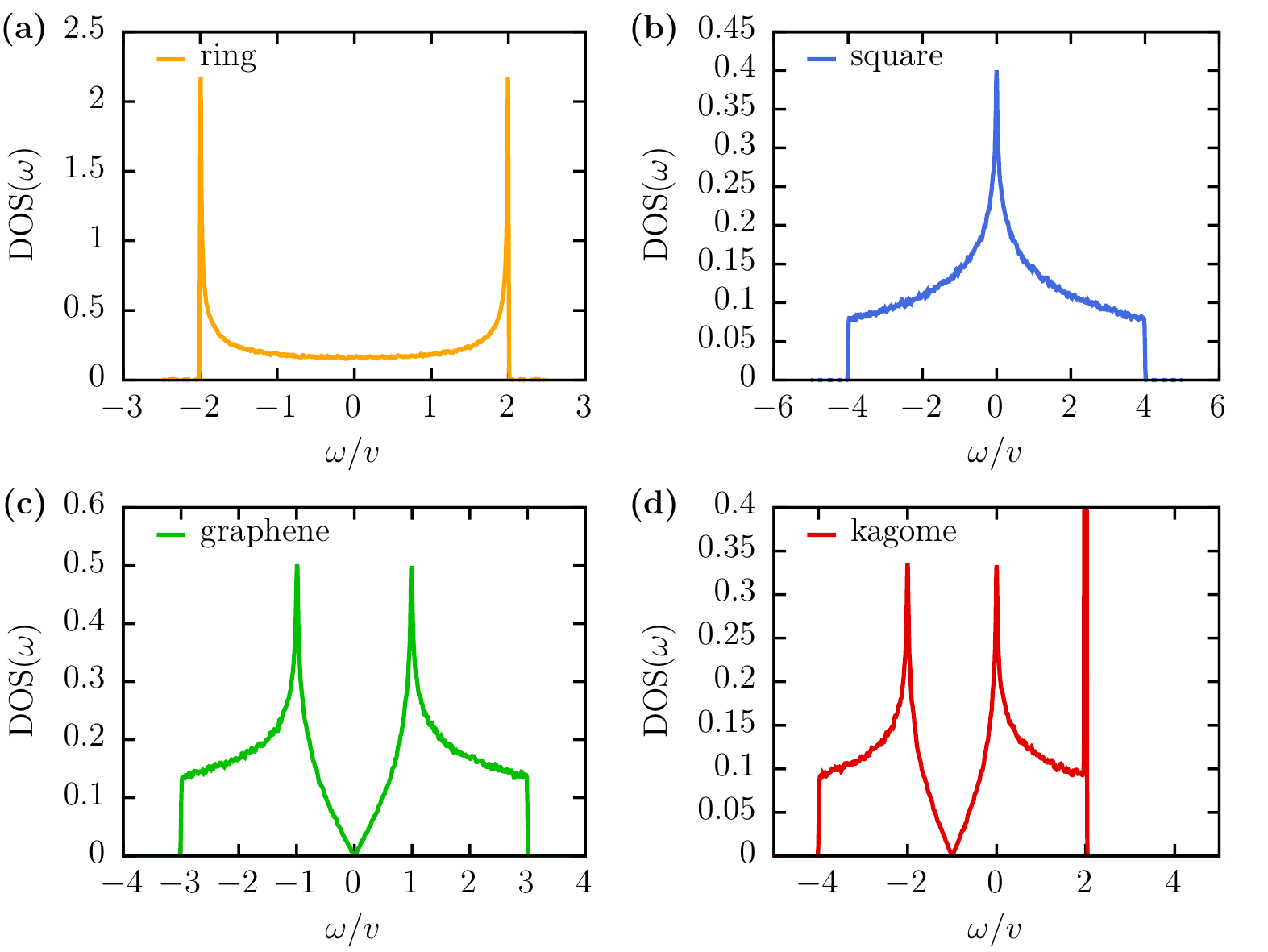} 
    \caption{(Color online) Density of states as a function of the frequency in units
    of the hopping constant $v_{mn}=v$ (see Eq.~(\ref{EX0})).
    (a) One-dimensional chain with $D=1000000$ sites;
    (b) two-dimensional square lattice with $D=1000000$ sites;
    (c) graphene lattice with $D=1080000$ sites;
    (d) kagome lattice with $D=1008200$ sites.
    All lattices have periodic boundary conditions.
    In all cases, $\mathrm{DOS}(\omega)$ has been obtained with the random state technology Eq.~(\ref{DOS1b}),
    where $N=1000$ and $T=N\tau$ with $\tau=0.8\pi/\Vert H\Vert_1$,
    and where $\Vert H\Vert_1=2v$ for the ring,
    $\Vert H\Vert_1=4v$ for the square and kagome lattices,
    and $\Vert H\Vert_1=3v$ for the graphene lattice.
    The TDSE~(\ref{TDSE}) is solved by using the second-order real-space product-formula algorithm~\cite{RAED87}
    with a time step $\tau/5$ (see Eq.~(\ref{EX1})).
    The source code to reproduce this figure can be downloaded at~\cite{JugitRST}.}
    \label{fig:dos_all}
\end{figure}

In practice, the most efficient method to compute $\langle\Psi|e^{-it H} |\Psi\rangle$ for $t=\tau,2\tau,\ldots,N\tau$ is to solve
the time-dependent Schr\"odinger equation (TDSE)
\begin{eqnarray}
i\frac{\partial}{\partial t} |\phi\rangle &=& H|\phi\rangle,
\label{TDSE}
\end{eqnarray}
using a suitable method such as the
Suzuki-Trotter product formula algorithm~\cite{RAED87,RAED06},
the Chebyshev polynomial algorithm~\cite{TALE84,DOBR03,WEIS06},
a Lanczos-iteration based method~\cite{PREL17,SCHN20}, or the fourth-order Runge-Kutta method~\cite{TARE13,STEIN14}.
The values of $\langle\Psi|e^{-it H} |\Psi\rangle$ for $t=-\tau,-2\tau,\ldots,-N\tau$ are obtained by using
$\langle\Psi|e^{+it H} |\Psi\rangle=\langle\Psi|e^{-it H} |\Psi\rangle^\ast$.

In numerical work, the matrix $H$ is bounded.
Denoting by $\Vert H\Vert$ the maximum absolute value
of the eigenvalues of $H$ (i.e., the spectral norm or $2$-norm of $H$), the function $\mathrm{LDOS}_\Psi(\omega)=0$ if $|\omega| > \Vert H \Vert$,
that is this function is band limited.
Then Nyquist's sampling theorem tells us that we should sample with a time step $\tau < \pi/\Vert H \Vert$
in order to cover the full range of eigenvalues and avoid aliasing~\cite{PRES03}.
As $\tau = T/N$, this means that the time interval should be chosen such that $T<N\pi/\Vert H\Vert$.
Eigenvalues that differ by less than $\pi/T$ cannot be resolved.

In summary, the number of samples $2N$ determines the resolving power of the method, and the time interval $2T$ has to satisfy $T<N\pi/\Vert H\Vert$ to cover the full range of eigenvalues.
For many problems of interest, $\Vert H \Vert$ is not known, but it can be replaced by a bound
such as $\Vert H \Vert\le \Vert H \Vert_1 = \max_{1\le j \le D} \sum_{i=1}^D |H_{i,j}|$
(i.e., the maximum absolute column sum, which is closely related to the concept of the Gershgorin disk~\cite{GERS31})
which is easy to compute.

We illustrate the application of the random state-based technique by computing the DOS for a single particle hopping on a one-dimensional ring, a square, a graphene, and a kagome lattice.
The Hamiltonian is given by
\begin{eqnarray}
H&=&\sum_{\langle m,n \rangle} v^{}_{mn}\left(a^+_m a^{}_n + \mathrm{h.c.}\right)+\sum_{m} w^{}_m a^+_m a^{}_m
,
\label{EX0}
\end{eqnarray}
where $a^+_m$ ($a^{}_n$) is the fermion creation (annihilation) operator of a particle
at site $m$ ($n$), $v^{}_{mn}$ are the hopping integrals, and $w^{}_m$ is an on-site potential.
The sum over $\langle m,n \rangle$ is over all nearest-neighbor bonds of the lattice with $m<n$.
Note that because of the restriction to a one-particle problem, it does not matter whether we use fermion or boson operators.

Being a one-particle problem, we can compute all physical properties if we can diagonalize the quadratic form defined by the matrix $\mathbf{V}$ with elements $v^{}_{mn}+\delta^{}_{mn}w^{}_m$.
In some cases (see below), $\mathbf{V}$ can be diagonalized analytically for any size or shape of the lattice. If that is not possible, we can diagonalize $\mathbf{V}$ numerically if its dimension is not too large, in practice limiting this approach to matrices with a dimension in the range 10000 to 100000. The random state technology can handle
(much) larger matrices~\cite{HAMS00}.

\begin{figure}[th]
    \centering
    \includegraphics{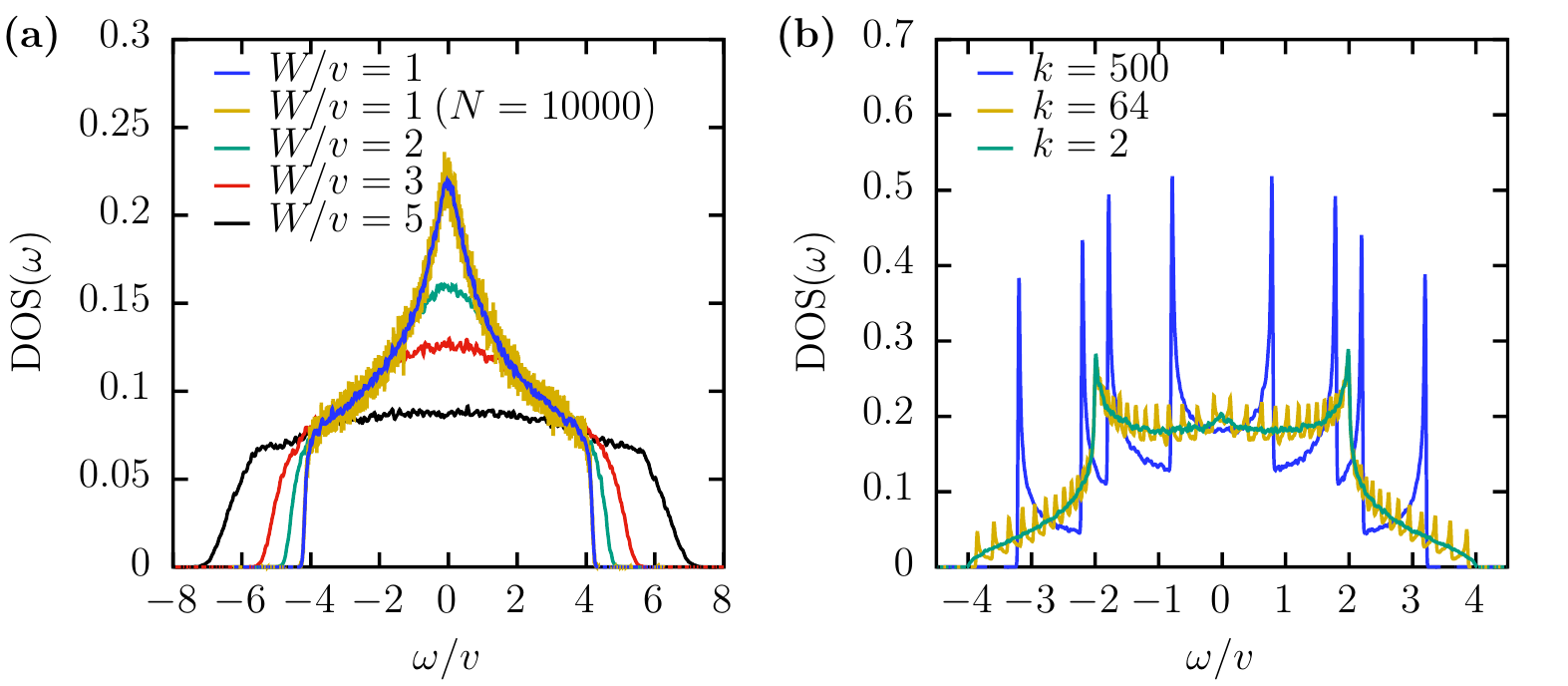} 
    \caption{(Color online) Density of states for square lattices of dimension $D=L\times L=1000000$
    for which all the eigenvalues of $\mathbf{V}$ cannot be computed analytically or by numerical diagonalization.
    (a) Anderson model with random energies $w_m$ drawn uniformly from $[-W,W]$;
    (b) sinusoidal bond model with anisotropic bond strengths $v_x = v \sin^2(\pi k x/2L)$
    and $v_y = v$ where $x=1,\ldots,L-1$ enumerates the horizontal bonds of the square lattice
    (the case $k=2$ corresponds to strong bonds in the center and weak bonds at the boundaries and
    the case $k=500$ corresponds to alternating weak and strong bonds with a period of four sites).
    In all cases, $\mathrm{DOS}(\omega)$ has been obtained with the random state technology Eq.~(\ref{DOS1b}),
    where $N=1000$ (or $N=10000$ if stated in the legend) and $T=N\tau$ with $\tau=0.8\pi/\Vert H\Vert_1$,
    where $\Vert H\Vert_1=4v+\vert W\vert$ for the Anderson model and $\Vert H\Vert_1=4v$ for the sinusoidal bond model.
    The TDSE~(\ref{TDSE}) is solved by using the second-order real-space product-formula algorithm~\cite{RAED87}
    with a time step $\tau/5$.
    The source code to reproduce this figure can be downloaded at~\cite{JugitRST}.}
    \label{fig:dos_special}
\end{figure}

The numerical procedure consists of three steps (see also the demonstration available at~\cite{JugitRST}):
\begin{enumerate}
\item
Generate a Gaussian random state $|\Phi\rangle=\sum_{m} c_m^{+}  a^+_m|0\rangle$ and a copy: $|\Psi\rangle = |\Phi\rangle$.
\item
Use the first-order real-space Suzuki-Trotter formula~\cite{RAED87}
\begin{eqnarray}
e^{-i\widetilde\tau H}&\approx&U_1(\widetilde\tau)=\prod_{\langle m,n \rangle} \exp\left[-i\widetilde\tau v^{}_{mn} (a_m^{+}a_n^{} + \mathrm{h.c.})\right]
\prod_{m} \exp\left[-i\widetilde\tau  w^{}_m a_m^{+}a_m^{}\right]
\;,
\label{EX1}
\end{eqnarray}
to construct the second-order approximation
$U_2(\widetilde\tau)=
U_1^{\mathrm{T}}(\widetilde\tau/2)U^{}_1(\widetilde\tau/2)$.
The order in which the matrix exponentials in Eq.~(\ref{EX1}) appear can be chosen freely but once chosen, this order has to be kept fixed.
Solve the TDSE by repeating $|\Psi(t+\tau)\rangle\leftarrow \left(U_2(\widetilde\tau)\right)^l|\Psi(t)\rangle$,
where $\widetilde\tau=\tau/l$ and with $l\ge1$ controlling  the accuracy of the product formula approximation.
During the time stepping, at each time $t=-\tau,-2\tau,\ldots,-N\tau$, store the values
$f(t)=\langle\Phi |\Psi\rangle$ and $f(-t)=f(t)^*$.
\item
Compute the DFT of $f(t)g(t)$ where $g(t)$ is a Gaussian window function.
\end{enumerate}
In Fig.~\ref{fig:dos_all}, we present the (well-known) results for the DOS of the four mentioned lattices with $v^{}_{mn}=v$ and $w^{}_{m}=0$, as obtained by the above procedure.

For problems for which all eigenvalues of $\mathbf{V}$ are known in analytical or numerical form, such as the examples for a particle moving on one of the lattices considered above, there exist much more efficient methods to compute the (L)DOS than the one based on solving the TDSE.
For instance, simply counting the number of times that the energies fall in bins $[E,E+\Delta E]$ gives the DOS in the form of a histogram.
However, the method based on random states in combination with the solution of the TDSE excels when the calculation of {\bf all} the eigenvalues of $H$ is prohibitive.
We illustrate this point by Fig.~\ref{fig:dos_special} where we present DOS results for the two-dimensional Anderson model of localization and a square lattice with rather exotic anisotropic hopping integrals.

\section{Application: quantum statistical physics}\label{QSP}

An important class of problems for which the random state technology can be put to good use
is the calculation of thermal equilibrium averages of observables of a quantum system, described in terms
of its Hamiltonian $H=H^\dagger$.
If $Y=Y^\dagger$ represents the matrix of such an observable, the task is to compute
\begin{eqnarray}
\langle Y \rangle
&=&\frac{\mathbf{Tr\;}e^{-\beta H}Y}{\mathbf{Tr\;}e^{-\beta H}}=
\frac{\mathbf{Tr\;}e^{-\beta H/2}Ye^{-\beta H/2}}{\mathbf{Tr\;}e^{-\beta H}}
,
\label{QSP0}
\end{eqnarray}
where $\beta$ denotes the inverse temperature in units of $1/k_{\mathrm{B}}$,
and we used invariance of the trace under cyclic permutation ($\mathbf{Tr\;} AB=\mathbf{Tr\;}BA$)
to bring $\langle Y \rangle$ into a form that is most suited for the application
of the random state technology.
According to the general recipe laid out in Section~\ref{PRE}, the idea is to replace the calculation of the traces by
the calculation of only one matrix element.

The first step in the procedure is to generate a random state $|\Phi\rangle$, by using
for instance one of the methods given in Appendix~\ref{PICK}.
Next, we compute the so-called random thermal state defined by
\begin{eqnarray}
|\Phi_\beta\rangle &\equiv& e^{-\beta H/2}|\Phi\rangle
,
\label{QSP1}
\end{eqnarray}
whereby it is implicitly understood that one has an efficient and accurate
algorithm to compute $e^{-\beta H/2}|\Phi\rangle$.
In practice, the Chebyshev polynomial representation of $e^{-\beta H/2}$ can be used to compute,
with close to optimal efficiency, $e^{-\beta H/2}|\Phi\rangle$ to almost machine precision~\cite{TALE84,LEFO91,IITA97,DOBR03}.
The final computational step is then to estimate the thermal equilibrium expectation value
according to
\begin{eqnarray}
\langle Y \rangle
&=&\frac{\mathbf{Tr\;}e^{-\beta H/2}Ye^{-\beta H/2}}{\mathbf{Tr\;}e^{-\beta H}}
\approx
\frac{\langle\Phi_\beta|Y|\Phi_\beta\rangle }{\langle\Phi_\beta|\Phi_\beta\rangle}
=
\frac{\langle\Phi|X|\Phi\rangle }{\langle\Phi|Z|\Phi\rangle}
,
\label{QSP2}
\end{eqnarray}
where $X=e^{-\beta H/2} Y e^{-\beta H/2}$ and $Z=e^{-\beta H}$.
As before, in order to show that the replacement of the traces by single matrix elements
makes sense, one has to show that the variance of the right-hand side of Eq.~(\ref{QSP2}) is small.
Because of the presence of $Z$ in the denominator, we are unable to calculate the average
or the variance of ${\langle\Phi|X|\Phi\rangle }/{\langle\Phi|Z|\Phi\rangle}$ exactly, except
if $\beta=0$ in which case the problem reduces to the one treated in Section~\ref{PRE}.
Therefore, we have to resort to the approximate treatment based on Eqs.~(\ref{AE0}) and~(\ref{AE1}).
To calculate the first three terms in Eqs.~(\ref{AE0}) and~(\ref{AE1}), we only need
\begin{eqnarray}
\mathrm{E}\left[\langle \Phi|X|\Phi\rangle\right]&=&\mathrm{E}\left[|c|^2 \right] \mathbf{Tr\;}X
= \mathrm{E}\left[|c|^2 \right]\mathbf{Tr\;}e^{-\beta H} Y = \mathrm{E}\left[|c|^2 \right]\mathbf{Tr\;} Z Y
\nonumber \\
\mathrm{E}\left[\langle \Phi|Z|\Phi\rangle\right]&=&\mathrm{E}\left[|c|^2 \right] \mathbf{Tr\;}Z
\nonumber \\
\mathrm{E}\left[\langle \Phi|X|\Phi\rangle^2\right]&=&
\mathrm{E}\left[|c|^2 |\widehat c|^2 \right] \left( \mathbf{Tr\;} (ZY)^2 +\left(\mathbf{Tr\;}ZY\right)^2\right)
 +
 \left(\mathrm{E}\left[|c|^4 \right] - 2\mathrm{E}\left[|c|^2 |\widehat c|^2 \right]\right)
 \sum_{i=1}^D  \langle i|X|i\rangle^2
\nonumber \\
\mathrm{E}\left[\langle \Phi|Z|\Phi\rangle^2\right]&=&
\mathrm{E}\left[|c|^2 |\widehat c|^2 \right] \left( \mathbf{Tr\;} Z^2 +\left(\mathbf{Tr\;}Z\right)^2\right)
 +
 \left(\mathrm{E}\left[|c|^4 \right] - 2\mathrm{E}\left[|c|^2 |\widehat c|^2 \right]\right)
 \sum_{i=1}^D  \langle i|Z|i\rangle^2
\nonumber \\
\mathrm{E}\left[\langle \Phi|X|\Phi\rangle\langle \Phi|Z|\Phi\rangle\right]&=&
\mathrm{E}\left[|c|^2 |\widehat c|^2 \right] \left( \mathbf{Tr\;} ZY\mathbf{Tr\;} Z +\mathbf{Tr\;}Z^2Y \right)
\nonumber \\
&& +
 \left(\mathrm{E}\left[|c|^4 \right] - 2\mathrm{E}\left[|c|^2 |\widehat c|^2 \right]\right)
 \sum_{i=1}^D  \langle i|X|i\rangle\langle i|Z|i\rangle
.
\label{QSP3}
\end{eqnarray}
It is now straightforward to use the expressions for the moments given in Table~\ref{tab1}
and obtain the expressions for the average and variance of ${\langle\Phi|X|\Phi\rangle }/{\langle\Phi|Z|\Phi\rangle}$.
For example, in Case B we find
\begin{eqnarray}
\mathrm{E}\left[
\frac{\langle\Phi|X|\Phi\rangle}{\langle\Phi|Z|\Phi\rangle}
\right]
&\approx&
\frac{\mathbf{Tr\;} Z Y}{\mathbf{Tr\;} Z}
-\frac{\mathbf{Tr\;} Z^2 Y}{\left(\mathbf{Tr\;} Z\right)^2}
+
\frac{\mathbf{Tr\;} Z Y}{\mathbf{Tr\;} Z}
\frac{\mathbf{Tr\;} Z^2}{\left(\mathbf{Tr\;} Z\right)^2}
\nonumber \\&=&
\frac{\mathbf{Tr\;} Z Y}{\mathbf{Tr\;} Z}
+\frac{\mathbf{Tr\;} Z^2}{\left(\mathbf{Tr\;} Z\right)^2}
\bigg\{
\frac{ \mathbf{Tr\;} Z Y}{\mathbf{Tr\;} Z}
-
\frac{\mathbf{Tr\;} Z^2 Y}{\mathbf{Tr\;} Z^2}
\bigg\}
,
\label{QSP4}
\end{eqnarray}
and
\begin{eqnarray}
\mathrm{Var}\left[
\frac{\langle\Phi|X|\Phi\rangle}{\langle\Phi|Z|\Phi\rangle}
\right]
&\approx&
\frac{\mathbf{Tr\;} Z^2}{\left(\mathbf{Tr\;} Z\right)^2}
\bigg\{
\frac{\mathbf{Tr\;} (Z Y)^2}{\mathbf{Tr\;} Z^2}
-2
\frac{\mathbf{Tr\;} Z Y}{\mathbf{Tr\;} Z}
\frac{\mathbf{Tr\;} Z^2 Y}{\mathbf{Tr\;} Z^2}
+
\left(\frac{\mathbf{Tr\;} Z Y}{\mathbf{Tr\;} Z}\right)^2
\bigg\}
.
\label{QSP5}
\end{eqnarray}
The absolute values of the contributions in the curly brackets are readily shown
to be bounded by $2\Vert Y \Vert$ (Eq.~(\ref{QSP4}))  and $4\Vert Y \Vert^2$ (Eq.~(\ref{QSP5})).
In practice, we are only interested in observables $Y$ for which $\Vert Y \Vert={\cal O}(\log D)$.
Therefore, the magnitude of correction (second term of Eq.~(\ref{QSP5})) to
and the variance of the thermal average of $Y$ is primarily determined by
the factor ${\mathbf{Tr\;} Z^2}/{\left(\mathbf{Tr\;} Z\right)^2}=1/D$.
As $\mathbf{Tr\;} Z=\mathbf{Tr\;} e^{-\beta H} = e^{-\beta F(\beta)}$ where
$F(\beta)$ is the free energy, we have, in general,
\begin{eqnarray}
\frac{1}{D}\le \frac{\mathbf{Tr\;} Z^2}{\left(\mathbf{Tr\;} Z\right)^2}
&=& e^{-2\beta (F(2\beta) - F(\beta))}\le 1
,
\label{QSP6}
\end{eqnarray}
where the first inequality follows from applying the Schwarz inequality to the inner product $\mathbf{Tr\;} A^\dagger B$.
The second inequality follows by writing $\mathbf{Tr\;} Z=\sum_{i=1}^D \exp(-\beta E_i)$,
$\mathbf{Tr\;} Z^2=\sum_{i=1}^D \exp(-2\beta E_i)$ and noting that
\begin{eqnarray}
\frac{\mathbf{Tr\;} Z^2}{\left(\mathbf{Tr\;} Z\right)^2}=
\frac{1}{1+\left(\sum_{i\not=j}^D \exp(-\beta( E_i+E_j)\right)/\left(\sum_{i=1}^D \exp(-2\beta E_i)\right) }\le1
.
\label{QSP7}
\end{eqnarray}
From the left-hand side of Eq.~(\ref{QSP6}) and Eq.~(\ref{QSP7}), it follows immediately that $F(2\beta) \geq F(\beta)$.
This is in concert with the second law of thermodynamics.
Furthermore, since the free energy is an extensive quantity,
${\mathbf{Tr\;} Z^2}/{\left(\mathbf{Tr\;} Z\right)^2}$ vanishes exponentially with increasing system size.
If we assume that $\Vert Y \Vert$ is at most proportional to the system size
then, using the fact that the terms in the curly brackets are bounded and
their prefactors vanish exponentially with increasing system size,
the correction to the average and the variance itself vanish in the same manner.

As $F(2\beta)\approx F(\beta)\approx E_0$ for $\beta\to\infty$, Eq.~(\ref{QSP6}) suggests
that the efficiency of the random state approach is lost when $\beta\to\infty$.
Indeed, when $\beta\to\infty$, the random thermal state turns into
the ground state $|0\rangle$ of $H$, that is $\lim_{\beta\to\infty}|\Phi_\beta\rangle \to |0\rangle$,
but we cannot recommend calculating the ground state by projection with $e^{-\beta H/2}$
because for large $\beta$, the efficiency of this projection is inferior to that of the Lanczos method, for instance.

It is instructive to illustrate these general features through a very simple example.
To this end, we compute $e^{-2\beta (F(2\beta) - F(\beta))}=\mathbf{Tr\;} Z^2/\left(\mathbf{Tr\;} Z\right)^2$
for a system of $N$ non-interacting spins, described by the Hamiltonian $H=-\Omega\sum_{i=1}^N \sigma^z$
where $\sigma^z$ is the $z$-component of the Pauli-spin matrices, having eigenvalues $\pm1$.
A simple calculation yields
\begin{eqnarray}
\frac{\mathbf{Tr\;} Z^2}{\left(\mathbf{Tr\;} Z\right)^2}
&=&\left(\frac{1+\tanh^2\beta\Omega}{2}\right)^N
,
\label{QSP6z}
\end{eqnarray}
which shows that $e^{-2\beta (F(2\beta) - F(\beta))}$
smoothly increases from $D^{-1}=2^{-N}$ to one when $\beta$ increases from zero to infinity.
For $\beta<\infty$, Eq.~(\ref{QSP6z}) vanishes exponentially with increasing $N$.

In summary, the factor ${\mathbf{Tr\;} Z^2}/{\left(\mathbf{Tr\;} Z\right)^2}$ determines
the rate at which the variance Eq.~(\ref{QSP5}) vanishes with the system size but this factor
can, depending on the temperature $1/\beta$, vary from $1/D$ to close to one.
In the latter case, the random state technology looses its efficiency.

\subsection{Specific heat}\label{SpH}

In this section, we use the random state technology to calculate
the specific heat of frustrated and non-frustrated spin-$1/2$, nearest-neighbor, antiferromagnetic Heisenberg models.
The Hamiltonian is defined by
\begin{equation}
H=-J\sum_{\left < i,j \right >} {S_i}\cdot{S_j},
\end{equation}
where ${S_i}=(S_i^x,S_i^y,S_i^z)$ is the spin-$1/2$ operator at the $i$-th site, $\left < i,j \right >$
refers to the nearest-neighbor sites on the lattice, and $J=-1$ is the antiferromagnetic coupling.
The specific heat is calculated from the fluctuation of the energy, i.e., as
$C=\frac{1}{N}\beta^2 \left(\left < H \right >^2-\left < H^2 \right >\right)$,
where $N$ is the number of lattice sites.

Numerically, there are only two important operations involved.
The first is the preparation of the random thermal state
(projection on a random state, Eq.~(\ref{QSP1}))
and the second is the operation $H|\Phi_\beta\rangle$.
There are several ways to do the projection, e.g., by the power method~\cite{WILK65},
a product formula algorithm~\cite{RAED87,RAED06},
or the Chebyshev polynomial algorithm~\cite{TALE84,DOBR03,WEIS06}.
These methods only need storage for a few state vectors, and the numerical errors are well under control.
Another method, widely used in the solid-state physics community, is the finite-temperature Lanczos method~\cite{PREL17}.
This method uses the eigenvalues and eigenstates obtained from the standard Lanczos method to calculate the matrix element
of the operator $Y$. It requires storage of wave functions proportional to the Lanczos iteration order $M$
and reproduces the results of the high-temperature expansion up to order $M$.
According to the theory given in the previous section, the errors due to the use of the random
state are bounded (simply replace $Y$ by $H$ or $H^2$) and vanish as $D\to\infty$.

\begin{figure}[t]
\includegraphics[width=0.6\columnwidth]{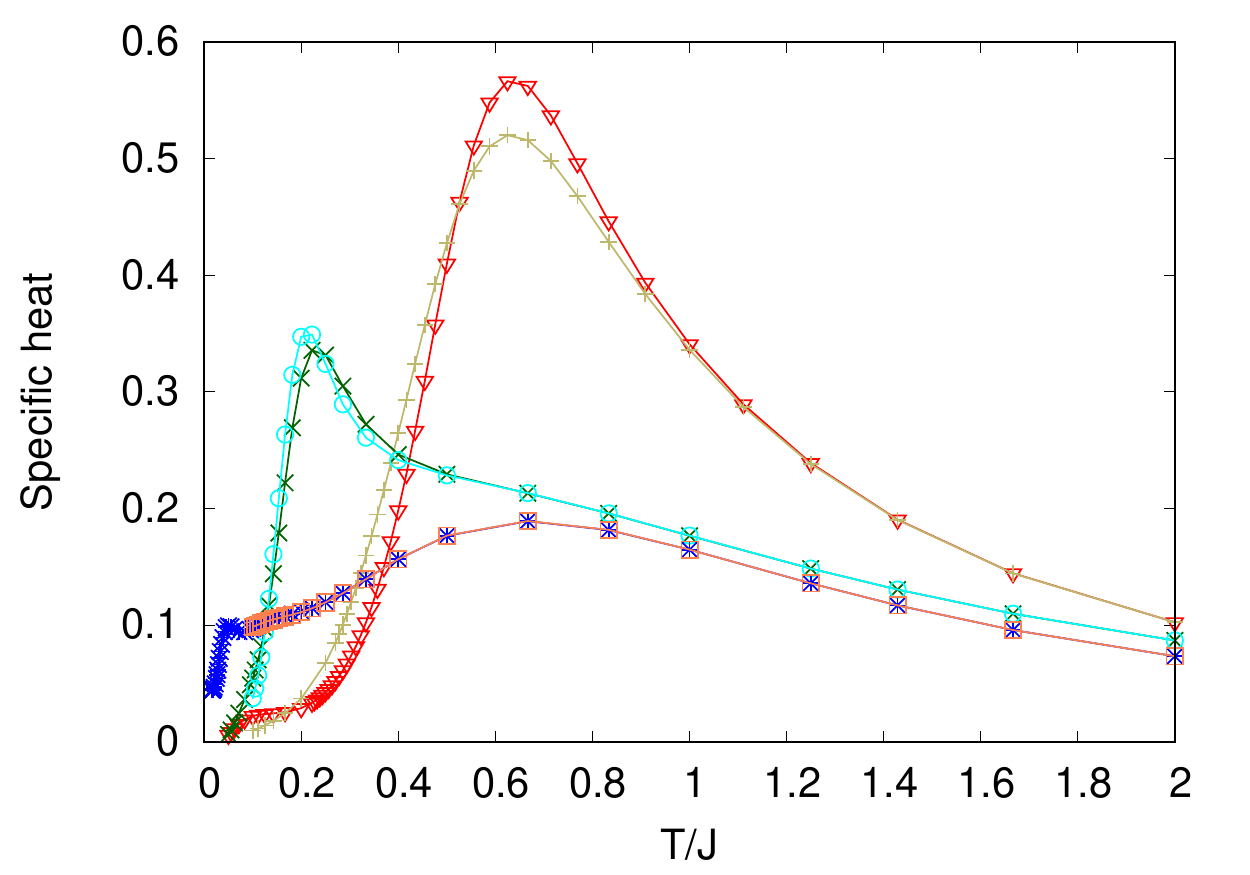} 
\caption{(Color online) Specific heat for the square (triangles and pluses), triangular (crosses and circles),
and kagome (stars and squares) lattice with $N=36$ spins. For each lattice, we show results obtained by using two different seeds to generate the initial random state.}
\label{figsph}
\end{figure}

Figure~\ref{figsph} shows the results for the specific heat for three different lattices,
namely, the square, triangular, and kagome lattice.
For all cases, the lattice size $N= 36$ and periodic boundaries are used.
For both triangular and kagome lattices, the shape of the lattice is a rhombus.
Within the range of temperatures covered,
we clearly see that for the square and triangular lattice,
there is only one peak in the specific heat, located around $T/J=0.6$ and $T/J=0.2$, respectively.
The specific heat for the kagome lattice shows multiple peaks.
The first peak is around $T/J=0.6$.
The second peak is around $T/J=0.1$, shoulder-like and greatly reduced
compared to results obtained for system sizes less than $N=36$ (data not shown).

Our results are consistent with those obtained from exact diagonalization~\cite{ELST94},
finite-temperature Lanczos method (FTLM)~\cite{SCHN18,PREL18},
and the transfer-matrix quantum Monte Carlo methods~\cite{NAKA95}.
Shimokawa and Kawamura used the random state technology
to calculate the finite-temperature properties of the
antiferromagnetic Heisenberg model on the kagome lattice up to size $36$~\cite{SHIM16}.
Sugiura and Shimizu developed the same method and demonstrated its use by
calculating the specific heat on the same model up to size $30$~\cite{SUGI13}.
At even lower temperatures, a third peak appears,
see Ref.~\cite{SCHN18}, in which the finite-temperature Lanczos method was used to calculate the thermal properties
for a kagome lattice with up to $N=42$ spins.

As the temperature decreases further,
the factor $\mathbf{Tr\;} Z^2/\left(\mathbf{Tr\;} Z\right)^2$ approaches one.
The corresponding loss of statistical accuracy can only be compensated for
by averaging over different realizations of the random states~\cite{HAMS00}, see also appendix~\ref{SRS}.
As this paper focuses on the basic principles of the random state technology,
we do not pursue the aspect of averaging over different realizations of the random states here
and therefore, we do not calculate the specific heat for $T/J<0.01$.
Instead, as an illustration,
we also include in Fig.~\ref{figsph} the results obtained by using different random states, for temperatures down to $T/J=0.1$.
Figure.~\ref{figsph} shows that the data for the kagome lattice, obtained
using two different initial random states,
agree with each other up to $T/J=0.1$, while the results
for the other two lattices only agree up to temperatures of about $T/J=1$.
This can be understood as follows.
At sufficiently low temperatures, we have
$\mathbf{Tr\;} Z^2/\left(\mathbf{Tr\;} Z\right)^2\approx(1+e^{-2\beta(E_1-E_0)})/(1+e^{-\beta(E_1-E_0)})^2$ where $E_1-E_0$ is the gap between the ground state and the first excited state.
The gaps for square, triangular, and kagome lattice with $N=36$ sites are $E_1-E_0=0.29$, $E_1-E_0=0.34$, and  $E_1-E_0=0.01$ respectively.
Therefore, for fixed $N$, the temperature at which
$\mathbf{Tr\;} Z^2/\left(\mathbf{Tr\;}Z\right)^2$
approaches one (and the variance may become large) is about 30 times higher for the
square and triangular lattice than for the
kagome lattice.
In other words, if we want to compute temperature-dependent averages with approximately the same accuracy and $\exp[-\beta(E_1-E_0)]$
becomes small, it is necessary to average over more than one realization of the random state.

For extremely low temperatures, the specific heat can be calculated
through the exact low-lying eigenvalues which can be obtained by several methods,
such as Lanczos-based algorithms~\cite{CULL02} and the Sakurai-Sugiura method~\cite{SAKU03}.
For moderately low temperatures, Morita and Tohyama proposed two algorithms improved
from FTLM by utilizing the exact low-lying eigenvalues and eigenvectors~\cite{MORI20}.
The first so-called replaced FTLM algorithm replaces the energies by the exact low-lying energies in the FTLM.
The second so-called orthogonalized FTLM is to start the FTLM by random states orthogonal to all the exact low-lying eigenvectors.
The effectiveness of these algorithms is illustrated by results for the specific heat and structure factor
for Kitaev-Heisenberg models on kagome and triangular lattices for systems with up to $N=36$ spins.
The second algorithm can be easily adapted to the projection method used to calculate the specific heat in the present paper.

\section{Application: quantum dynamics}\label{QDYN}

Random state technology can be used for the calculation of the expectation value of time- or frequency
dependent observables as well.
In this section, we illustrate this fact by showing calculations of the current-current correlation, density-density
correlations, and ESR spectrum for spin-$1/2$ models.
For examples of the application of the random state technology to the
calculations of the dynamic properties of 2D materials see Refs.~\cite{YUAN10d,YUAN10e}

\subsection{Current-current correlation}

In this subsection we focus on the current-current autocorrelation function defined by
\begin{equation}
C(t)=\frac{\mathbf{Tr}e^{-\beta H}j(t)j(0)}{\mathbf{Tr}e^{-\beta H}},
\end{equation}
where $j$ is the (problem-dependent) current and $j(t)=e^{itH}je^{-itH}$.
If we set $Y=j(t)j(0)$, it is obvious that the random state technology can be applied to calculate the current-current
correlation, Eq.~(\ref{QSP5}) guaranteeing
that the variance of $C(t)$ vanishes exponentially with increasing system size.
In numerical work, we introduce two auxiliary states
\begin{eqnarray}
\left |\phi(t)\right> &=& e^{-iHt}e^{-\beta H/2} \left |\Phi\right>, \\
\left |\varphi(t)\right> &=& e^{-iHt}j e^{-\beta H/2} \left |\Phi\right>,
\end{eqnarray}
and obtain the autocorrelation function $C(t)$ by computing
\begin{equation}
C(t)=\frac{\left <\phi(t)\right | j\left |\varphi(t)\right>} {\left <\Phi\right | e^{-\beta H} \left |\Phi\right>}.
\end{equation}
As already mentioned, the action of the time evolution operator $e^{-iHt}$
on a state vector can be computed by algorithms such as the
fourth-order Runge-Kutta scheme~\cite{TARE13,STEIN14}, the Suzuki-Trotter product formula algorithm~\cite{RAED87,RAED06},
and the Chebyshev polynomial algorithm~\cite{TALE84,DOBR03,WEIS06}.

We take as an example the 1D spin-$1/2$ Heisenberg XXZ ring defined by
\begin{equation}
H=-J\sum_i S_i^x S_{i+1}^x +S_i^y S_{i+1}^y + \Delta S_i^z S_{i+1}^z,
\label{Hxxz}
\end{equation}
where $J=-1$ sets the energy scale and $\Delta=1.5$ is the anisotropy.
From the lattice version of the continuity equation, it follows that the spin current operator $j$
is given by
\begin{equation}
j=-J\sum_i S_i^x S_{i+1}^y -S_i^y S_{i+1}^x .
\end{equation}
Spin transport properties can be obtained by Fourier transform of the current-current correlation function.

\begin{figure}[t]
\includegraphics[width=0.6\columnwidth]{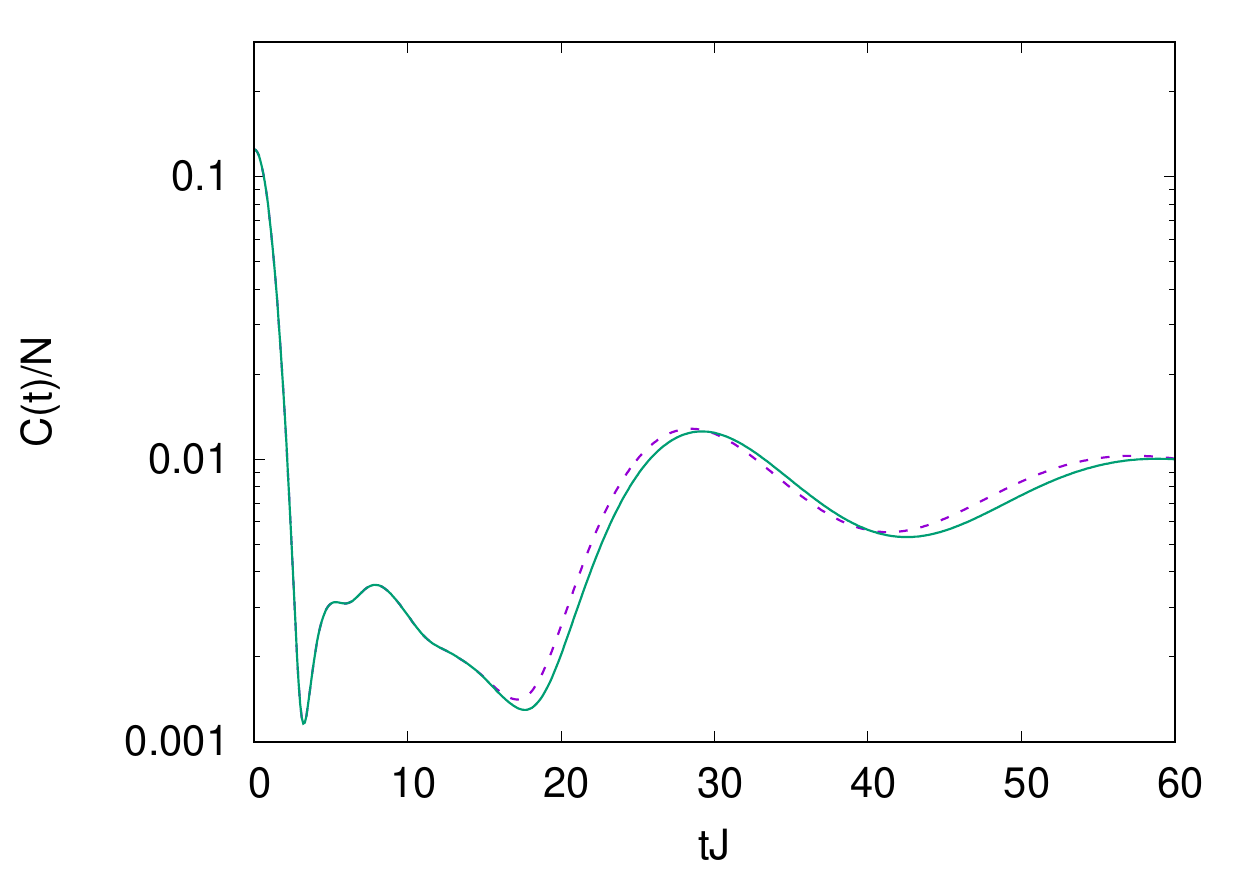} 
\caption{(Color online) Current-current autocorrelation function normalized by the lattice size
for the 1D XXZ model with $\Delta=1.5$ for system sizes $N=37$ (dashed line) and $N=38$ (solid line).}
\label{figcurrent}
\end{figure}

Figure~\ref{figcurrent} shows the results of the current-current autocorrelation function
for two system sizes $N=37$, $38$ obtained by using the random state technology.
The Suzuki-Trotter product formula algorithm~\cite{RAED87,RAED06} was used
to compute the time evolution of the two state vectors.
For simplicity, we choose $\beta=0$.
The data for the two different system sizes follow the same curve up to $tJ\approx 15$ and
then start to deviate from each other.
This suggests that the current-current autocorrelation function shows strong finite size effects, except
for relatively short times.
The data presented here are in concert with earlier results~\cite{STEI14b,KARR12,KARR13}.
A way to alleviate this effect is to apply the idea of numerical
linked cluster expansion, which allows the accurate estimation of $C(t)$
for much longer times~\cite{RICH19b,RICH19c}.

\subsection{Density-density correlation}

The density-density correlation is intimately related to the current-current correlation via the continuity equation
but is of interest by itself. It is defined by
\begin{equation}
C(t)=\frac{\mathbf{Tr}\;e^{-\beta H}n(t)n(0)}{\mathbf{Tr}e^{-\beta H}},
\end{equation}
where $n$ is the density operator and $n(t)=e^{itH} n(0) e^{-itH}$.
We only have to set $Y=n(t)n(0)$ to see that also
in the case of the density-density correlation function, the random state technology can safely be applied.
Numerically we can calculate this quantity by exactly the same technique as
the one used for the current-current correlation, namely from the time evolution of two projected state vectors.
For $\beta=0$, we can eliminate one of the time evolutions.
For a local density operator, it is always possible to find its square root, either analytically or if necessary
numerically.
Then, it is straightforward to show that
\begin{equation}
C(t)=\frac{\mathbf{Tr}\;\sqrt{n(0)}n(t)\sqrt{n(0)}}{D}
\approx  \left < \Phi\right | \sqrt{n(0)}n(t)\sqrt{n(0)} \left |\Phi\right>
= \left < \psi(t)\right | n(0)  \left | \psi(t)\right >,
\end{equation}
where $\left | \psi(t)\right >=e^{-itH}\sqrt{n(0)} \left |\Phi\right>$ is an unnormalized state vector.
Note that if the density operator has negative eigenvalues, the square root of this operator is obtained by shifting
the eigenvalues by the lowest eigenvalue, and the above formula should be changed accordingly.

\begin{figure}[t]
\includegraphics[width=0.6\columnwidth]{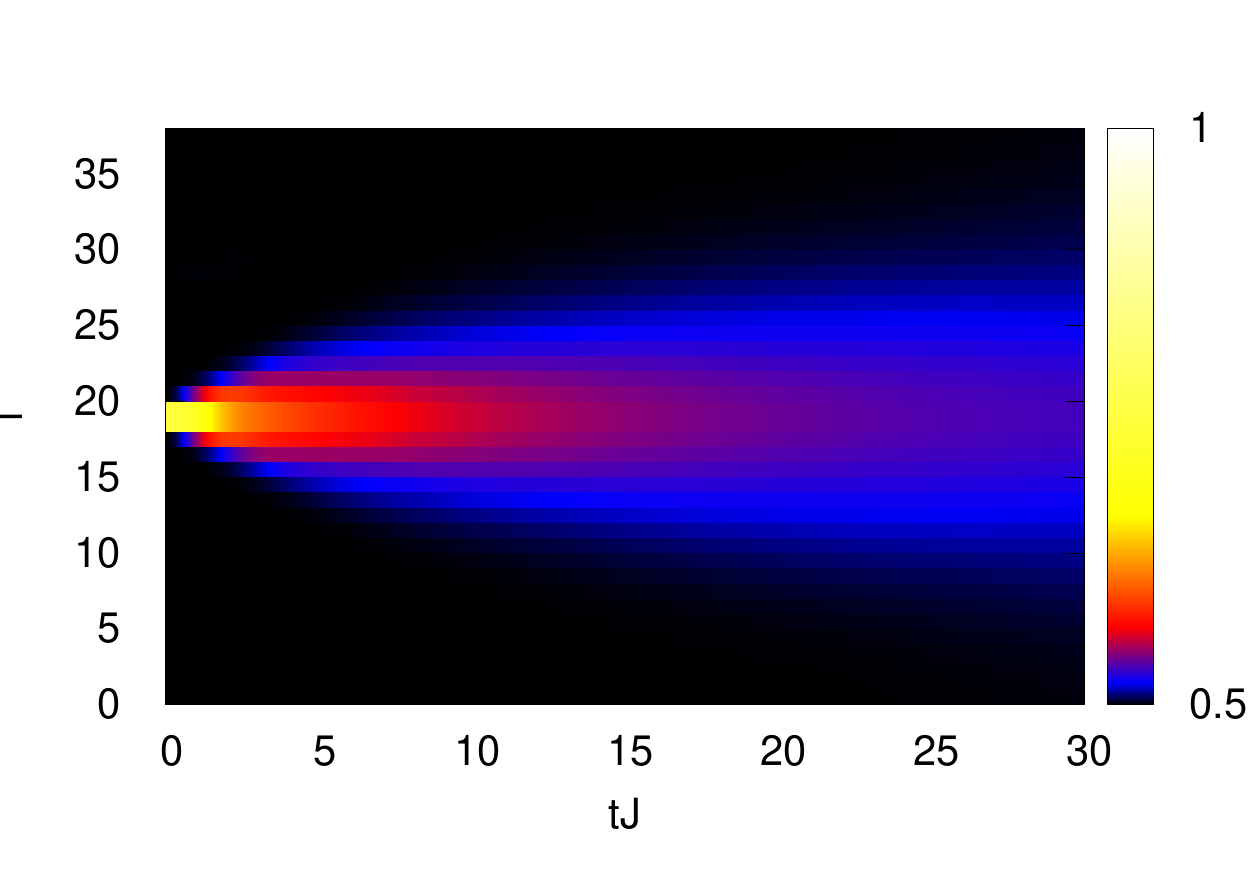} 
\caption{(Color online) Density plot for the time evolution of the local occupation number $p_l$
for the 1D XXZ model (see Eq.~(\ref{Hxxz}) with $\Delta=1.5$ and $N=39$. }
\label{figdensity}
\end{figure}

We illustrate the method by considering once more the Heisenberg XXZ model Eq.~(\ref{Hxxz}).
As the density operator, we take the local occupation number $n_l=S_l^z+1/2$.
Note that $\sqrt{n_l}=n_l$ here.
The density-density correlation function at $\beta=0$ is given by
\begin{equation}
C(t)=\frac{\mathbf{Tr}\;n_l(t)n_{L/2}}{D} = \frac{\mathbf{Tr}\;n_{L/2}n_l(t)n_{L/2}}{D}
\approx  \left < \Phi\right | n_{L/2}n_l(t)n_{L/2} \left |\Phi\right>
= \left < \psi(t)\right | n_l  \left | \psi(t)\right >=p_l/2,
\end{equation}
where $\left | \psi(t)\right >=e^{-itH} n_{L/2} \left |\Phi\right>$,
$\left < \psi(t)\right . \left | \psi(t)\right > = 1/2$, and $p_l$ is the expectation
value of $n_l$ (after proper normalization of the state vector).
As only one time evolution of the state vector $\left | \psi(t)\right >$ is needed,
using the same resources as in the case of the current-current correlations
allows us to simulate a system with one extra spin, that is the 1D XXZ model with $N=39$ spins.

Figure~\ref{figdensity} shows the results $p_l$.
The initial state produces a peak in the center of the chain, i.e., $p_{20}=1$.
As a function of time, this peak then spreads over the neighbors and the density profile
shows a tendency to become stationary. Within the maximum time $tJ=30$ shown,
the profile does not reach the boundary yet, indicating that up to this time
finite size effects are not yet relevant.
It is clear that the width of the profile develops slowly, which, in fact, fits well to
a square-root increase after an initial linear increase for short time scales $tJ\leq 1$.
The data  provides clear evidence for diffusion in the integrable XXZ model with large anisotropy $\Delta>1$.
Our results agree with results obtained from time-dependent density matrix
renormalization group calculations~\cite{KARR14}.
Further detailed discussions about diffusion and data for other models
can be found in Refs.~\cite{STEI17a,STEI17b,RICH18,RICH19}.

\subsection{Electron spin resonance}

According to linear response theory, the ESR signal is related to the Fourier transform of
the autocorrelation correlation function~\cite{KUBO57}.
The ESR signal averaged over a period $2T$ is proportional to~\cite{KUBO57}
\begin{eqnarray}
C(\omega)&\equiv&\frac{1}{2T}\int_{-T}^{T}
\frac{\mathbf{Tr\,} e^{-\beta H} C(t)}{\mathbf{Tr\,} e^{-\beta H}}\, \cos\omega t \,dt
,
\label{AC0}
\end{eqnarray}
where
\begin{eqnarray}
C(t)&=&
\frac{1}{2}\{e^{it H}M^x e^{-it H}, M^x\}
,
\label{AC1}
\end{eqnarray}
and $M^x$ is the $x$-component of the total magnetization.
We set $X(\omega)=e^{-\beta H/2}  Y(\omega) e^{-\beta H/2}$ and
$Y(\omega)= \frac{1}{2T}\int_{-T}^{T} C(t) \cos\omega t \,dt$.
Because $Y(\omega)=Y^\dagger(\omega)$ and $X(\omega)=X^\dagger(\omega)$,
we can use Eq.~(\ref{QSP5}) to show
that the variance of $C(\omega,T)$ vanishes exponentially with increasing system size.

\begin{figure}[tb]
\includegraphics[width=0.6\columnwidth]{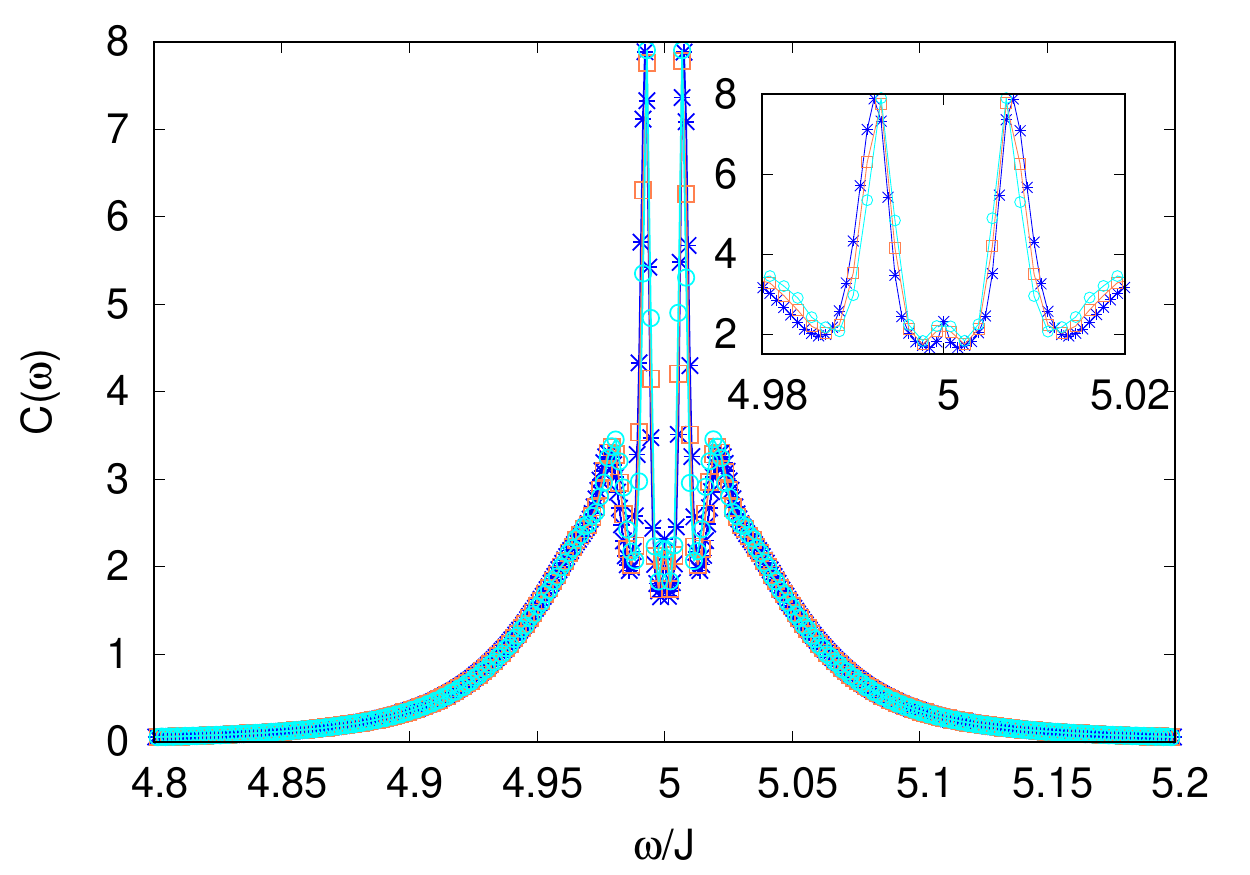} 
\caption{(Color online) ESR spectrum (see Eq.~(\ref{AC0})) for $N=30$ (stars), $N=32$ (squares), $N=34$ (circles), and $\beta=0$ for the 1D XXZ model given by Eq.~(\ref{Hxxz}) with an additional term $H_m=-h\sum_i S_i^z$ to represent the external magnetic field. The inset shows the spectrum $C(\omega)$ for $4.98\le\omega/J\le5.02$.}
\label{figesr}
\end{figure}

Again, we take the XXZ model as an example, except that now we add an extra term
to Eq.~(\ref{Hxxz}), given by $H_m=-h\sum_i S_i^z$, to represent the external magnetic field.
For simplicity, we do not include the dipole-dipole interaction term~\cite{MIYA99}.
The parameters are taken to be $J=-1$, $\Delta=0.84$, and $h=5$ and we adopt open boundary condition.

Figure~\ref{figesr} presents the results for the ESR spectra obtained from simulating
the given XXZ model up to times $tJ=4096$ for $N=30$ and $tJ=2048$ for $N=32$ and $34$, respectively.
As the scale of the $x$-axis of Fig.~\ref{figesr} indicates, this particular
calculation requires a high resolution in the frequency domain.
Accordingly, we need simulation data for a very large time interval, effectively
limiting the system sizes we can study within a reasonable amount of elapsed time.
Therefore, we computed data for system sizes $N=30$, $32$, $34$ and $\beta=0$ only.

As Fig.~\ref{figesr} shows, the ESR line shape clearly displays a four-peak structure,
which fits well to a sum of four Lorentzians~\cite{ELSH10}.
The separation between the two central peaks seems to decrease
as $N$ increases from $N=30$ to $N=34$.
This may suggest that the central double-peak structure disappears in the thermodynamic limit.
However, on the basis of numerical data, it is difficult to draw a definite
conclusion about the vanishing of the central double-peak structure.
In fact, this is a subtle issue and we refer the interested reader to Refs.~\cite{OSHI99,IKEU15,IKEU17} for additional data on the size and
temperature dependence of the ESR line shape.

\section{Application: quantum supremacy}\label{AQSUP}

Recently, Gaussian random states (see Table~\ref{tab1}, case A)
found new applications in the field of quantum information processing.
In this rather long section, we scrutinize the possibility of letting a universal quantum computer
generate Gaussian random states.
We address the conceptual differences of realizing such states
on a conventional digital computer and a hypothetical, universal quantum computer.
We also discuss the difficulties that arise in designing and testing quantum circuits
which are tailored to generate states that exhibit features characteristic of Gaussian random states.

A demonstration that a programmable quantum device
can perform a (not necessarily useful) task which is out of reach for present-day and
near-future supercomputers is called quantum supremacy~\cite{PRES18,BOIX18,GOOG19}.
Assessing the potential of such a quantum device involves
performing a series of established tests, so-called benchmarks.
In this section and in section~\ref{QINF}, the term ``benchmarking'' refers to measuring
the quality of a quantum information processor as a computing device
by comparing experimental data produced by the processor
with results obtained from (a computer simulation on a digital computer
of) the quantum theoretical model for the device.
More specifically, this section discusses benchmarking
of the Sycamore superconducting processor~\cite{GOOG19} using the
cross entropy~\cite{BOIX18} (see section~\ref{QSUP}) as a measure.
In section~\ref{QINF}, we present theoretical results related to the theory of ``randomized'' benchmarking~\cite{emerson2005randomunitaryoperatorstwirling,
Knill2008randomizedbenchmarking}, which involves averaging the fidelity~\cite{nielsen2002gatefidelity} of experiments
performed with different, randomly chosen, quantum circuits.

We start by giving an outline of the content of this section.
In subsection~\ref{AQSUPa}, we analyze the general problem of sampling from a probability
distribution that is specified through the amplitudes of a pure quantum state such as the Gaussian random
state of Table~\ref{tab1}, case A.
We compare the resources that are needed to perform this sampling
on a digital computer and on the theoretical model of a quantum computer.
We also address the complexity of constructing and validating quantum gate circuits
that are designed to generate states that exhibit features characteristic of Gaussian random states.

Subsection~\ref{qs_gen} discusses general aspects of the recent quantum
supremacy experiment with the Sycamore superconducting processor~\cite{GOOG19}.
Subsection~\ref{QSUP} introduces the measure, the cross entropy, that is used
to discriminate between
outcomes that are distributed uniformly over all possibilities
and outcomes reflecting the quantum gate sequence
that the device was instructed to carry out.

In subsection~\ref{MAX}, we use the maximum entropy method
and the knowledge of the measured cross entropy to establish a relation between
the theoretically expected probability distribution
and the unknown probability distribution that describes the observed outcomes.

In subsection~\ref{QST}, we adopt the Gaussian random state (see Table~\ref{tab1}, case A)
as the model of the theoretically expected probability distribution. We show that, given a numerical estimate of the cross entropy, the maximum entropy method~\cite{JAYN03}
yields a unique solution for the probability distribution that describes the observed outcomes.
This solution indeed allows us to discriminate between
a device that generates outcomes reflecting the applied quantum gate sequence
and a device that produces uniformly distributed outcomes.

Finally, in section~\ref{QSES}, we combine supercomputer simulations with the
experimental data produced by the Sycamore superconducting processor~\cite{GOOG19} to assess the claim that quantum supremacy has been demonstrated.

\subsection{Using a quantum computer to sample from a random state}\label{AQSUPa}

Up to this date, applications of the random state technology employ digital computers to solve problems.
Obviously, for this purpose, we need to be able to
store at least one pure state $|\Phi\rangle=\sum_{j=0}^{D-1} c_j |j\rangle$
in the computer's memory.
This means that the computer should have enough memory to store the $D$ complex numbers $c_j$.
A supercomputer with 1 PiB ($2^{50}$ bytes) of random access memory (RAM) can store
$2^{46}$ complex double-precision coefficients $c_j$.
Translating this into how many qubits (or $S=1/2$ spins) we can represent with
such a large amount of RAM, we find the disappointingly low number $L=46$.
Every time we add a qubit to the system, we have to double the amount of memory,
just to be able to store the state $|\Phi\rangle$.
The required amount of memory $D=2^L$ grows exponentially with the number of qubits $L$.
Clearly, the cost of such a digital computer, its power consumption, and its speed of operation
severely limit the number of quantum objects one is able to study by application of
the otherwise very efficient random state technology.

Random states $|\Phi\rangle$ that are drawn from
the uniform distribution on the $2D$-dimensional unit sphere, Case A of Table~\ref{tab1},
are not only very convenient for many applications of the random state technology but,
as shown in section~\ref{QINF} and will become clear later in this section,
they are important for the theory of randomized benchmarking of real quantum processors too.
Therefore, in this section, we limit the discussion to random states
that are drawn from the uniform distribution on the $2D$-dimensional unit sphere and,
to simplify the writing, will refer to such states as random states in the following.

Applications of the random state technology generically involve the calculation of
\begin{eqnarray}
\langle X \rangle = \langle \Phi|X|\Phi\rangle
\;,
\label{AQS0}
\end{eqnarray}
that is the expectation value of a matrix $X$ for a random state $|\Phi\rangle$.
Let us simplify the discussion by considering matrices $X$ which are diagonal
with respect to the (computational) basis states $\{|0\rangle,\ldots,|D-1\rangle\}$, $X|j\rangle=x_j|j\rangle$. Using this, Eq.~(\ref{AQS0}) becomes
\begin{eqnarray}
\langle X \rangle = \sum_{j=0}^{D-1} x_j |c_j|^2
\approx \frac{1}{m}\sum_{j\in{\cal J}} x_j
\;,
\label{AQS1}
\end{eqnarray}
were ${\cal J}$ is a set of $m\ll D$ states drawn with probability $|c_j|^2$
and we have used the law of large numbers~\cite{GRIM01} to approximate the exact expressions by a sum of $m$ instead of $D$ terms.
The problem of computing $\langle X \rangle$ has now been replaced
by the problem of sampling integers $j$ with probabilities $|c_j|^2$ (recall $\sum_{j=0}^{D-1}|c_j|^2=1$.)
The generic way for performing this sampling on a digital computer is to
compute the cumulative probability distribution $P(j)=\sum_{i=0}^j|c_i|^2$ for all $j=0,\ldots,D-1$,
generate a uniform pseudo-random number $0<r<1$, and output the largest value of
$j$ for which $P(j)\le r$.

Let us now ask ourselves how difficult, in terms of memory and computation time, it is to
compute and store the cumulative probability distribution $P(j)$.
It may not be too far off to assume that
the total amount of storage available in the world at this time is of the order of 10 EiB ($\approx 2^{63}$ bytes, although it is not easy to find a reliable number).
For a value of $D$ of that order, the calculation of $P(j)$ needs to be performed
with 8-byte floating point arithmetic. This implies that with 10 EiB of storage, we may be able
to store $2^{59}$ values of the $P(j)$'s, the corresponding quantum system is described in terms of 59 qubits.

Using the algorithm described in section~\ref{ALTMUL}, we can generate $c_0,c_1,\dots, c_{D-1}$ independently.
Thus, to fill the table of $P(j)$'s, we can simply generate $c_j$ and set $P(j)=P(j-1)+|c_j|^2$ for $j$ running from $0$ to $D-1$
(with $P(-1)=0$). This takes of the order of $D$ floating-point operations.
To reduce the elapsed time, we may distribute this calculation over many processors.
Suppose that we can really get exclusive access to
the huge amount of storage that our application needs, can we then sample
from the cumulative distribution $P(j)$?
Disregarding the fact that this storage is distributed over the whole world and access times to
memory locations may be relatively long, we probably can
because for a given random number $r$, we can find $j$ by binary search, that
is in a number of steps that is proportional to $L$, not to $D=2^L$.
The upshot of these considerations is
that the limit on the value of $D$ of the random state from which we would like to sample
is dictated by the amount of available memory.

Can a gate-based quantum computer do better than a digital computer in terms of problem size?
According to the mathematical model of gate-based quantum computing,
the state of an $L$-qubit quantum computer is described by a linear
superposition of the $D=2^L$ computational basis states~\cite{NIEL00}.
As the state of the quantum computer changes, all $D=2^L$ coefficients
change in parallel~\cite{NIEL00}.
Thus, the mathematical model of a quantum computer provides us with
a machine that exhibits an unprecedented level of intrinsic parallelism
and a tremendous amount of memory (with each additional qubit, we double the amount
of coefficients describing the state).
The relevant question then is ``can we exploit these features in practice?''.

Returning to the conceptually simple task of sampling from a random state,
with a quantum computer at our disposal we would proceed as follows:
\begin{enumerate}
\item
Design a gate circuit C, that is a sequence of unitary operations~\cite{NIEL00}, that changes the state
$|j=0\rangle$ into the random state $|\Phi\rangle=\sum_{j=0}^{D-1} c_j |j\rangle$.
\item
Apply C to the initial state $|j=0\rangle$.
\item
With the quantum computer in the state $|\Phi\rangle=\sum_{j=0}^{D-1} c_j |j\rangle$,
perform a measurement of all the qubits.
This measurement yields values $q_k=0,1$ for each of the $k=0,\ldots,L-1$ qubits.
The sequence of values $q_{L-1}\cdots q_0$ is called a bitstring
and is equivalent to the integer $j=\sum_{k=0}^{L-1} 2^kq_k$,
corresponding to the index of the state $|j\rangle\in\{|0\rangle,\ldots,|D-1\rangle\}$.
By Born's rule, the probability for this event is given by $|c_j|^2$.
\item
Store the $m$ values of $j$, obtained by $m$ repetitions of steps 2 and 3, to form the set ${\cal M}$.
\item
Compute $m^{-1}\sum_{j\in{\cal M}} x_j$ to obtain the desired approximation to $\langle X \rangle$.
\end{enumerate}
Note that designing such circuits C is by no means trivial, as C
should act on all $D$ coefficients to create the random state.
To the best of our knowledge, there is no rigorous proof that such circuits can be constructed
with a depth polynomial in the number of qubits.
Instead, Boixo {\it et al.} constructed so-called random circuits
from single- and two-qubit gates~\cite{BOIX18}.
For those circuits which can be simulated numerically, they demonstrated that they
have a depth which grows polynomially with the number of qubits
and produce states that yield the Porter-Thomas distribution,
a direct consequence of the state being a Gaussian random state~\cite{BOIX18}.
We make the plausible assumption that the length/depth of the circuit C
and the time it takes to measure all qubits are of order $L$, not of order $D$.
Then in the mathematical realm, the gate-based quantum
computer can sample from a random state of dimension $D$, for values of $D$
which are out of reach for a digital computer. Memory is not an issue.

Although the mathematical model holds great promise for very fast and very large computation,
the technological hurdles to build a quantum information processor that
realizes (part of) these promises seem enormous.
In fact, there still is a wide gap between the mathematical model and its physical realizations.
For instance, experiments with publicly accessible quantum processors
show that they are not yet capable of reliably performing
simple computational tasks such as adding two small integers~\cite{MICH17b}.
Moreover, the performance of the current generation of gate-based quantum information processors
is adversely affected by various sources of noise.
For this reason, they are often referred to as Noisy Intermediate-Scale Quantum (NISQ) devices.
The prospects of building a fully error-corrected quantum processor with NISQ technology are rather dim.
Therefore, in order for quantum information processing to become a practical reality on short terms,
it is necessary to
\begin{enumerate}[(i)]
\item
characterize the performance of NISQ devices by a well-defined procedure
\item
develop algorithms for solving practically relevant, nontrivial problems using these error-prone NISQ devices.
\end{enumerate}
The recent quantum supremacy experiment~\cite{GOOG19} targets point (i), not point (ii).

\subsection{Quantum supremacy: general aspects}\label{qs_gen}
As mentioned earlier, current NISQ devices have difficulties to perform e.g. simple arithmetic operations.
However, because of their noisy operation, they excel at producing ``random'' output.
The latter feature may be put to good use to demonstrate
that a NISQ device can perform a computational task which is
prohibitive for state-of-the-art digital (super)computers.
Such a demonstration is coined ``quantum supremacy''.
Quoting Boixo {\it et al.}: ``we propose the task of sampling from the output distribution
of random quantum circuits as a demonstration of quantum supremacy''~\cite{BOIX18}.
The output distribution Boixo {\it et al.} have in mind corresponds to a realization of a state
drawn from the uniform distribution on the $2D$-dimensional unit sphere, Case A of Table~\ref{tab1}, that is a random state.
The Case A distribution is exceptional among multidimensional distributions
because for any $D$, many of its properties can be studied analytically through closed form expressions.
For instance, one can show (see the derivation of Eq.~(\ref{PT2})) that
the exact expression for probability density $p(z)$ of ``the probability $z=|c_j|^2$
to observe the bitstring $j$'' is given by
$p(z)=(D-1)(1-z)^{D-2}$ (see Eq.~(\ref{PT2}) in Appendix~\ref{SOM}). For large $D$, $p(z)\approx e^{-Dz}$, a result
which is often referred to as the Porter-Thomas law~\cite{PORT56}.

Referring to item 1 in section~\ref{AQSUPa},
the first step is to design circuits C that perform the desired task.
To this end, Boixo {\it et al.} use what they call a ``random circuit'', denoted by R in the following,
These R's are specifically designed to generate states that are close to random states.
Heuristics and simulation are used to guide the construction of the R's~\cite{BOIX18,GOOG19,Zlokapa2005BoundariesQS}.
Boixo {\it et al.} have constructed a large number of such R's~\cite{BOIX18,GOOG19}.

The quantum supremacy demonstration~\cite{BOIX18,GOOG19} has four different components:
\begin{enumerate}[(a)]
\item
Design of a random circuit R, operating on $L$ qubits, for which the output distribution is known
through simulation of the same (or approximately the same) circuit on a digital computer.
\item
A real quantum processor that can execute the random circuit R and can produce a set of measured bitstrings ${\cal M}$.
\item
A measure for the correlation between the set of observed bitstrings and the random circuit
executed on the ideal quantum computer.
\item
An error model to account for the NISQ character of the quantum processor
and a procedure to extrapolate the results to values of $L$ that cannot be
simulated by a state-of-the-art digital computer.
\end{enumerate}

Regarding (b), recall that
simulating an $L$-qubit circuit C on a digital computer requires resources of the order $D$.
At the time of writing, this prohibits the simulation (with an accuracy of 10 digits or better)
of a universal quantum computer with to $L>45$~\cite{RAED19a}.
Therefore, merely executing C on a real quantum processor having say, $L=53$ qubits, is
a trivial kind of ``quantum supremacy''.
One has to demonstrate that the output distribution is indeed the one expected for the circuit C.
In principle, this requires taking of the order of $S\times D$ samples where $S$ is the number
of repetitions (say of the order of $S=10000$)
required to estimate each $|c_j|^2$ with sufficient statistical accuracy.
Obviously, for large $L$ (even for $L=45$), this sampling task is prohibitive, even with a
noise-free quantum computer.
Therefore, in order to proceed, one has to be less demanding and feel content with showing that
the measured set of bitstring ${\cal M}$ complies with the hypothesis that
these bitstrings are samples drawn from the distribution corresponding to R.
This then requires the specification of one or more measures for the correlation, as mentioned in item (c),
which is the topic of the following subsection.
Item (d) is essential for the interpretation of the data produced in the quantum supremacy experiment~\cite{GOOG19}
but is outside the scope of this paper.

\subsection{Cross entropy}\label{QSUP}

Consider an ideal quantum computer with $L$ qubits executing a gate circuit U represented by a unitary matrix $U$.
This quantum computer generates states labeled by $j=0,\ldots,D-1$ ($D=2^L$) with probabilities $p_U(j)$.
Assume that we have a physical realization of the quantum computer.
As discussed above, currently only NISQ devices are able to execute the gate circuit $U$.
This physical device produces bitstrings $j$ with a-priori unknown probabilities denoted by $p_V(j)$.
We denote the collection of bitstrings (represented by integers) generated by this device by
${\cal J}=\{j_1,\ldots,j_m\}$ where each $j_k\in \{0,\ldots,D-1\}$.
The key question is now
\begin{center}
\framebox{
\parbox[t]{0.7\hsize}{%
\centerline{To what extent are the probability distributions $p_U$ and $p_V$ similar?}
}}
\end{center}
If we had enough data to build a histogram that approximates $p_V$
and we knew $p_U$, we could resort to standard statistical tests for comparing the two distributions~\cite{PRES03}.
However, even for small $L$ (say $L=20$, $D=2^L=1048576$), a large number of bitstrings (say $m\gg 100D$)
is required in order to estimate the histogram with some confidence, rendering this approach useless for
practical purposes if $L$ is large.
Boixo et al.~\cite{BOIX18} proposed to circumvent this ``sampling problem'' by using the cross entropy,
\begin{eqnarray}
C(V,U)&=&-\sum_{j=0}^{D-1} p_V(j) \log p_U(j) 
,
\label{MAD0}
\end{eqnarray}
as a measure for the difference between the observed ($p_V$) and expected ($p_U$) distribution of bitstrings.
Note that the cross entropy $C(V,U)$ is closely related to
the Kullback-Leibler divergence $D(V,U)=\sum_{j=0}^{D-1} p_V(j)\log [p_V(j)/p_U(j)]$,
i.e., $C(V, U)=S(V)+D(V,U)$, where $S(V)=-\sum_{j=0}^{D-1} p_V(j)\log {p_V(j)}$ is the Shannon entropy.
By invoking the central limit theorem, Boixo et al. use the sample average
\begin{eqnarray}
c_U&=&-\frac{1}{m}\sum_{j\in{\cal J}} \log p_U(j)
,
\label{MAD0a}
\end{eqnarray}
of the elements in the set ${\cal J}$ (i.e., the $m$ bitstrings generated by the device) to approximate the sum over all $j$ in Eq.~(\ref{MAD0}).
This circumvents the problem of not being able, in practice, to collect enough data to estimate
all the $p_V(j)$'s.
For completeness, we mention that the recent quantum supremacy demonstration
used, in addition to Eq.~(\ref{MAD0}), also the ``linearized'' cross entropy
$L(V,U)=\sum_{j=0}^{D-1} p_V(j) (D p_U(j)-1)\approx (1/m)\sum_{j\in{\cal J}}(D p_U(j)-1)$~\cite{GOOG19}.
For the purpose of discussing the ideas behind these recent quantum supremacy experiments,
it does not seem to matter if one uses $C(V,U)$ or $L(U,V)$~\cite{GOOG19}.
Therefore, in the remainder of this section, we only consider the cross entropy $C(U,V)$.

For a fixed $p_U$, $c_U$ is a positive number, obtained by computing the cross entropy Eq.~(\ref{MAD0a})
using a data set of bitstrings ${\cal J}$, generated by a device.
The key question can now be reformulated as
\begin{center}
\framebox{
\parbox[t]{0.8\hsize}{%
Given $p_U$ and the numerical value $c_U>0$, what can be said about the unknown distribution $p_V$?
}}
\end{center}
In order to make a statement, one has to adopt a model.
In the next subsection we appeal to the principle of maximum entropy,
a recipe to obtain a probability distribution with specified averages of
functions of variables (and nothing else) and which otherwise is as uniform as possible~\cite{JAYN03}.

\subsection{Maximum entropy method}\label{MAX}
We follow Jaynes (Ref.~\cite{JAYN03}, section 11.6) and search for the extrema of
\begin{eqnarray}
F=\sum_{j=0}^{D-1} \left[ -p_V(j) \log p_V(j)+ \lambda \left(p_V(j)-\frac{1}{D}\right)
+ \mu \left(p_V(j) \log p_U(j)+\frac{c_U}{D}\right)\right]
,
\label{MAX0}
\end{eqnarray}
where the Lagrange multipliers $\lambda$ and $\mu$ account for the constraints
$\sum_{j=1}^{N} p_V(j) =1$ and $c_U=-(1/m)\sum_{j\in{\cal J}} \log p_U(j)$, respectively.
There is only one extremum, namely the maximum of $F$~\cite{JAYN03}.
Hence, we solve the problem of maximizing the first term in Eq.~(\ref{MAX0}) which
is the entropy (or Shannon information), subject to the named constraints.
The solution for the maximum is given by $p_V(j) = \exp( \lambda +\mu -1 + \mu \log p_U(j))$.
Using the named constraints we find $\exp(1-\lambda-\mu)=\sum_{j=0}^{D-1} p_U^\mu(j)$ and
\begin{eqnarray}
p_V(j) &=&\frac{e^{\mu \log p_U(j)}}{\sum_{j=0}^{D-1} e^{\mu \log p_U(j)}}=
\frac{p_U^\mu(j)}{\sum_{j=0}^{D-1} p_U^\mu(j)}
,
\label{MAX4}
\end{eqnarray}
We can find $\mu$ by solving
\begin{eqnarray}
-\sum_{j=0}^{D-1} p_U^\mu(j)\log p_U(j) = c_U\sum_{j=0}^{D-1} p_U^\mu(j)
.
\label{MAX5}
\end{eqnarray}
In summary, if we combine the knowledge that the cross entropy $C(V,U)=c_U$
with the principle of maximum entropy,
there is a definitive relation between the known probabilities $p_U(j)$ and the unknown probabilities $p_V(j)$ from which the samples ${\cal J}$ are drawn.
Given the value of $c_U$ and probabilities $p_U(j)$'s
the maximally non-committal probability distribution is given by Eq.~(\ref{MAX4}) with
$\mu$ being the solution of Eq.~(\ref{MAX5}).
Phrased differently, there are many more probabilities to be assigned than
there are constraints (the normalization and the value of $c_U$) and the
maximum entropy principle yields the broadest distribution that is
compatible with these constraints.
Therefore Eq.~(\ref{MAX4}) may be viewed as a minimally prejudiced assignment
complying with the named constraints.

Assume that an experiment with a device yields a value of $c_U$
for which the solution of Eq.~(\ref{MAX5}) yields $\mu=1$,
Then, from Eq.~(\ref{MAX4}), it follows that $p_V(j)=p_U(j)$.
In other words, excluding all other knowledge we might have,
the principle of maximum entropy suggests (but not guarantees) that the device is working properly,
meaning that it generates bitstrings according to the probabilities $p_U(j)$ that
correspond to the circuit $U$.
Similarly, if the device produces a value of $c_U$
for which the solution of Eq.~(\ref{MAX5}) yields $\mu=0$,
the principle of maximum entropy suggests (but not guarantees) that
$p_V(j)=1/D$, that is the states are sampled from a uniform superposition of basis states
(see Case C in Table~\ref{tab1}).

Up to this point, the discussion has been entirely general,
independent of a particular choice of the gate sequence $U$.
Clearly, in combination with the maximum entropy principle,
the cross entropy estimate Eq.~(\ref{MAD0a})
is a useful measure for the correspondence between the observed
bitstrings ${\cal J}$ produced by a quantum device and the ideal gate sequence $U$ that
one would like this device to carry out.

\subsection{Quantum supremacy: theory}\label{QST}

A key advantage of using the cross entropy is that
it can be estimated without having to perform a prohibitively large number of measurements.
To be useful, the only requirement is that $p_U(j)$ is known,
through simulation on a digital computer~\cite{RAED07x,RAED19a,GOOG19}
or through a model that can be treated analytically.
For an arbitrary sequence $U$, memory requirements limit the calculation of
the $p_U(j)$'s to less than 50 qubits ($D<2^{50}$) on current supercomputers~\cite{RAED19a,GOOG19}.
To demonstrate that a quantum processor can perform a task which cannot be performed by a digital computer,
one has to resort to circuits U for which the cumulative probability distribution cannot be determined analytically and is not amenable to computation by a digital computer.

It is instructive to first consider the case in which the circuit E generates a set of bitstrings ${\cal E}$ for the uniform
superposition state by applying a Hadamard gate to each qubit~\cite{NIEL00},
that is Case C of Table~\ref{tab1} (with all $a_j$'s equal to zero)
for which $p_E(j)=1/D$ and, from Eq.~(\ref{MAD0a}), $c_{E}=\log D$.
Then Eq.~(\ref{MAX5}) is an identity for arbitrary $\mu$ and Eq.~(\ref{MAX4}) yields $p_V(j)=p_E(j)=1/D$.
Thus, assuming the device to work properly,
the circuit E is expected to generate $m$ bitstrings ${\cal E}$ for which $c_{E}=\log D$ (up to statistical fluctuations).

Next, we consider somewhat less trivial states, namely
those that are distributed uniformly on the $2D$-dimensional unit sphere.
These are the states A of Table~\ref{tab1} that are at the heart of the random state technology.
In the recent quantum supremacy experiment~\cite{GOOG19},
the quantum processor executes a so-called random circuit R and produces sets of bitstrings ${\cal R}$ for
each instance of the circuit R.
The basic idea is that the state produced by R is an instance of
the uniform distribution on the $2D$-dimensional unit sphere.
Each sample ${\cal R}$ of $m$ bitstrings carries a ``fingerprint''
of the particular circuit R that produced these bitstrings.
In symbols, for a particular random circuit R, the state of the ideal quantum
computer executing R reads $|\Psi_R\rangle=\sum_{j=0}^{D-1} z_j|j\rangle$
where $(z_1,\ldots,z_{D-1})$ is a sample drawn from a uniform distribution over the $2D$-dimensional unit sphere.
However, if we assume that the state generated by such a circuit
is indeed an instance of the uniform random state $|\Phi\rangle$, we can use random state technology to
explore its features analytically.

In the remainder of this subsection, we consider the case that the quantum processor
executes the circuit R without producing any error and that the resulting state
seems to have been drawn from the uniform distribution over the $2D$-dimensional unit sphere.
The first issue to address is whether, for the application to cross-entropy benchmarking,
we can obtain accurate estimates from one realization of the state $|\Phi\rangle$ if $D$ is large.
From Eq.~(\ref{SUB3}) it follows that
\begin{eqnarray}
\mathrm{Var}\left[\sum_{j=0}^{D-1} p_R(j)\log p_R(j)\right]&=&\frac{\pi^2-9}{3D} +{\cal O}(\frac{1}{D^2})
,
\label{MAX6}
\end{eqnarray}
such that also in the case of random sampling from $|\Phi\rangle$, we can expect an
accurate estimate of the entropy from one realization (if $D$ is large).
Similarly, it follows that for large $D$, the variance
of the difference between entropy and its estimate computed from the set ${\cal R}$ having $J$ elements
is given by
\begin{eqnarray}
\mathrm{E}\left[\left(\frac{D}{J}\sum_{j\in{\cal R}} p_R(j)\log p_R(j)
- \sum_{j=0}^{D-1}  p_{R}(j)\log p_{R}(j) \right)^2\right]
=\frac{1-J/D}{J}\left[\frac{\pi^2-9}{3} +(\log D +\gamma-2)^2\right]
\;,
\nonumber \\
\label{MAX7}
\end{eqnarray}
where $\gamma\approx0.577$ is Euler's constant.
As $\log D \approx 0.69L$, Eq.~(\ref{MAX7}) shows that if $J\gg L^2$,
any subset ${\cal R}$ will yield an estimate of the entropy which, on average,
will be an accurate approximation to the entropy $-\sum_{j=1}^D  p_{R}(j)\log p_{R}(j)$.

Finally, we consider the solution of the maximum entropy problem Eq.~(\ref{MAX5}).
According to Eq.~(\ref{MAX6}), if $D$ is large, we may replace the
calculation based on one realization of the random state by the average over all equivalent random states.
This amounts to performing integrals over multidimensional Gaussians, which is straightforward.
It then follows from Eq.~(\ref{SUB2}) that we have to find the value of $\mu$ that satisfies the equation
\begin{eqnarray}
c_{R}&=&\log D -\mathrm{PolyGamma}(0, \mu+1)  \quad,\quad c>0
.
\label{MAX8}
\end{eqnarray}
The function $\mathrm{PolyGamma}(0, \mu+1)$ is the logarithmic derivative of the Gamma function, that is $\mathrm{PolyGamma}(0, \mu+1) = \Gamma'(\mu+1)/\Gamma(\mu+1)$. This function is monotonically increasing function for $\mu>-1$ with a divergence at $-1$,
negative for $-1 <\mu < 0.46163$, and positive for $\mu> 0.46164$.
For large $\mu$, we have $\mathrm{PolyGamma}(0, \mu+1)=\log \mu + 1/2\mu + {\cal O}(1/\mu^2)$.
Therefore, Eq.~(\ref{MAX8}) has a unique solution for $\mu$.

Instead of solving Eq.~(\ref{MAX8}) for $\mu$,
it is easier to compute, for a chosen value of $\mu$,
the difference between the value of $c_{E}=\log D$
of the uniform distribution $p_E(j)=1/D$ and the theoretical value
$c_{R}$ given by Eq.~(\ref{MAX8}).
We find
\begin{eqnarray}
c_{E}+\gamma-c_{R}&=&\left\{
\begin{array}{lrcl}
0 &,\; \mu=0&\leftarrow & \hbox{uniform distribution over all $D$ states}\\
1 &,\; \mu=1&\leftarrow & \hbox{uniform distribution on the $2D$-dimensional unit sphere}\\
3/2 &,\; \mu=2& &
\end{array}
\right.
.
\label{MAX9}
\end{eqnarray}
In the next subsection, we use the analytical results of this subsection
to interpret the experimental results of a recent quantum supremacy experiment~\cite{GOOG19}.

\subsection{Quantum supremacy: experiment and simulation}\label{QSES}

In this subsection, we scrutinize the results for the cross entropy,
obtained by combining the bitstrings ${\cal M}$
produced by the 53-qubit Sycamore superconducting processor~\cite{GOOG19}
and probabilities $p_R(j)$ calculated with the
universal quantum computer simulator JUQCS-E~\cite{RAED07x,RAED19a,JUQCSNIC}.

For a given quantum circuit R, designed to generate a random state, JUQCS-E~\cite{RAED19a} executes R
and computes the probability distribution $p_R(j)$ for each quantum state $j\in\{0,\ldots,D-1\}$.
JUQCS-E also computes the cumulative distribution function
$P_R(k)=\sum_{j=0}^k p_R(j)$ and samples states from this distribution,
yielding a set of bitstrings ${\cal R}$.
All these calculations are numerically exact (up to at least 10 digits).
A feature of JUQCS-E, not documented in Ref.~\cite{RAED19a},
allows the user to specify a set ${\cal M}$ of $m$ bitstrings
for which JUQCS-E calculates $p_R(j)$ for all $j\in{\cal M}$.
The latter feature allows us to compute the estimate
$-(1/m)\sum_{j\in{\cal M}} \log p_R(j)$ for the cross entropy.

Following the methodology for cross-entropy benchmarking~\cite{BOIX18,GOOG19},
the quantities of interest are
\begin{eqnarray}
\alpha_{R,R}&\equiv&c_{E}+\gamma-c_{R}=\log D +\gamma + \sum_{j=0}^{D-1} p_R(j) \log p_R(j)
\;,
\label{alpha0}
\\
\alpha_{R,{\cal X}}&\equiv&\log D +\gamma + \frac{1}{m}\sum_{j\in{\cal X}} \log p_R(j)
\;,
\label{alpha1}
\end{eqnarray}
where ${\cal X}={\cal R},{\cal M},{\cal E}$ and
${\cal R}$ and ${\cal M}$ are sets of bitstrings generated by JUQCS-E
and by the Sycamore processor~\cite{GOOG19}, respectively.
${\cal E}$ is a set of bitstrings sampled from the uniform distribution ($p_E(j)=1/D)$.
From the theoretical model presented in section~\ref{QST}, it follows that
$\alpha_{R,{\cal M}}\approx0$ if a quantum processor produces bitstrings that are distributed uniformly.
The larger the value of $\alpha_{R,{\cal M}}$, the more likely it is that
the quantum processor has been sampling from the correct distribution.
In this sense, the uniform distribution corresponding to $\alpha_{R,{\cal M}}\approx0$
provides a baseline for the assessment of the quality of a NISQ device.

If $m$ is sufficiently large ($m=500000$ for the experimental data sets~\cite{GOOG19}),
we may expect that $\alpha_{R,{\cal R}}\approx \alpha_{R,R}$.
If the circuit R produces a genuine random state, averaging over all such R's yields $\langle \alpha_{R,R}\rangle_R =1$, see
Eq.~(\ref{MAX9}).
Note that the last term in Eq.~(\ref{alpha0}) (Eq.~(\ref{alpha1})) is equal
to minus the cross entropy $C(R,R)$ ($C(R,{\cal X})$).

\begin{table*}[t]
\caption{%
Results for the $\alpha$'s (directly related to the cross entropies) defined by Eqs.~(\ref{alpha0})--(\ref{alpha1}).
The probabilities $p_R(j)$ for the circuit $R$ have been obtained by JUQCS-E~\cite{RAED19a}.
The corresponding sets of $m=500000$  bitstrings ${\cal M}$ have been obtained from experiments~\cite{GOOG19}.
In the first column, the letter in brackets
identifies the instance of the different random circuits R used in the experiments~\cite{GOOG19}.
The results obtained by using the circuit $39[b]$ and the measured data generated by the circuit indicated in the corresponding row
are listed as $\alpha_{39[b],{\cal M}}$.
R: pseudo-random circuit;
${\cal R}$: sampled states, obtained by executing $R$ on the simulator JUQCS-E;
${\cal M}$: sampled states, produced by the Sycamore processor executing $R$~\cite{GOOG19};
${\cal E}$: states sampled from a uniform distribution.
}
\begin{ruledtabular}
\begin{tabular}{cccccc}
qubits[circuit ID]&
$\alpha_{R,R}        $ &
$\alpha_{R,{\cal R}} $ &
$\alpha_{R,{\cal M}} $ &
$\alpha_{39[b],{\cal M}} $ &
$\alpha_{R,{\cal E}} $ \\
\hline\noalign{\smallskip}
$30[a]$ & $1.0000$ &  $0.9997$ & $0.0708$ &                        &  $\phantom{-}0.0026$ \\
$39[a]$ & $1.0000$ &  $0.9992$ & $0.0281$ &                        &  $-0.0003$ \\
$39[b]$ & $1.0000$ &  $1.0002$ & $0.0350$ &                        &  $\phantom{-}0.0006$ \\
$39[c]$ & $1.0000$ &  $0.9996$ & $0.0351$ & $-0.0013$ &  $\phantom{-}0.0034$ \\
$39[d]$ & $1.0000$ &  $0.9999$ & $0.0375$ & $-0.0007$ &  $\phantom{-}0.0036$ \\
$42[a]$ & $1.0000$ &  $0.9998$ & $0.0287$ &                        &  $-0.0024$ \\
$42[b]$ & $1.0000$ &  $1.0027$ & $0.0254$ &                        &  $\phantom{-}0.0014$ \\
$43[a]$ & $1.0000$ &  $1.0013$ & $0.0182$ &                        &  $-0.0010$ \\
\end{tabular}
\end{ruledtabular}
\label{QSUP1}
\end{table*}

In Table~\ref{QSUP1}, we present results for the $\alpha$'s defined by Eqs.~(\ref{alpha0})--(\ref{alpha1}).
Most of these results were included in the original report
on the quantum supremacy experiment~\cite{GOOG19}.
Note that if we use Eq.~(\ref{MAX7}) to estimate the standard deviation of $\alpha_{R,{\cal X}}$
we find that for $m=500000$ samples, the standard deviation is about $0.0014$.
The second and third column 
show that the results
produced by JUQCS-E are in excellent agreement with the theoretical prediction Eq.~(\ref{MAX9}) for $\mu=1$.
Recall that the latter has been obtained by averaging over all states
distributed uniformly over the $2D$-dimensional unit sphere whereas the former
is obtained from a simulation with a single instance of the random circuits which have been ``engineered''~\cite{GOOG19}
to generate instances of such a state.
The fact that $\alpha_{R,{\cal R}}\approx 1$ may suggest, but is not a proof, that the circuit R,
executed by JUQCS-E, produces a random state of the type A (see Table~\ref{tab1}).
In the fourth column 
we present the results
obtained by using the sets of bitstrings ${\cal M}$ measured in
the recent quantum supremacy experiment~\cite{GOOG19}.
The fifth column 
shows two results
for $\alpha_{R,{\cal M}}$, obtained by using the JUQCS-E data for $p_R(j)$, computed for circuit $39[b]$, and
the experimental data generated by circuits $39[c]$ and $39[d]$, respectively.
The $\alpha$'s obtained by replacing the measured bitstrings ${\cal M}$
by bitstrings ${\cal E}$ sampled from a uniform distribution
are given in the sixth column of Table~\ref{QSUP1}.

The numerical results presented in Table~\ref{QSUP1} can be summarized as follows:
\begin{enumerate}[(i)]
\item
For $L=39,42,43$, the experimental bitstrings yield values of
$\alpha_{R,{\cal M}}$ in the range $0.018$--$0.038$ (column four).
The value of $\alpha_{R,{\cal M}}$ decreases exponentially as the number of qubits $L$ increases
(analysis not shown, see Ref.~\cite{GOOG19}).
\item
The values of $|\alpha_{R,{\cal E}}|$ (column six) are at least one order of magnitude
smaller that those of $|\alpha_{R,{\cal M}}|$ (column four).
\item
Replacing bitstrings
produced by the experiment implementing circuit $39[b]$
by bitstrings ${\cal M}$
generated by an experiment performing a different circuit $R=39[c],39[d]$
 yields values of $|\alpha_{39[b],{\cal M}}|$ (column five). They are at least one order of magnitude
smaller than those of $|\alpha_{R,{\cal M}}|$ (column four).
\end{enumerate}
On the basis of (i) alone, it seems that there is little evidence
in support of the hypothesis that the states of the set ${\cal M}$
have been sampled from the distribution that is characteristic for the circuit R.
In fact, one might even be tempted to conclude that the Sycamore processor
executing circuit R samples bitstrings from a uniform distribution.
However, even though the numerical values of $\alpha_{R,{\cal M}}$
are small, they are still significantly larger than the values of $\alpha_{R,{\cal E}}$
obtained by sampling from the uniform distribution, see (ii).

We can make this tentative conclusion more quantitative by using the model derived by the maximum entropy method and by performing a hypothesis test.
First, given a value of $c_R$, the solution of Eq.~(\ref{MAX8}) is unique. Therefore,
$\alpha_{R,{\cal M}}\approx 0$ implies that $\mu\approx0$.
From Eq.~(\ref{MAX4}) it then follows that $p_V(j)\approx1/D$.
Second, following Jaynes (Ref.~\cite{JAYN03}, section 9.11), we consider the expression
\begin{eqnarray}
\psi_X=\sum_{j=0}^{D-1} n_j \log \left(\frac{n_j}{m p_X(j)}\right)
\ge 0\;,
\label{HYP0}
\end{eqnarray}
where $n_j$ is the number of times a bitstring $j$ has been observed in $m$ repetitions of the experiment and we have omitted constants that are irrelevant for the present purpose.
The larger the value of $\psi_X$, the less weight the hypothesis X has relative to
any alternative hypothesis belonging to the same Bernoulli class
(represented by probabilities for the bitstrings that are independent and stationary)~\cite{JAYN03}.
Thus, according to this hypothesis test, the data gives more weight to
hypothesis R than to hypothesis E if $\psi_E>\psi_R$.
From Eq.~(\ref{HYP0}), we have $\psi_E-\psi_R\propto\alpha_{R,{\cal M}}-\gamma$.
Therefore, we should opt for hypothesis R if $\alpha_{R,{\cal M}}>\gamma$,
an inequality which is obviously not supported by the data
(since $\alpha_{R,{\cal M}}\approx0$). Note that this test does not rule out that
there are other, better hypotheses than the two, E and R, that we have considered.
At any rate, on the basis of the values of $\alpha_{R,{\cal M}}$ alone, we should conclude that the processor is more likely to execute E than R. However, this conclusion is premature because the preceding analysis focuses on the sampling and hypothesized distributions without taking  into account what the circuit R actually does when it is executed by Sycamore processor.

According to (ii), feeding JUQCS-E with the bitstrings ${\cal E}$, generated on a digital computer using a uniform distribution, does not support the latter conclusion.
Indeed, the values of $|\alpha_{R,{\cal E}}|$ (column six) are at least one order of magnitude smaller than those of $|\alpha_{R,{\cal M}}|$ (column four).
Furthermore, from (iii) it follows that
the $|\alpha_{R,{\cal M'}}|$'s of bitstrings ${\cal M'}$ produced by another circuit R' are also much smaller that those obtained by using bitstrings ${\cal M}$ produced by circuit R.
This observation strongly suggests that the bitstrings ${\cal M}$ produced by
the Sycamore processor carry a definite (albeit weak) signature of the circuit R that was used to generate bitstrings ${\cal M}$.
An important missing element in the above analysis is mentioned in (d) above in Sec.~\ref{qs_gen}, namely
the fact that the NISQ processor is error-prone.
If fact, these errors are substantial~\cite{GOOG19}.
For a detailed discussion of the analysis that incorporates an error model, see Ref.~\cite{GOOG19}.

To refute a claim that quantum supremacy has been demonstrated,
the relevant question is then ``for each R,
can we construct an approximation that uses
only a (non-exponential) fraction of the $D$-dimensional Hilbert space and
yield values of $\alpha_{R,{\cal M}}$ similar to those shown in Table~\ref{QSUP1}.
A recent paper suggests that the answer to this question may be affirmative
although the results presented in that paper do not refute quantum supremacy yet~\cite{ZHOU20}.
By restricting state vectors of the quantum computer to a class of matrix-product states,
the paper shows that it is possible to generate states that carry signatures
(expressed through fidelities rather than cross entropies) of the generating circuit
with quality similar to those observed in the quantum supremacy experiments~\cite{GOOG19}.
However, using an approximation based on matrix-product states allows
computation on digital computer at a cost that does not increase
exponentially with the number of qubits~\cite{ZHOU20}.
Clearly, the last word about a quantum processor surpassing the
power of a digital computer for a specific computational task has not been said yet.


\section{Application: quantum information theory}\label{QINF}

It has become common practice to characterize NISQ devices
through what is called randomized benchmarking~\cite{emerson2005randomunitaryoperatorstwirling,
Knill2008randomizedbenchmarking}.
In essence, randomized benchmarking characterizes a quantum processor in terms of averages over many
different (random) instances of sequences of gates.

In this section, we use elements of the random state technology
to address some theoretical aspects of characterizing NISQ devices through randomized benchmarking. The result will be a generalization of a well-known formula for the average fidelity~\cite{nielsen2002gatefidelity,Gilchrist2005fidelities} to non-trace-preserving quantum operations.

Let $\rho$ and $\sigma$ denote the density matrices describing the state
of the mathematical idealization of a quantum processor and of a real device, respectively.
The fidelity defined by~\cite{Jozsa1994fidelity}
\begin{align}
  \label{eq:fidelityMixedStates}
  F(\rho,\sigma) = \left(\mathbf{Tr}\sqrt{\sqrt\rho\sigma\sqrt\rho}\right)^2
\end{align}
is a measure of the difference between the density matrices $\rho$ and $\sigma$.
In the case that $\rho=|\psi\rangle\langle\psi|$ is a pure state, we have $\rho=|\psi\rangle\langle\psi|$, $\sqrt\rho=\rho=\rho^2$, and the
fidelity simplifies to the overlap, that is $F(\rho,\sigma)=\langle\psi|\sigma|\psi\rangle$.

A mathematically convenient way to discuss the result of a sequence of
quantum gate operations applied to the initial state $|\psi\rangle$ is through
the language of linear maps and quantum operations~\cite{NIEL00,BREU02}.
In short, a quantum operation transforms an initial density matrix $\rho$
as $\rho'=\mathcal E(\rho)=\sum_\alpha E_\alpha\rho E_\alpha^\dagger$
where the $E_\alpha$ are the so-called operation elements (also known as Kraus operators; not to be confused with gate operations)~\cite{NIEL00}.
The number of different $E_\alpha$'s is at most $D^2$~\cite{NIEL00}.
In general, $\rho\mapsto\mathcal E(\rho)$ defines a completely positive map~\cite{NIEL00,BREU02} which need not be
trace-preserving, meaning that $\mathbf{Tr}\,\sum_\alpha E_\alpha^\dagger E_\alpha$
may be less than one~\cite{NIEL00,BREU02}.
In the special case that $\mathcal E$ is trace-preserving,
we have $\mathbf{Tr}\,\mathcal E(\rho) = \mathbf{Tr}\,\mathcal \rho=1$.

In this language, an ideal quantum gate operation corresponding to a unitary matrix $U$ is described by the map $\mathcal E_{\mathrm{id}}(\rho) = U\rho U^\dagger$.
We imagine that a real quantum device, prepared in a pure state $\rho=|\psi\rangle\langle\psi|$,
actually carries out a slightly different operation denoted by the linear map
$\mathcal E_{\mathrm{ac}}(\rho)$. We consider the quantum operation $\mathcal E(\rho) = \mathcal E_{\mathrm{ac}}(\mathcal E_{\mathrm{id}}^{-1}(\rho)) = \mathcal E_{\mathrm{ac}}(U^\dagger\rho U)$. If the device's implementation of $U$ was perfect, the operation $\mathcal E$ would be equal to the identity operation. The fidelity of the operation is given by
\begin{align}
  \label{QINF0}
  F(\rho,\mathcal E(\rho)) =
  \langle\psi|\mathcal E(|\psi\rangle\langle\psi|)|\psi\rangle.
\end{align}
Clearly, the fidelity Eq.~(\ref{QINF0}) is a numerical measure for how
well a quantum processor performs the gate sequence
represented by the unitary matrix $U$. If the fidelity is equal to one (zero),
the quantum processor is performing perfectly (producing output that has no relation to $U$).

Instead of estimating the fidelities of NISQ devices for a collection
of different states $|\psi\rangle$, randomized benchmarking is motivated by drawing a sample of $|\psi\rangle$'s
from a uniform distribution on the unit sphere and computing the average over the fidelity Eq.~(\ref{QINF0}),
\begin{align}
  \label{eq:fidelityAvg}
  F_{\text{avg}}(\mathcal E)
  = \int \langle\psi|{\mathcal E}(|\psi\rangle\langle\psi|)|\psi\rangle\;  d\!\psi,
\end{align}
where the integral in Eq.~(\ref{eq:fidelityAvg}) is over all pure states or, equivalently,
over all points on the surface of a unit sphere of dimension $2D$. These integrals can be evaluated by the same method that we have used before.
Writing $|\psi\rangle=\sum_{j=1}^D c_i|j\rangle$ (see Eq.~(\ref{PRE0})), we have
\begin{eqnarray}
  \label{eq:fidelityAvgIntegral1}
  F_{\text{avg}}(\mathcal E)
  &=& \sum_\alpha \sum_{j,k=1}^D\sum_{l,m=1}^D \langle j|E_\alpha|k\rangle\langle l|E_\alpha^\dagger|m\rangle
  \int c_j^*c_k^{}c_l^*c_m^{}\;d\mathbf{c}
  \nonumber \\
  &=& \sum_\alpha \sum_{j,k=1}^D\sum_{l,m=1}^D \langle j|E_\alpha|k\rangle\langle l|E_\alpha^\dagger|m\rangle
  \mathrm{E}\left[c_j^*c_k^{}c_l^*c_m^{}\right]
  \;.
\end{eqnarray}
where the integral is over all complex-valued $c$'s such that $\sum_{j=1}^D c_j^*c_j^{}=1$.
Using Eq.~(\ref{PRE1c}) and the results given in the first row of Table~\ref{tab1},
we have $\mathrm{E}\big[c_j^*c_k^{}c_l^*c_m^{}\big]=
(\delta_{j,k}\delta_{l,m} + \delta_{j,m}\delta_{k,l})/D(D+1)$, and Eq.~(\ref{eq:fidelityAvgIntegral1})
becomes
\begin{align}
 \label{eq:fidelityFavgEvaluatedClosedForm}
 F_{\mathrm{avg}}(\mathcal E) = \sum_\alpha
 \frac{\left|\mathbf{Tr}\,E_\alpha\right|^2 +  \mathbf{Tr}\,E_\alpha^\dagger E_\alpha}{D(D+1)}
=
 \frac{D^2F_{\mathrm{ent}}(\mathcal E) +  \sum_\alpha\mathbf{Tr}\,E_\alpha^\dagger E_\alpha}{D(D+1)},
\end{align}
where $F_{\mathrm{ent}}(\mathcal E)=\sum_\alpha \left|\mathbf{Tr}\,E_\alpha\right|^2/D^2$ is
the so-called entanglement fidelity~\cite{SCHU96,NIEL00}.
In the case that $\mathcal E$ is trace preserving we have $\sum_\alpha\mathbf{Tr}\,E_\alpha^\dagger E_\alpha=D$
and Eq.~(\ref{eq:fidelityFavgEvaluatedClosedForm}) reduces to~\cite{horodecki1999fidelity}
\begin{align}
 \label{QINF1}
 F_{\mathrm{avg}}(\mathcal E) =  \frac{DF_{\mathrm{ent}}(\mathcal E) + 1}{D+1}.
\end{align}
The relation Eq.~(\ref{eq:fidelityFavgEvaluatedClosedForm})
generalizes an earlier result~\cite{nielsen2002gatefidelity,Gilchrist2005fidelities}
to the case of non-trace-preserving quantum operations
and also yields simple expressions
for the average and entanglement fidelity in terms of the operation elements $E_\alpha$.

\section{Final remarks}\label{frm}
As mentioned in the introduction, for several decades
many scientists have used random states in their simulation work. In this paper, we have presented a systematic and rigorous analysis of the mathematical foundation of the random state technology as it is being used in numerical simulation.
Applications of this technology in areas of
quantum statistical physics, quantum dynamics, and quantum information processing have been given, with the primary aim to illustrate the power and versatility of the random state technology.

In essence, the random state technology as applied
in the numerical simulation arena reduces the computational burden from ${\cal O}(D^2)$ to ${\cal O}(D)$, where $D$ is the dimension of the Hilbert space used to describe the quantum system. For such applications, there are rigorous bounds on the errors and statistical fluctuations that result
from the use of random states.

We have also shown how the random state technology can help to analyze numerical simulations and experiments that aim for establishing quantum supremacy on a NISQ processor, as well as to prove a statement for non-trace-preserving maps in quantum information theory.

In view of the computational power, flexibility and versatility of the random state technology and the fact that it is based on solid principles, we may expect to see many new applications in the near future.

\section*{Acknowledgements}\label{ACK}
We have profited from discussions with Seiji Miyashita and J\"urgen Schack.
The authors gratefully acknowledge the computing time granted through JARA on
the supercomputer JURECA~\cite{JURECA} at Forschungszentrum J\"ulich.

\appendix
\section{Some useful integrals}\label{SOM}

We first consider the Gaussian random states listed in the first two rows of Table~\ref{tab1}.
Since the expressions of interest
involve averages of $\langle\Phi|X|\Phi\rangle/\langle \Phi|\Phi\rangle$ only,
both choices for $c_j(\bm\xi)$ yield the same expressions
for $\langle\Phi|X|\Phi\rangle/\langle \Phi|\Phi\rangle$.
In other words, the exact expressions to be derived apply to both choices of $c_j(\bm\xi)$.
Some of the integrals used below can also be found in Ref.~\cite{ULLA64}.

Using the symmetries of the probability densities
and the fact that both the real and imaginary part of $c_j(\bm\xi)=c_j(a_1,b_1,\ldots,a_D,b_D)$
are odd functions of each of the $a$'s and $b$'s, we have
\begin{eqnarray}
\mathrm{E}\left[
\frac{\langle\Phi|X|\Phi\rangle}{\langle \Phi|\Phi\rangle}
\right]
&=&
\mathrm{E}\left[
\frac{\sum_{i,j=1}^D
c^{\ast}_i c^{}_j \langle i|X|j\rangle}{
\sum_{i=1}^D
c^{\ast}_i c^{}_i }
\right]
=
\mathrm{E}\left[
\frac{
c^{\ast}_1 c^{}_1}{
\sum_{i=1}^D
c^{\ast}_i c^{}_i }
\right] \mathbf{Tr\;}X
=
\mathrm{E}\left[
\frac{
\xi^{\ast}_1 \xi^{}_1}{
\sum_{i=1}^D
\xi^{\ast}_i \xi^{}_i }
\right] \mathbf{Tr\;}X
\;,
\label{SOM0a}
\end{eqnarray}
and
\begin{eqnarray}
\mathrm{E}\left[
\frac{
\xi^{\ast}_1 \xi^{}_1}{
\sum_{i=1}^D
\xi^{\ast}_i \xi^{}_i }
\right]
&=&
\frac{1}{\pi^D}
\int_{-\infty}^{+\infty}
\frac{a_1^2+b_1^2}{a_1^2+b_1^2+a_2^2+b_2^2+\ldots+a_D^2+b_D^2}e^{-(a_1^2+b_1^2+a_2^2+b_2^2+\ldots+a_D^2+b_D^2)}
da_1db_1\ldots
da_Ddb_D
\nonumber  \\
&=&
\frac{1}{\pi^D}
\int_{S_2}\int_{S_{2D-2}}
\int_{0}^{\infty}
\int_{0}^{\infty}
\frac{r^2}{r^2+R^2}e^{-(r^2+R^2)}
r R^{2D-3}dr\;dR\;d\Omega_2\;d\Omega_{2D-2}
\;.
\label{SOM0b}
\end{eqnarray}
Using $\int_{S_{p}} d\Omega_{p}= 2\pi^{p/2}/\Gamma(p/2)$ and making the change of variables
$r=x\cos\theta$ and $R=x\sin\theta$, we obtain
\begin{eqnarray}
\mathrm{E}\left[
\frac{
\xi^{\ast}_1 \xi^{}_1}{
\sum_{i=1}^D
\xi^{\ast}_i \xi^{}_i }
\right]
&=&
\frac{4}{\Gamma(1)\Gamma(D-1)}
\int_{0}^{\infty}
\int_{0}^{\pi/2}
x^{2D-1} e^{-x^2} \cos^3\theta \sin^{2D-3}\theta
\;dx\;
d\theta
\nonumber \\
&=&
\frac{4}{\Gamma(1)\Gamma(D-1)}\frac{\Gamma(2)\Gamma(D-1)}{4\Gamma(D+1)}
\int_{0}^{\infty}
y^{D-1} e^{-y}\;dy
=
\frac{\Gamma(D)}{\Gamma(D+1)}=\frac{1}{D}
\;,
\label{SOM0}
\end{eqnarray}
where we used the identity
$\int_{0}^{\pi/2} \cos^{2p-1}\theta \sin^{2q-1}\theta  d\theta = \Gamma(p)\Gamma(q)/2\Gamma(p+q)$
and made the substitution $y=x^2$.

Similarly, we have
\begin{eqnarray}
\mathrm{E}\left[
\frac{
(c^{\ast}_1 c^{}_1)^2}{
\left(\sum_{i=1}^D
c^{\ast}_i c^{}_i\right)^2 }
\right]
&=&
\mathrm{E}\left[
\frac{
(\xi^{\ast}_1 \xi^{}_1)^2}{
\left(\sum_{i=1}^D
\xi^{\ast}_i \xi^{}_i\right)^2 }
\right]
\nonumber \\
&=&
\frac{1}{\pi^D}
\int_{-\infty}^{+\infty}
\frac{(a_1^2+b_1^2)^2}{(a_1^2+b_1^2+a_2^2+b_2^2+\ldots+a_D^2+b_D^2)^2}e^{-(a_1^2+b_1^2+a_2^2+b_2^2+\ldots+a_D^2+b_D)^2}
da_1db_1\ldots
da_Ddb_D
\nonumber  \\
&=&
\frac{1}{\pi^D}
\int_{S_2}\int_{S_{2D-2}}
\int_{0}^{\infty}
\int_{0}^{\infty}
\frac{r^4}{(r^2+R^2)^2}e^{-(r^2+R^2)}
r R^{2D-3}dr\;dR\;d\Omega_2\;d\Omega_{2D-2}
\nonumber  \\
&=&
\frac{4}{\Gamma(1)\Gamma(D-1)}
\int_{0}^{\infty}
\int_{0}^{\pi/2}
x^{2D-1} e^{-x^2} \cos^5\theta \sin^{2D-3}\theta
\;dx\;
d\theta
=\frac{2}{D(D+1)}
\;,
\label{SOM2}
\end{eqnarray}
and
\begin{eqnarray}
\mathrm{E}\left[
\frac{
c^{\ast}_1 c^{}_1c^{\ast}_2 c^{}_2}{
\left(\sum_{i=1}^D
c^{\ast}_i c^{}_i\right)^2 }
\right]
&=&
\mathrm{E}\left[
\frac{
\xi^{\ast}_1 \xi^{}_1\xi^{\ast}_2 \xi^{}_2}{
\left(\sum_{i=1}^D
\xi^{\ast}_i \xi^{}_i\right)^2 }
\right]
\nonumber \\
&=&
\frac{1}{\pi^D}
\int_{-\infty}^{+\infty}
\frac{(a_1^2+b_1^2)(a_2^2+b_2^2)}{(a_1^2+b_1^2+a_2^2+b_2^2+\ldots+a_D^2+b_D^2)^2}e^{-(a_1^2+b_1^2+a_2^2+b_2^2+\ldots+a_D^2+b_D^2)^2}
da_1db_1\ldots
da_Ddb_D
\nonumber  \\
&=&
\frac{1}{\Gamma^2(1)\Gamma(D-2)}
\int_{0}^{\infty}
\int_{0}^{\infty}
\int_{0}^{\infty}
\frac{r^2s^2}{(r^2+s^2+R^2)^2}rsR^{2D-5}e^{-(r^2+s^2+R^2)}
dxdydz
\nonumber  \\
&=&
\frac{1}{\Gamma^2(1)\Gamma(D-2)}
\int_{0}^{\pi/2}
\int_{0}^{\pi/2}
\int_{0}^{\infty}
\cos^3\phi\sin^3\phi \sin^5\theta\cos^{2D-5}\theta x^{2D-1}e^{-x^2}
dxd\theta d\phi
\nonumber  \\
&=&
\frac{1}{D(D+1)}
\;,
\label{SOM1}
\end{eqnarray}
from which it follows that
\begin{eqnarray}
\mathrm{E}\left[
\left|\frac{D\langle\Phi|X|\Phi\rangle}{\langle \Phi|\Phi\rangle}\right|^2
\right]
&=&
\frac{\mathbf{Tr\;}XX^\dagger+\left|\mathbf{Tr\;}X\right|^2
}{D(D+1)}
.
\label{SOM3}
\end{eqnarray}
Combining Eqs.~(\ref{SOM0a}), ~(\ref{SOM0}), and~(\ref{SOM3}), we
obtain the exact expressions
\begin{eqnarray}
\mathrm{E}\left[
\frac{D\langle\Phi|X|\Phi\rangle}{\langle \Phi|\Phi\rangle}
\right]=\mathbf{Tr\;}X
\quad,\quad
\mathrm{Var}\left[
\frac{D\langle\Phi|X|\Phi\rangle}{\langle \Phi|\Phi\rangle}
\right]
=
\frac{D\mathbf{Tr\;}XX^\dagger-\left|\mathbf{Tr\;}X\right|^2
}{(D+1)}
.
\label{SOM4}
\end{eqnarray}

In the case of the random phase state (Case C in Table~\ref{tab1}),
$\langle \Phi|\Phi\rangle=1$ by construction.
Then, from Eq.~(\ref{PRE2a}), it follows directly that $\mathrm{E}\left[
{D\langle\Phi|X|\Phi\rangle}/{\langle \Phi|\Phi\rangle}
\right]=\mathbf{Tr\;}X$.
Using Eq.~(\ref{PRE4}), $\langle \Phi|\Phi\rangle=1$, and the expressions
for the moments given in the third line of Table~\ref{tab1}
we find
\begin{eqnarray}
\mathrm{Var}\left[
\frac{D\langle\Phi|X|\Phi\rangle}{\langle \Phi|\Phi\rangle}
\right]
&=&\mathbf{Tr\;}XX^\dagger-\sum_{j=1}^D|\langle j|X|j\rangle|^2
=\sum_{j,k=1}^D (1-\delta_{j,k}) |\langle j|X|k\rangle|^2
.
\label{SOM5}
\end{eqnarray}
Note that unlike in the case of the Gaussian random states,
for the random phase state the variance depends on the choice of the basis states
$\{|j\rangle\,|\,j=1,\ldots,D\}$.

Given a real number $z$, we may ask for the probability density $p(z|D)$ that the $j$-th basis state occurs with probability $p(j)=z$. For a given random state, we have $p(j)=(a_j^2+b_j^2)/(a_1^2+b_1^2+...+a_D^2+b_D^2)$.
By symmetry, all basis states are equivalent and it suffices to consider only $j=1$, for instance.
Because $0\le p(j)\le 1$ we have $p(z|D)=0$ if $z<0$ or $z>1$.
For $0\le z\le 1$ we can use the same tricks as before
and by some elementary algebra we obtain
\begin{eqnarray}
p(z|D)&=&\int\delta\left(z-\frac{a_j^2+b_j^2}{a_1^2+b_1^2+...+a_D^2+b_D^2}\right)
p(a_1,b_1\ldots,a_D,b_D) da_1 \ldots db_D
\nonumber \\
&=&\frac{1}{(2\pi)^D} \frac{2\pi\; 2\pi^{D-1}}{\Gamma(D-1)}
\int_{0}^{\infty}\int_{0}^{\infty}\delta\left(z-\frac{r^2}{r^2+R^2}\right)
e^{-r^2/2} e^{-R^2/2}\; rdr \; R^{2N-3} dR
\nonumber \\
&=&\frac{1}{(2\pi)^D} \frac{2\pi \; 2\pi^{D-1}}{4\Gamma(D-1)}
\int_{0}^{\infty}\int_{0}^{\infty}\delta\left(z-\frac{x}{x+y}\right)  y^{D-2} e^{-x/2} e^{-y/2}\; dx \; dy
\nonumber \\
&=&\frac{1}{2^D} \frac{1}{\Gamma(D-1)}
\int_{0}^{\infty} \frac{y^{D-1}}{(1-z)^2} e^{-y/[2(1-z)]}\; dy
\nonumber \\
&=&\frac{1}{2^D} \frac{2^D\Gamma(D)}{\Gamma(D-1)} (1-z)^{D-2} = (D-1) (1-z)^{D-2}
.
\label{PT0}
\end{eqnarray}
More generally, we can ask for the probability density that $p(j_1),\ldots, p(j_k)$ (all $j$'s different
from each other) take the values $z_1,\ldots,z_k$, respectively.
A calculation similar to the one used to obtain Eq.~(\ref{PT0}) yields for $1\le k<D$
\begin{eqnarray}
p(z_1,z_2,\ldots,z_k|D)&=&(D-1)(D-2)\ldots(D-k)(1-z_1-z_2-\ldots-z_k)^{D-k-1}
\nonumber \\
&&\times\Theta(z_1)\ldots\Theta(z_k)\Theta(1-z_1-z_2-\ldots-z_k)
.
\label{PT1}
\end{eqnarray}
The results of Eqs.~(\ref{SOM0}),~(\ref{SOM2}), and~(\ref{SOM1}) are now readily
obtained by calculating $\int_{0}^1 dz\, zp(z|D)$, $\int_{0}^1 dz\, z^2p(z|D)$, and
$\int_{0}^1 dz\, z_1z_2p(z_1,z_2|D)$, respectively.

If $D$ is large and $zD$ is not, we have
\begin{eqnarray}
p(z|D)&=&(D-1) (1-z)^{D-2}\approx D \left(1-\frac{Dz}{D}\right)^{D}\approx  e^{-Dz},
\label{PT2}
\end{eqnarray}
which is known as Porter-Thomas distribution~\cite{PORT56}.
Note that for evaluating averages over the unit hypersphere there is no advantage of using the (approximate)
Porter-Thomas distribution in place of the exact distribution Eq.~(\ref{PT0}).

With Eqs.~(\ref{PT0}) and Eq.~(\ref{PT1}), it is straightforward to compute the averages,
denoted by $\langle.\rangle$, over all random states $|\Phi\rangle$.
In general, we have
$\left\langle F(p(j))\right\rangle = \int_{0}^1 dz\, F(z)p(z|D)$,
$\left\langle F(p(j))G(p(k\not=j))\right\rangle = \int_{0}^1 \int_{0}^1 dz_1 dz_2\, F(z_1)G(z_2)p(z_1z_2|D)$,  etc.
Specifically we have
\begin{eqnarray}
\frac{\sum_{j=0}^{D-1} \left\langle p^\mu(j)\log p(j)\right\rangle}{\sum_{j=0}^{D-1} \left\langle p^\mu(j)\right\rangle}
&=&
\mathrm{PolyGamma}(0,D+\mu) - \mathrm{PolyGamma}(0, \mu+1)
\nonumber \\
&\overset{D\to\infty}{\approx}&\log D - \mathrm{PolyGamma}(0, \mu+1)
=\left\{
\begin{array}{lr}
\log D + \gamma &,\; \mu=0\\
\log D + \gamma - 1 &,\; \mu=1\\
\log D + \gamma - 3/2 &,\; \mu=2
\end{array}
\right.
,
\label{SUB2}
\end{eqnarray}
where $\gamma$ is Euler's constant.
For large $D$, the variance on the entropy is found to be
\begin{eqnarray}
\langle (\sum_{j=0}^{D-1}p(j)\log p(j))^2 \rangle - \langle \sum_{j=0}^{D-1}p(j)\log p(j) \rangle^2=\frac{\pi^2/3 +
(\log D +\gamma)^2-4(\log D +\gamma)+1
}{D^2} + {\cal O}(\frac{1}{D^3})
\;.
\nonumber \\
\label{SUB3}
\end{eqnarray}

\section{Approximate treatment}

\begin{table*}[t]
\caption{%
Expressions for the various terms that appear in Eqs.~(\ref{AE0}) and~(\ref{AE1}),
corresponding to the combination of probabilities $P(\bm\xi)$ and amplitudes $c_j(\bm\xi)$ listed in Table~\ref{tab1}.
We use the shorthand $x=D\langle\Phi|X|\Phi\rangle$ and $y=\langle\Phi|\Phi\rangle$.
The last two columns show the result of keeping the first three terms in the expansions
Eqs.~(\ref{AE0}) and~(\ref{AE1}).
}
\begin{ruledtabular}
\begin{tabular}{cccccccc}
$\mathrm{E}\left[y\right]$  &  $\mathrm{E}\left[x\right]$ & $\mathrm{E}\left[xy\right]$  &
$\mathrm{cov}\left[x,y\right]$ & $\mathrm{Var}\left[y\right]$ & $\mathrm{Var}\left[x\right]$ &
$\mathrm{E}\left[\frac{x}{y}\right]$ & $\mathrm{Var}\left[\frac{x}{y}\right]$\\
\noalign{\medskip}\hline\noalign{\medskip}
$1$ & $\mathbf{Tr\;}X$ & $\mathbf{Tr\;}X$ & $0$ & $0$ & $\frac{D\mathbf{Tr\;}XX^\dagger-\left|\mathbf{Tr\;}X\right|^2}{D+1}$ & $\mathbf{Tr\;}X$ & $\frac{D\mathbf{Tr\;}XX^\dagger-\left|\mathbf{Tr\;}X\right|^2}{D+1}$ \\
$1$ & $\mathbf{Tr\;}X$ & $\frac{D+1}{D}\mathbf{Tr\;}X$ & $\frac{1}{D}\mathbf{Tr\;}X$ & $\frac{1}{D}$ & $\mathbf{Tr\;}XX^\dagger$ & $\mathbf{Tr\;}X$ & $\frac{D\mathbf{Tr\;}XX^\dagger-\left|\mathbf{Tr\;}X\right|^2}{D}$ \\
$1$ & $\mathbf{Tr\;}X$ & $\mathbf{Tr\;}X$ & $0$ & $0$ & $\mathbf{Tr\;}XX^\dagger-\sum_{i=1}^D |\langle i|X|i\rangle|^2$ & $\mathbf{Tr\;}X$ & $\mathbf{Tr\;}XX^\dagger-\sum_{i=1}^D |\langle i|X|i\rangle|^2$ \\
\end{tabular}
\end{ruledtabular}
\label{tab2}
\end{table*}

The presence of the fraction in Eq.~(\ref{PRE3z}) makes it difficult to derive a generally valid
expression for the average and the variance.
However, generally valid approximations can be obtained through the
use of the multivariate Taylor expansion for the average
\begin{eqnarray}
\mathrm{E}\left[
\frac{x}{y}\right]
\approx
\frac{\mathrm{E}[x]}{\mathrm{E}[y]}
-\frac{\mathrm{E}[xy]-\mathrm{E}[x]\mathrm{E}[y]}{\mathrm{E}^2[y]}
+\frac{\mathrm{E}[x]\mathrm{Var}[y]}{\mathrm{E}^3[y]}+\ldots
,
\label{AE0}
\end{eqnarray}
and a similar expansion for the variance
\begin{eqnarray}
\mathrm{Var}\left[\frac{x}{y}\right]
\approx
\frac{\mathrm{Var}[x]}{\mathrm{E}^2[y]}-2\frac{\mathrm{E}[x](\mathrm{E}[xy]-\mathrm{E}[x]\mathrm{E}[y])}{\mathrm{E}^3[y]}
+\frac{\mathrm{E^2}[x]\mathrm{Var}[y]}{\mathrm{E}^4[y]}+\ldots
.
\label{AE1}
\end{eqnarray}
where, $x$ and $y$ can be any functions of the random variables $\bm\xi$.

In Section~\ref{PRE}, we give explicit expression for the expectation values of $x$, $y$, etc., in terms
of the moments that are listed in Table~\ref{tab1}.
Table~\ref{tab2} summarizes the results of the calculations.
It directly follows that up to the three terms of the expansions shown
in Eq.~(\ref{AE0}), $\mathrm{E}\left[x/y\right]=\mathbf{Tr\;}X$, for each of the
choices listed in Table~\ref{tab1}
and that the expressions for the variances agree with those given by Eq.~(\ref{PRE4}).

\section{Sampling over random states}\label{SRS}
In the main text, we focus on the calculation of the trace by using only one random state.
Here we consider the case that further averaging over $R>1$ different random states is
necessary to produce results with good statistics.
This is necessary for small systems or at very low temperature.

We assume that samples of pairs of variables $(x_i, y_i)$ for $i=1,\ldots,R$
have been obtained from different independent realizations of a random variable $\bm\xi$.
Obviously, for $i\neq j$, $\mathrm{E}[x_i y_j]=\mathrm{E}[x_i]\mathrm{E}[y_j]=\mathrm{E}[x]\mathrm{E}[y]=\mu_x\mu_y$ and
the correlation vanishes.
For $i=j$, $\mathrm{E}[x_i y_i]\neq\mathrm{E}[x]\mathrm{E}[y]=\mu_x\mu_y$, i.e., there can be a non-zero correlation
because $x_i$ and $y_i$ are obtained from the same realization of the random variable $\bm\xi$.
Furthermore, $\mathrm{E}[x_i^2]=\mathrm{E}[x^2]$ and $\mathrm{E}[y_i^2]=\mathrm{E}[y^2]$.

There are two ways to average over the $R$ samples~\cite{SUGI13,SCHN20}, namely
\begin{eqnarray}
M1 &=& \frac{1}{R} \sum_{i=1}^R \frac{x_i} {y_i} ,
\label{M1}
\end{eqnarray}
or
\begin{eqnarray}
M2 &=& \frac{ \sum_{i=1}^R x_i} { \sum_{i=1}^R y_i}.
\label{M2}
\end{eqnarray}
Before we use Eqs.~(\ref{AE0}) and~(\ref{AE1}) to analyze the mean and variance for
the two different ways of computing the average of the $R$ samples,
we first recall the well-known formulas for the ensemble mean and variance.
We have
\begin{eqnarray}
\mathrm{E}\left[\frac{1}{R}\sum_{i=1}^R x_i \right] &=& \frac{1}{R}\sum_{i=1}^R \mathrm{E}[x_i] = \mu_x, \\
\mathrm{E}\left[\frac{1}{R}\sum_{i=1}^R y_i \right] &=& \frac{1}{R}\sum_{i=1}^R \mathrm{E}[y_i] = \mu_y, \\
\mathrm{Var}\left[\frac{1}{R}\sum_{i=1}^R x_i\right] &=& \mathrm{E}\left[(\frac{1}{R}\sum_{i=1}^R x_i)^2\right] - \mu_x^2
= \mathrm{E}\left[\frac{1}{R^2}\sum_{i=1}^R x_i^2+\frac{1}{R^2}\sum_{i\neq j}^R x_i x_j\right] -\mu_x^2  \cr
&=& \frac{1}{R^2}\sum_{i=1}^R \mathrm{E}[x_i^2]+\frac{1}{R^2}\sum_{i\neq j}^R \mathrm{E}[x_i x_j] -\mu_x^2 
=\frac{1}{R}\mathrm{E}[x^2] + \frac{R-1}{R} \mathrm{E^2}[x] -\mu_x^2
\nonumber \\
&=&
\frac{\mathrm{E}[x^2]-\mu_x^2}{R} 
=\frac{\mathrm{Var}[x]}{R},    \\
\mathrm{Var}\left[\frac{1}{R}\sum_{i=1}^R y_i\right] &=& \frac{\mathrm{Var}[y]}{R} ,  \\
\mathrm{Cov}\left[\frac{1}{R}\sum_{i=1}^R x_i\frac{1}{R}\sum_{j=1}^R y_j \right]
&=& \mathrm{E}\left[\frac{1}{R}\sum_{i=1}^R x_i\frac{1}{R}\sum_{j=1}^R y_j\right]
-\mathrm{E}\left[\frac{1}{R}\sum_{i=1}^R x_i\right]\mathrm{E}\left[\frac{1}{R}\sum_{j=1}^R y_j\right]   \cr
&=& \mathrm{E}\left[\frac{1}{R^2}\sum_{i=1}^R x_i y_i + \frac{1}{R^2}\sum_{i\neq j}^R x_i y_j\right]  - \mu_x\mu_y   
=\frac{1}{R} \mathrm{E}[xy] +  \frac{R-1}{R}\mathrm{E}[x]\mathrm{E}[y]-\mu_x\mu_y  
\nonumber \\
&=&
\frac{\mathrm{Cov[xy]}}{R}.
\end{eqnarray}
Therefore, for the first way of averaging Eq.~(\ref{M1}), we have
\begin{eqnarray}
\mathrm{E}[M1] &=& \mathrm{E} \left[\frac{x}{y}\right]
\approx
\frac{\mathrm{E}[x]}{\mathrm{E}[y]}
-\frac{\mathrm{E}[xy]-\mathrm{E}[x]\mathrm{E}[y]}{\mathrm{E}^2[y]}
+\frac{\mathrm{E}[x]\mathrm{Var}[y]}{\mathrm{E}^3[y]}+\ldots
,
\label{C9a}
\nonumber  \\
\mathrm{Var}[M1] &=&  \frac{\mathrm{Var}\left[\frac{x}{y}\right] }{R},
\label{C10}
\end{eqnarray}
whereas for the second way of averaging Eq.~(\ref{M2}), we obtain
\begin{eqnarray}
\mathrm{E}[M2] &=& \mathrm{E} \left[\frac{ \frac{1}{R}\sum_{i=1}^R x_i} { \frac{1}{R}\sum_{i=1}^R y_i}\right]
\approx
\frac{\mathrm{E}[x]}{\mathrm{E}[y]}
-\frac{1}{R}\frac{\mathrm{E}[xy]-\mathrm{E}[x]\mathrm{E}[y]}{\mathrm{E}^2[y]}
+\frac{1}{R}\frac{\mathrm{E}[x]\mathrm{Var}[y]}{\mathrm{E}^3[y]}+\ldots,
\label{C11a}
\nonumber \\
\mathrm{Var}[M2] &=&  \mathrm{Var}\left[\frac{ \frac{1}{R}\sum_{i=1}^R x_i} { \frac{1}{R}\sum_{i=1}^R y_i}\right]
\nonumber \\
&\approx&
\frac{1}{R}\frac{\mathrm{Var}[x]}{\mathrm{E}^2[y]}-2\frac{1}{R}\frac{\mathrm{E}[x](\mathrm{E}[xy]-\mathrm{E}[x]\mathrm{E}[y])}{\mathrm{E}^3[y]}
+\frac{1}{R}\frac{\mathrm{E^2}[x]\mathrm{Var}[y]}{\mathrm{E}^4[y]}+\ldots  
=\frac{\mathrm{Var}\left[\frac{x}{y}\right] }{R}.
\label{C11}
\end{eqnarray}
Comparing Eqs.~(\ref{C10}) and~(\ref{C11}),
it is clear that the variances show the same $1/R$ dependence (as usual for independent samples).
However, the presence of the factor $1/R$ in the second and third term of the expression for the mean Eq.~(\ref{C11a})
obtained using Eq.~(\ref{M2}) implies that these correction terms vanish as $R\to \infty$,
unlike in the case the mean Eq.~(\ref{C9a}) is computed according to Eq.~(\ref{M1}).
In other words, if possible, we should use Eq.~(\ref{M2}) to compute the average of independent realizations of
the random states.

\section{Numerical methods for uniformly picking points from the unit sphere}\label{PICK}

\subsection{Muller's method}\label{MULL}
Muller's method\cite{MULL59} for generating a vector $\mathbf{x}$ that is distributed uniformly on the $2D$-dimensional unit sphere consists of the following steps:
\begin{enumerate}
    \item
    Using the Box-Muller method~\cite{PRES03} to generate $D$ pairs of real-valued pseudo-random numbers with a normal (Gaussian) distribution and use these pairs to assign values to the elements of $\mathbf{x}$.
    \item
    Compute the Euclidean norm $\Vert\mathbf{x}\Vert$ of $\mathbf{x}$
    and replace $\mathbf{x}$ by $\mathbf{x}/\Vert\mathbf{x}\Vert$
\end{enumerate}
We now show that this procedure indeed generates points that are distrubuted uniformly over the $2D$-dimensional unit sphere.

Specifying an unnormalized vector in the $D$-dimensional Hilbert space
requires $2D$ real numbers. In this subsection, we simplify the notation
by introducing the symbol $d=2D$.
We start from the normalized Gaussian distribution
\begin{eqnarray}
p(x_{1},\ldots,x_{d})=\frac{1}{\pi^{d/2}} e^{-x_{1}^2-\ldots-x_{d}^2}
,
\label{gauss0}
\end{eqnarray}
and write it in spherical coordinates
\begin{eqnarray}
x_{1}&=&r\cos\theta_1\sin\theta_2\sin\theta_{3}\ldots \sin\theta_{d-1}
\nonumber \\
x_{2}&=&r\sin\theta_1\sin\theta_2\sin\theta_{3}\ldots \sin\theta_{d-1}
\nonumber \\
x_{3}&=&\phantom{\cos\theta_1}r\cos\theta_2\sin\theta_{3}\ldots \sin\theta_{d-1}
\nonumber \\
x_{4}&=&\phantom{\cos\theta_1\cos\theta_2}r\cos\theta_{3}\ldots \sin\theta_{d-1}
\nonumber \\
\ldots&=& \ldots
\nonumber \\
x_{d-1}&=&\phantom{\cos\theta_1\cos\theta_2}r\cos\theta_{d-2}\sin\theta_{d-1}
\nonumber \\
x_{d}&=&\phantom{\cos\theta_1\cos\theta_2\sin\theta_{d-2}}r\cos\theta_{d-1}
,
\label{gauss4}
\end{eqnarray}
where $0\le \theta_1 \le 2\pi$, $0\le \theta_j \le \pi$ for $j>1$, and the condition $x_1^2+\ldots+x_d^2=r^2$
is automatically satisfied.
The Jacobian $|J|$ of this transformation reads~\cite{MILL64}
\begin{equation}
|J|=r^{d-1} \sin^{d-2}\theta_{d-1} \sin^{d-3}\theta_{d-2}\ldots\sin\theta_{2}
,
\label{gauss5}
\end{equation}
and we have
\begin{equation}
p(r,\theta_1,\ldots,\theta_d)
=\frac{1}{\pi^{d/2}} r^{d-1}e^{-r^2}  \sin^{d-2}\theta_{d-1} \sin^{d-3}\theta_{d-2}\ldots\sin\theta_{2}
.
\label{gauss6}
\end{equation}
Using the identity
\begin{equation}
\int_0^{\infty} r^{d-1}e^{-r^2} dr = \frac{\Gamma(d/2)}{2}
,
\label{gauss7}
\end{equation}
and $\int p(x_1,\ldots,x_d) dx_1\ldots dx_d=1$, we can write Eq.~(\ref{gauss6}) as
\begin{equation}
p(r,\theta_1,\ldots,\theta_d) = p(r)p(\theta_1,\ldots,\theta_d)
,
\label{gauss8}
\end{equation}
where
\begin{eqnarray}
p(r)&=& \frac{2}{\Gamma(d/2)}  r^{d-1}e^{-r^2/2},
\\
\noalign{\noindent and}
p(\theta_1,\ldots,\theta_n)&=&\frac{\Gamma(d/2)}{2\pi^{d/2}} \sin^{d-2}\theta_{d-1} \sin^{d-3}\theta_{d-2}\ldots\sin\theta_{2}
.
\label{gauss9}
\end{eqnarray}

In these spherical coordinates, the infinitesimal volume element of the unit hypersphere reads
\begin{eqnarray}
\Omega_{d}&=&\sin^{d-2}\theta_{d-1} \sin^{d-3}\theta_{d-2}\ldots\sin\theta_{2}\;
d\theta_{d-1}\ldots
d\theta_{2}
d\theta_{1}
.
\label{gauss9a}
\end{eqnarray}
As $p(x_{1},\ldots,x_{d})$ is a properly normalized density and
$\int_{0}^{\infty} p(r) =1$, if follows from Eq.~(\ref{gauss8}) that
the integral of $p(\theta_1,\ldots,\theta_d)$ over all angles
must be equal to one.
Integrating Eq.~(\ref{gauss9}) over the surface of the $d$-dimensional unit hypersphere $S_{d}$
we find that $\int_{S_{d}} d\Omega_{d}= 2\pi^{d/2}/\Gamma(d/2)$.

From Eqs.~(\ref{gauss8}) and Eq.~(\ref{gauss9})
it immediately follows that $r$ and the $\theta$'s are all independent random numbers.
Denoting $\Vert x\Vert^2=x_1^2+\ldots+x_d^2$, the probability density for picking a point
$x=(x_1,\ldots,x_d)$ on a hypersphere of radius R is
\begin{eqnarray}
{\widehat p}(R)&=&\int \delta( \Vert x \Vert -R) p(x_1,\ldots,x_d) dx_1\ldots dx_d
\nonumber \\
&=&\int \delta(\Vert x\Vert -R) p(r)p(\theta_1,\ldots,\theta_d)
\nonumber \\
&=&\int \delta(r-R) p(r) =  \frac{2\pi^{d/2}}{\Gamma(d/2)}  R^{d-1}e^{-d^2/2}
,
\label{gauss10}
\end{eqnarray}
showing that because of the independence of $r$ and the $\theta$'s, ${\widehat p}(R)$ is independent
of the direction of the vector $x=(x_1,\ldots,x_n)$.
Therefore, the distribution of points $x_i/\Vert x\Vert$ is uniform over the sphere.

\subsection{An alternative to Muller's method}\label{ALTMUL}
The following is based on material presented in Ref.~\cite{BENG06}, pages 111 and 182.
In contrast to representation Eq.~(\ref{gauss4}) where we use $2D$ spherical coordinates to encode
pairs of real numbers that determine the amplitude of a basis state,
the representation that we adopt in this section uses $D-1$ spherical coordinates
to encode the square root of the probabilities of the basis states and another $D$ random numbers $\{\nu_1,\ldots,\nu_D\}$ to encode the phases.
We represent the coefficients of the state vector by $\{x_1 e^{i\phi_1},\ldots,x_D e^{i\phi_D}\}$ where
for all $1\le j\le D$, $0\le\phi_j\le 2\pi$, $x_j\ge0$ and $x_1^2+\ldots+x_D^2=r^2$.
We will set $r=1$ later.
The $x_j$'s and $\phi_j$ are called  octant coordinates~\cite{BENG06}.
In terms of the spherical coordinates, we have
\begin{eqnarray}
x_{1}&=&r\cos\theta_1\sin\theta_2\sin\theta_{3}\ldots \sin\theta_{D-1}
\nonumber \\
x_{2}&=&r\sin\theta_1\sin\theta_2\sin\theta_{3}\ldots \sin\theta_{D-1}
\nonumber \\
x_{3}&=&\phantom{\cos\theta_1}r\cos\theta_2\sin\theta_{3}\ldots \sin\theta_{D-1}
\nonumber \\
x_{4}&=&\phantom{\cos\theta_1\cos\theta_2}r\cos\theta_{3}\ldots \sin\theta_{D-1}
\nonumber \\
\ldots&=& \ldots
\nonumber \\
x_{D-1}&=&\phantom{\cos\theta_1\cos\theta_2}r\cos\theta_{D-2}\sin\theta_{D-1}
\nonumber \\
x_{D}&=&\phantom{\cos\theta_1\cos\theta_2\sin\theta_{D-2}}r\cos\theta_{D-1}
,
\label{OCT0}
\end{eqnarray}
where $0\le \theta_j \le \pi/2$ and the conditions $0\le x_j$ and $x_1^2+\ldots+x_D^2=r^2$
are automatically satisfied.

We can find the volume element of the $2D$-dimensional sphere in these coordinates as follows.
First consider a single complex number $z=u+iv=xe^{i\phi}$ and compute the
Jacobian for the transformation $(u,v)\to(x,\phi)$ to find that $du\,dv = x dx\, d\phi$.
For $D$ complex variables, the volume element in Cartesian coordinates
is $du_1\,dv_1\ldots du_D\,dv_D=x_1\ldots x_D dx_1\ldots dx_D\; d\phi_1\ldots d\phi_D$.
Next, we use the spherical coordinates (restricted to the first octant)
for $x_1,\ldots, x_D$, see Eq.~(\ref{OCT0}).
The Jacobian of this transformation is given by Eq.~(\ref{gauss5}) with $n=D$.
Collecting all sines and cosines we find
\begin{eqnarray}
d\Omega_D&=&r^{2D-1}\prod_{k=1}^{D-1} \cos\theta_{k}(\sin\theta_{k})^{2k-1}\;d\theta_k\;d\phi_k \;d\phi_D
=r^{2D-1}\prod_{k=1}^{D-1} y^{2k-1}\;d y_k\;d\phi_k\;d\phi_D
\label{OCT1b}
\end{eqnarray}
where $y_k=\sin\theta_{k}$.
Integrating Eq.~(\ref{OCT1}) over $0\le r \le R$, $0\le y_k\le 1$ and $0 \le \phi_k \le 2\pi$
we find
\begin{eqnarray}
V_D=\int d\Omega_D&=& \frac{R^{2D}}{2D} \frac{(2\pi)^D}{2^{D-1} (D-1)!}= \frac{\pi^D R^{2D}}{D!},
\label{OCT1a}
\end{eqnarray}
which is the volume of a $2D$-dimensional sphere of radius $R$, as expected.

The probability density for the random variables $\{y_1,\ldots,y_{D-1},\phi_1,\ldots,\phi_D\}$
reads
\begin{eqnarray}
p(y_1,\ldots,y_{D-1},\phi_1,\ldots,\phi_D)&=&
\frac{1}{(2\pi)^D} \prod_{k=1}^{D-1} (2k)y_k^{2k-1}
.
\label{OCT1}
\end{eqnarray}
From Eq.~(\ref{OCT1}), it is easy to find the probability density for any of the $x$'s.
For instance, the probability that $x_{D}^2$ is less than $z$ is given by
\begin{eqnarray}
P(x_{D}^2<z) = P(1-y_{D-1}^2<z)&=&
\left(
\prod_{k=1}^{D-2} \int_0^1 (2k)y_k^{2k-1}\;d y_k
\right)
\int_{\sqrt{1-z}}^1 2(D-1) y_{D-1}^{2D-3}\;d y_{D-1}
\nonumber \\&=&
(D-1)\int_{1-z}^1 x^{D-2}\;d x
,
\label{OCT1z}
\end{eqnarray}
from which, by differentiation with respect to $z$, we find $p(z|D)=(D-1)(1-z)^{D-2}$, see Eq.~(\ref{PT0}).

Unlike with the standard spherical coordinates, see Eq.~(\ref{gauss9}),
the expression Eq.~(\ref{OCT1}) allows us to sample the angles $\theta_{k}$'s, or equivalently, the $y_k$'s independently.
The probability density and probability for $y_k$ are given by
\begin{eqnarray}
p(y_k)&=&2k y^{2k-1}
\nonumber \\
P(y_k \le Y_k)&=&2k \int_0^{Y_k}  y^{2k-1}\;d y_k = Y^{2k}_k
,
\label{OCT2}
\end{eqnarray}
respectively, from which it follow that in order to generate a random number $Y_k$ with
the correct distribution, we simply have to generate a uniform random variable $r_k$ and put $Y_k=r_k^{1/2k}$.
In practice, $k$ can be very large and numerically, we should use $Y_k^2=\exp(\log(r)/k)\approx 1-\log(r)/k$
if $\log(r)/k$ is very small.
For numerical purposes, we find that M\"uller's method is more convenient, in particular if $D$ is large.

\bibliography{90044}

\begin{thebibliography}{124}%
\makeatletter
\providecommand \@ifxundefined [1]{%
 \@ifx{#1\undefined}
}%
\providecommand \@ifnum [1]{%
 \ifnum #1\expandafter \@firstoftwo
 \else \expandafter \@secondoftwo
 \fi
}%
\providecommand \@ifx [1]{%
 \ifx #1\expandafter \@firstoftwo
 \else \expandafter \@secondoftwo
 \fi
}%
\providecommand \natexlab [1]{#1}%
\providecommand \enquote  [1]{``#1''}%
\providecommand \bibnamefont  [1]{#1}%
\providecommand \bibfnamefont [1]{#1}%
\providecommand \citenamefont [1]{#1}%
\providecommand \href@noop [0]{\@secondoftwo}%
\providecommand \href [0]{\begingroup \@sanitize@url \@href}%
\providecommand \@href[1]{\@@startlink{#1}\@@href}%
\providecommand \@@href[1]{\endgroup#1\@@endlink}%
\providecommand \@sanitize@url [0]{\catcode `\\12\catcode `\$12\catcode
  `\&12\catcode `\#12\catcode `\^12\catcode `\_12\catcode `\%12\relax}%
\providecommand \@@startlink[1]{}%
\providecommand \@@endlink[0]{}%
\providecommand \url  [0]{\begingroup\@sanitize@url \@url }%
\providecommand \@url [1]{\endgroup\@href {#1}{\urlprefix }}%
\providecommand \urlprefix  [0]{URL }%
\providecommand \Eprint [0]{\href }%
\providecommand \doibase [0]{http://dx.doi.org/}%
\providecommand \selectlanguage [0]{\@gobble}%
\providecommand \bibinfo  [0]{\@secondoftwo}%
\providecommand \bibfield  [0]{\@secondoftwo}%
\providecommand \translation [1]{[#1]}%
\providecommand \BibitemOpen [0]{}%
\providecommand \bibitemStop [0]{}%
\providecommand \bibitemNoStop [0]{.\EOS\space}%
\providecommand \EOS [0]{\spacefactor3000\relax}%
\providecommand \BibitemShut  [1]{\csname bibitem#1\endcsname}%
\let\auto@bib@innerbib\@empty
\bibitem [{\citenamefont {Wigner}(1955)}]{WIGN55}%
  \BibitemOpen
  \bibfield  {author} {\bibinfo {author} {\bibfnamefont {E.P.}\ \bibnamefont
  {Wigner}},\ }\bibfield  {title} {\enquote {\bibinfo {title} {Characteristic
  vectors of bordered matrices with infinite dimensions},}\ }\href
  {http://www.jstor.org/stable/1970079} {\bibfield  {journal} {\bibinfo
  {journal} {Ann. Math.}\ }\textbf {\bibinfo {volume} {62}},\ \bibinfo {pages}
  {548--564} (\bibinfo {year} {1955})}\BibitemShut {NoStop}%
\bibitem [{\citenamefont {Bohigas}\ \emph {et~al.}(1984)\citenamefont
  {Bohigas}, \citenamefont {Giannoni},\ and\ \citenamefont {Schmit}}]{BOHI84}%
  \BibitemOpen
  \bibfield  {author} {\bibinfo {author} {\bibfnamefont {O.}~\bibnamefont
  {Bohigas}}, \bibinfo {author} {\bibfnamefont {M.~J.}\ \bibnamefont
  {Giannoni}}, \ and\ \bibinfo {author} {\bibfnamefont {C.}~\bibnamefont
  {Schmit}},\ }\bibfield  {title} {\enquote {\bibinfo {title} {Characterization
  of chaotic quantum spectra and universality of level fluctuation laws},}\
  }\href {\doibase 10.1103/PhysRevLett.52.1} {\bibfield  {journal} {\bibinfo
  {journal} {Phys. Rev. Lett.}\ }\textbf {\bibinfo {volume} {52}},\ \bibinfo
  {pages} {1--4} (\bibinfo {year} {1984})}\BibitemShut {NoStop}%
\bibitem [{\citenamefont {Russell}\ \emph {et~al.}(2017)\citenamefont
  {Russell}, \citenamefont {Chakhmakhchyan}, \citenamefont {{O'Brien}},\ and\
  \citenamefont {Laing}}]{RUSS17}%
  \BibitemOpen
  \bibfield  {author} {\bibinfo {author} {\bibfnamefont {N.J.}\ \bibnamefont
  {Russell}}, \bibinfo {author} {\bibfnamefont {L.}~\bibnamefont
  {Chakhmakhchyan}}, \bibinfo {author} {\bibfnamefont {J.L.}\ \bibnamefont
  {{O'Brien}}}, \ and\ \bibinfo {author} {\bibfnamefont {A.}~\bibnamefont
  {Laing}},\ }\bibfield  {title} {\enquote {\bibinfo {title} {Direct dialling
  of {Haar} random unitary matrices},}\ }\href {\doibase
  10.1088/1367-2630/aa60ed} {\bibfield  {journal} {\bibinfo  {journal} {New J.
  Phys.}\ }\textbf {\bibinfo {volume} {19}},\ \bibinfo {pages} {033007}
  (\bibinfo {year} {2017})}\BibitemShut {NoStop}%
\bibitem [{\citenamefont {Boixo}\ \emph {et~al.}(2018)\citenamefont {Boixo},
  \citenamefont {Isakov}, \citenamefont {Smelyanskiy}, \citenamefont {Ding},
  \citenamefont {Jiang}, \citenamefont {Bremner}, \citenamefont {Martinis},\
  and\ \citenamefont {Neven}}]{BOIX18}%
  \BibitemOpen
  \bibfield  {author} {\bibinfo {author} {\bibfnamefont {S.}~\bibnamefont
  {Boixo}}, \bibinfo {author} {\bibfnamefont {S.V.}\ \bibnamefont {Isakov}},
  \bibinfo {author} {\bibfnamefont {R.}~\bibnamefont {Smelyanskiy},
  \bibfnamefont {V.~N.~Babbush}}, \bibinfo {author} {\bibfnamefont
  {N.}~\bibnamefont {Ding}}, \bibinfo {author} {\bibfnamefont {Z.}~\bibnamefont
  {Jiang}}, \bibinfo {author} {\bibfnamefont {M.J.}\ \bibnamefont {Bremner}},
  \bibinfo {author} {\bibfnamefont {J.M.}\ \bibnamefont {Martinis}}, \ and\
  \bibinfo {author} {\bibfnamefont {H.}~\bibnamefont {Neven}},\ }\bibfield
  {title} {\enquote {\bibinfo {title} {Characterizing quantum supremacy in
  near-term devices},}\ }\href@noop {} {\bibfield  {journal} {\bibinfo
  {journal} {Nat. Phys.}\ }\textbf {\bibinfo {volume} {14}},\ \bibinfo {pages}
  {595} (\bibinfo {year} {2018})}\BibitemShut {NoStop}%
\bibitem [{\citenamefont {Edelman}\ and\ \citenamefont
  {Rao}(2005)}]{edelman2005randomMatrixTheory}%
  \BibitemOpen
  \bibfield  {author} {\bibinfo {author} {\bibfnamefont {A.}~\bibnamefont
  {Edelman}}\ and\ \bibinfo {author} {\bibfnamefont {N.R.}\ \bibnamefont
  {Rao}},\ }\bibfield  {title} {\enquote {\bibinfo {title} {Random matrix
  theory},}\ }\href {\doibase 10.1017/S0962492904000236} {\bibfield  {journal}
  {\bibinfo  {journal} {Acta Numer.}\ }\textbf {\bibinfo {volume} {14}},\
  \bibinfo {pages} {233--297} (\bibinfo {year} {2005})}\BibitemShut {NoStop}%
\bibitem [{\citenamefont {Deutsch}(1991)}]{DEUT91}%
  \BibitemOpen
  \bibfield  {author} {\bibinfo {author} {\bibfnamefont {J.~M.}\ \bibnamefont
  {Deutsch}},\ }\bibfield  {title} {\enquote {\bibinfo {title} {Quantum
  statistical mechanics in a closed system},}\ }\href@noop {} {\bibfield
  {journal} {\bibinfo  {journal} {Phys. Rev. A}\ }\textbf {\bibinfo {volume}
  {43}},\ \bibinfo {pages} {2046} (\bibinfo {year} {1991})}\BibitemShut
  {NoStop}%
\bibitem [{\citenamefont {Srednicki}(1994)}]{SRED94}%
  \BibitemOpen
  \bibfield  {author} {\bibinfo {author} {\bibfnamefont {M.}~\bibnamefont
  {Srednicki}},\ }\bibfield  {title} {\enquote {\bibinfo {title} {Chaos and
  quantum thermalization},}\ }\href {\doibase 10.1103/PhysRevE.50.888}
  {\bibfield  {journal} {\bibinfo  {journal} {Phys. Rev. E}\ }\textbf {\bibinfo
  {volume} {50}},\ \bibinfo {pages} {888--901} (\bibinfo {year}
  {1994})}\BibitemShut {NoStop}%
\bibitem [{\citenamefont {Rigol}\ \emph {et~al.}(2008)\citenamefont {Rigol},
  \citenamefont {Dunjko},\ and\ \citenamefont {Olshanii}}]{REGO08}%
  \BibitemOpen
  \bibfield  {author} {\bibinfo {author} {\bibfnamefont {M.}~\bibnamefont
  {Rigol}}, \bibinfo {author} {\bibfnamefont {V.}~\bibnamefont {Dunjko}}, \
  and\ \bibinfo {author} {\bibfnamefont {M.}~\bibnamefont {Olshanii}},\
  }\bibfield  {title} {\enquote {\bibinfo {title} {Thermalization and its
  mechanism for generic isolated quantum systems},}\ }\href@noop {} {\bibfield
  {journal} {\bibinfo  {journal} {Nature}\ }\textbf {\bibinfo {volume} {452}},\
  \bibinfo {pages} {854--858} (\bibinfo {year} {2008})}\BibitemShut {NoStop}%
\bibitem [{\citenamefont {Goldstein}\ \emph {et~al.}(2006)\citenamefont
  {Goldstein}, \citenamefont {Lebowitz}, \citenamefont {Tumulka},\ and\
  \citenamefont {Zangh{\`{i}}}}]{GOLD06}%
  \BibitemOpen
  \bibfield  {author} {\bibinfo {author} {\bibfnamefont {S.}~\bibnamefont
  {Goldstein}}, \bibinfo {author} {\bibfnamefont {J.~L.}\ \bibnamefont
  {Lebowitz}}, \bibinfo {author} {\bibfnamefont {R.}~\bibnamefont {Tumulka}}, \
  and\ \bibinfo {author} {\bibfnamefont {N.}~\bibnamefont {Zangh{\`{i}}}},\
  }\bibfield  {title} {\enquote {\bibinfo {title} {{Canonical typicality}},}\
  }\href@noop {} {\bibfield  {journal} {\bibinfo  {journal} {Phys. Rev. Lett.}\
  }\textbf {\bibinfo {volume} {96}},\ \bibinfo {pages} {050403} (\bibinfo
  {year} {2006})}\BibitemShut {NoStop}%
\bibitem [{\citenamefont {Goldstein}\ \emph {et~al.}(2010)\citenamefont
  {Goldstein}, \citenamefont {Lebowitz}, \citenamefont {Mastrodonato},
  \citenamefont {Tumulka},\ and\ \citenamefont {Zangh{\`{i}}}}]{GOLD10}%
  \BibitemOpen
  \bibfield  {author} {\bibinfo {author} {\bibfnamefont {S.}~\bibnamefont
  {Goldstein}}, \bibinfo {author} {\bibfnamefont {J.~L.}\ \bibnamefont
  {Lebowitz}}, \bibinfo {author} {\bibfnamefont {C.}~\bibnamefont
  {Mastrodonato}}, \bibinfo {author} {\bibfnamefont {R.}~\bibnamefont
  {Tumulka}}, \ and\ \bibinfo {author} {\bibfnamefont {N.}~\bibnamefont
  {Zangh{\`{i}}}},\ }\bibfield  {title} {\enquote {\bibinfo {title} {Approach
  to thermal equilibrium of macroscopic quantum systems},}\ }\href@noop {}
  {\bibfield  {journal} {\bibinfo  {journal} {Phys. Rev. E}\ }\textbf {\bibinfo
  {volume} {81}},\ \bibinfo {pages} {011109} (\bibinfo {year}
  {2010})}\BibitemShut {NoStop}%
\bibitem [{\citenamefont {Goldstein}\ \emph {et~al.}(2015)\citenamefont
  {Goldstein}, \citenamefont {Hara},\ and\ \citenamefont {Tasaki}}]{GOLD15}%
  \BibitemOpen
  \bibfield  {author} {\bibinfo {author} {\bibfnamefont {S.}~\bibnamefont
  {Goldstein}}, \bibinfo {author} {\bibfnamefont {T.}~\bibnamefont {Hara}}, \
  and\ \bibinfo {author} {\bibfnamefont {H.}~\bibnamefont {Tasaki}},\
  }\bibfield  {title} {\enquote {\bibinfo {title} {Extremely quick
  thermalization in a macroscopic quantum system for a typical nonequilibrium
  subspace.}}\ }\href@noop {} {\bibfield  {journal} {\bibinfo  {journal} {New.
  J. Phys.}\ }\textbf {\bibinfo {volume} {17}},\ \bibinfo {pages} {045002}
  (\bibinfo {year} {2015})}\BibitemShut {NoStop}%
\bibitem [{\citenamefont {Popescu}\ \emph {et~al.}(2006)\citenamefont
  {Popescu}, \citenamefont {Short},\ and\ \citenamefont {Winter}}]{POPE06}%
  \BibitemOpen
  \bibfield  {author} {\bibinfo {author} {\bibfnamefont {S.}~\bibnamefont
  {Popescu}}, \bibinfo {author} {\bibfnamefont {A.~J.}\ \bibnamefont {Short}},
  \ and\ \bibinfo {author} {\bibfnamefont {A.}~\bibnamefont {Winter}},\
  }\bibfield  {title} {\enquote {\bibinfo {title} {{Entanglement and the
  foundations of statistical mechanics}},}\ }\href@noop {} {\bibfield
  {journal} {\bibinfo  {journal} {Nature Phys.}\ }\textbf {\bibinfo {volume}
  {2}},\ \bibinfo {pages} {754 -- 758} (\bibinfo {year} {2006})}\BibitemShut
  {NoStop}%
\bibitem [{\citenamefont {Reimann}(2007)}]{REIM07}%
  \BibitemOpen
  \bibfield  {author} {\bibinfo {author} {\bibfnamefont {P.}~\bibnamefont
  {Reimann}},\ }\bibfield  {title} {\enquote {\bibinfo {title} {Typicality for
  generalized microcanonical ensembles},}\ }\href@noop {} {\bibfield  {journal}
  {\bibinfo  {journal} {Phys. Rev. Lett.}\ }\textbf {\bibinfo {volume} {99}},\
  \bibinfo {pages} {160404} (\bibinfo {year} {2007})}\BibitemShut {NoStop}%
\bibitem [{\citenamefont {Reimann}(2010)}]{REIM10}%
  \BibitemOpen
  \bibfield  {author} {\bibinfo {author} {\bibfnamefont {P.}~\bibnamefont
  {Reimann}},\ }\bibfield  {title} {\enquote {\bibinfo {title} {Canonical
  thermalization},}\ }\href@noop {} {\bibfield  {journal} {\bibinfo  {journal}
  {New. J. Phys.}\ }\textbf {\bibinfo {volume} {12}},\ \bibinfo {pages}
  {055027} (\bibinfo {year} {2010})}\BibitemShut {NoStop}%
\bibitem [{\citenamefont {Reimann}(2015)}]{REIM15}%
  \BibitemOpen
  \bibfield  {author} {\bibinfo {author} {\bibfnamefont {P.}~\bibnamefont
  {Reimann}},\ }\bibfield  {title} {\enquote {\bibinfo {title} {Typical fast
  thermalization processes in closed many-body systems},}\ }\href@noop {}
  {\bibfield  {journal} {\bibinfo  {journal} {Nat. Comm.}\ }\textbf {\bibinfo
  {volume} {7}},\ \bibinfo {pages} {10821} (\bibinfo {year}
  {2015})}\BibitemShut {NoStop}%
\bibitem [{\citenamefont {Bocchieri}\ and\ \citenamefont
  {Loinger}(1959)}]{BOCC59}%
  \BibitemOpen
  \bibfield  {author} {\bibinfo {author} {\bibfnamefont {P.}~\bibnamefont
  {Bocchieri}}\ and\ \bibinfo {author} {\bibfnamefont {A.}~\bibnamefont
  {Loinger}},\ }\bibfield  {title} {\enquote {\bibinfo {title} {Ergodic
  foundation of quantum statistical mechanics},}\ }\href@noop {} {\bibfield
  {journal} {\bibinfo  {journal} {Phys. Rev.}\ }\textbf {\bibinfo {volume}
  {114}},\ \bibinfo {pages} {948 -- 951} (\bibinfo {year} {1959})}\BibitemShut
  {NoStop}%
\bibitem [{\citenamefont {Tasaki}(1998)}]{TASA98}%
  \BibitemOpen
  \bibfield  {author} {\bibinfo {author} {\bibfnamefont {H.}~\bibnamefont
  {Tasaki}},\ }\bibfield  {title} {\enquote {\bibinfo {title} {{From quantum
  dynamics to the canonical distribution: General picture and a rigorous
  example}},}\ }\href@noop {} {\bibfield  {journal} {\bibinfo  {journal} {Phys.
  Rev. Lett.}\ }\textbf {\bibinfo {volume} {80}},\ \bibinfo {pages} {1373 --
  1376} (\bibinfo {year} {1998})}\BibitemShut {NoStop}%
\bibitem [{\citenamefont {Gemmer}\ \emph {et~al.}(2004)\citenamefont {Gemmer},
  \citenamefont {Michel},\ and\ \citenamefont {Mahler}}]{GEMM04}%
  \BibitemOpen
  \bibfield  {author} {\bibinfo {author} {\bibfnamefont {J.}~\bibnamefont
  {Gemmer}}, \bibinfo {author} {\bibfnamefont {M.}~\bibnamefont {Michel}}, \
  and\ \bibinfo {author} {\bibfnamefont {G.}~\bibnamefont {Mahler}},\
  }\href@noop {} {\emph {\bibinfo {title} {Quantum Thermodynamics: Emergence of
  Thermodynamic Behavior Within Composite Quantum Systems}}}\ (\bibinfo
  {publisher} {Springer Berlin Heidelberg},\ \bibinfo {address} {Berlin,
  Heidelberg},\ \bibinfo {year} {2004})\BibitemShut {NoStop}%
\bibitem [{\citenamefont {Sugiura}\ and\ \citenamefont
  {Shimizu}(2012)}]{SUGI12}%
  \BibitemOpen
  \bibfield  {author} {\bibinfo {author} {\bibfnamefont {S.}~\bibnamefont
  {Sugiura}}\ and\ \bibinfo {author} {\bibfnamefont {A.}~\bibnamefont
  {Shimizu}},\ }\bibfield  {title} {\enquote {\bibinfo {title} {Thermal pure
  quantum states at finite temperature},}\ }\href@noop {} {\bibfield  {journal}
  {\bibinfo  {journal} {Phys. Rev. Lett.}\ }\textbf {\bibinfo {volume} {108}},\
  \bibinfo {pages} {240401} (\bibinfo {year} {2012})}\BibitemShut {NoStop}%
\bibitem [{\citenamefont {Sugiura}\ and\ \citenamefont
  {Shimizu}(2013)}]{SUGI13}%
  \BibitemOpen
  \bibfield  {author} {\bibinfo {author} {\bibfnamefont {S.}~\bibnamefont
  {Sugiura}}\ and\ \bibinfo {author} {\bibfnamefont {A.}~\bibnamefont
  {Shimizu}},\ }\bibfield  {title} {\enquote {\bibinfo {title} {Canonical
  thermal pure quantum state},}\ }\href@noop {} {\bibfield  {journal} {\bibinfo
   {journal} {Phys. Rev. Lett.}\ }\textbf {\bibinfo {volume} {111}},\ \bibinfo
  {pages} {010401} (\bibinfo {year} {2013})}\BibitemShut {NoStop}%
\bibitem [{\citenamefont {Lloyd}(2013)}]{LLOY13}%
  \BibitemOpen
  \bibfield  {author} {\bibinfo {author} {\bibfnamefont {S.}~\bibnamefont
  {Lloyd}},\ }\bibfield  {title} {\enquote {\bibinfo {title} {Pure state
  quantum statistical mechanics and black holes},}\ }\href@noop {} {\bibfield
  {journal} {\bibinfo  {journal} {arXiv:1307.0378}\ } (\bibinfo {year}
  {2013})}\BibitemShut {NoStop}%
\bibitem [{\citenamefont {Hammersley}\ and\ \citenamefont
  {Handscomb}(1964)}]{HAMM64}%
  \BibitemOpen
  \bibfield  {author} {\bibinfo {author} {\bibfnamefont {J.~H.}\ \bibnamefont
  {Hammersley}}\ and\ \bibinfo {author} {\bibfnamefont {D.~C.}\ \bibnamefont
  {Handscomb}},\ }\href@noop {} {\emph {\bibinfo {title} {Monte Carlo
  Methods}}}\ (\bibinfo  {publisher} {John Wiley},\ \bibinfo {address} {New
  York},\ \bibinfo {year} {1964})\BibitemShut {NoStop}%
\bibitem [{\citenamefont {Landau}\ and\ \citenamefont {Binder}(2000)}]{LAND00}%
  \BibitemOpen
  \bibfield  {author} {\bibinfo {author} {\bibfnamefont {D.~P.}\ \bibnamefont
  {Landau}}\ and\ \bibinfo {author} {\bibfnamefont {K.}~\bibnamefont
  {Binder}},\ }\href@noop {} {\emph {\bibinfo {title} {A Guide to Monte Carlo
  Simulation in Statistical Physics}}}\ (\bibinfo  {publisher} {Cambridge
  University Press},\ \bibinfo {address} {Cambridge},\ \bibinfo {year}
  {2000})\BibitemShut {NoStop}%
\bibitem [{\citenamefont {Press}\ \emph {et~al.}(2003)\citenamefont {Press},
  \citenamefont {Flannery}, \citenamefont {Teukolsky},\ and\ \citenamefont
  {Vetterling}}]{PRES03}%
  \BibitemOpen
  \bibfield  {author} {\bibinfo {author} {\bibfnamefont {W.~H.}\ \bibnamefont
  {Press}}, \bibinfo {author} {\bibfnamefont {B.~P.}\ \bibnamefont {Flannery}},
  \bibinfo {author} {\bibfnamefont {S.~A.}\ \bibnamefont {Teukolsky}}, \ and\
  \bibinfo {author} {\bibfnamefont {W.~T.}\ \bibnamefont {Vetterling}},\
  }\href@noop {} {\emph {\bibinfo {title} {{Numerical Recipes}}}}\ (\bibinfo
  {publisher} {Cambridge University Press},\ \bibinfo {address} {Cambridge},\
  \bibinfo {year} {2003})\BibitemShut {NoStop}%
\bibitem [{\citenamefont {Gemmer}\ and\ \citenamefont
  {Mahler}(2003)}]{GEMM03a}%
  \BibitemOpen
  \bibfield  {author} {\bibinfo {author} {\bibfnamefont {J.}~\bibnamefont
  {Gemmer}}\ and\ \bibinfo {author} {\bibfnamefont {G.}~\bibnamefont
  {Mahler}},\ }\bibfield  {title} {\enquote {\bibinfo {title} {Distribution of
  local entropy in the {Hilbert} space of bi-partite quantum systems: origin of
  {Jaynes}' principle},}\ }\href {\doibase 10.1140/epjb/e2003-00029-3}
  {\bibfield  {journal} {\bibinfo  {journal} {Eur. Phys. J. B}\ }\textbf
  {\bibinfo {volume} {31}},\ \bibinfo {pages} {249--257} (\bibinfo {year}
  {2003})}\BibitemShut {NoStop}%
\bibitem [{\citenamefont {Bartsch}\ and\ \citenamefont
  {Gemmer}(2009)}]{BART09}%
  \BibitemOpen
  \bibfield  {author} {\bibinfo {author} {\bibfnamefont {C.}~\bibnamefont
  {Bartsch}}\ and\ \bibinfo {author} {\bibfnamefont {J.}~\bibnamefont
  {Gemmer}},\ }\bibfield  {title} {\enquote {\bibinfo {title} {Dynamical
  typicality of quantum expectation values},}\ }\href@noop {} {\bibfield
  {journal} {\bibinfo  {journal} {Phys. Rev. Lett}\ }\textbf {\bibinfo {volume}
  {102}},\ \bibinfo {pages} {110403} (\bibinfo {year} {2009})}\BibitemShut
  {NoStop}%
\bibitem [{\citenamefont {Steinigeweg}\ \emph
  {et~al.}(2014{\natexlab{a}})\citenamefont {Steinigeweg}, \citenamefont
  {Gemmer},\ and\ \citenamefont {Brenig}}]{STEIN14}%
  \BibitemOpen
  \bibfield  {author} {\bibinfo {author} {\bibfnamefont {R.}~\bibnamefont
  {Steinigeweg}}, \bibinfo {author} {\bibfnamefont {J.}~\bibnamefont {Gemmer}},
  \ and\ \bibinfo {author} {\bibfnamefont {W.}~\bibnamefont {Brenig}},\
  }\bibfield  {title} {\enquote {\bibinfo {title} {Spin-current
  autocorrelations from single pure-state propagation},}\ }\href {\doibase
  10.1103/PhysRevLett.112.120601} {\bibfield  {journal} {\bibinfo  {journal}
  {Phys. Rev. Lett.}\ }\textbf {\bibinfo {volume} {112}},\ \bibinfo {pages}
  {120601} (\bibinfo {year} {2014}{\natexlab{a}})}\BibitemShut {NoStop}%
\bibitem [{\citenamefont {Steinigeweg}\ \emph
  {et~al.}(2016{\natexlab{a}})\citenamefont {Steinigeweg}, \citenamefont
  {Herbrych}, \citenamefont {Pollmann},\ and\ \citenamefont
  {Brenig}}]{STEIN16h}%
  \BibitemOpen
  \bibfield  {author} {\bibinfo {author} {\bibfnamefont {R.}~\bibnamefont
  {Steinigeweg}}, \bibinfo {author} {\bibfnamefont {J.}~\bibnamefont
  {Herbrych}}, \bibinfo {author} {\bibfnamefont {F.}~\bibnamefont {Pollmann}},
  \ and\ \bibinfo {author} {\bibfnamefont {W.}~\bibnamefont {Brenig}},\
  }\bibfield  {title} {\enquote {\bibinfo {title} {Typicality approach to the
  optical conductivity in thermal and many-body localized phases},}\ }\href
  {\doibase 10.1103/PhysRevB.94.180401} {\bibfield  {journal} {\bibinfo
  {journal} {Phys. Rev. B}\ }\textbf {\bibinfo {volume} {94}},\ \bibinfo
  {pages} {180401} (\bibinfo {year} {2016}{\natexlab{a}})}\BibitemShut
  {NoStop}%
\bibitem [{\citenamefont {Alben}\ \emph {et~al.}(1975)\citenamefont {Alben},
  \citenamefont {Blume}, \citenamefont {Krakauer},\ and\ \citenamefont
  {Schwartz}}]{ALBE75}%
  \BibitemOpen
  \bibfield  {author} {\bibinfo {author} {\bibfnamefont {R.}~\bibnamefont
  {Alben}}, \bibinfo {author} {\bibfnamefont {M.}~\bibnamefont {Blume}},
  \bibinfo {author} {\bibfnamefont {H.}~\bibnamefont {Krakauer}}, \ and\
  \bibinfo {author} {\bibfnamefont {L.}~\bibnamefont {Schwartz}},\ }\bibfield
  {title} {\enquote {\bibinfo {title} {Exact results for a three-dimensional
  alloy with site diagonal disorder: comparison with the coherent potential
  approximation},}\ }\href@noop {} {\bibfield  {journal} {\bibinfo  {journal}
  {Phys. Rev. B}\ }\textbf {\bibinfo {volume} {12}},\ \bibinfo {pages} {4090}
  (\bibinfo {year} {1975})}\BibitemShut {NoStop}%
\bibitem [{\citenamefont {Imada}\ and\ \citenamefont
  {Takahashi}(1986)}]{IMAD86}%
  \BibitemOpen
  \bibfield  {author} {\bibinfo {author} {\bibfnamefont {M.}~\bibnamefont
  {Imada}}\ and\ \bibinfo {author} {\bibfnamefont {M.}~\bibnamefont
  {Takahashi}},\ }\bibfield  {title} {\enquote {\bibinfo {title} {Quantum
  transfer {Monte Carlo} method for finite temperature properties and quantum
  molecular dynamics method for dynamical correlation functions},}\ }\href
  {\doibase 10.1143/JPSJ.55.3354} {\bibfield  {journal} {\bibinfo  {journal}
  {J. Phys. Soc. Jpn.}\ }\textbf {\bibinfo {volume} {55}},\ \bibinfo {pages}
  {3354--3361} (\bibinfo {year} {1986})}\BibitemShut {NoStop}%
\bibitem [{\citenamefont {Skilling}(1988)}]{SKIL88}%
  \BibitemOpen
  \bibinfo {editor} {\bibfnamefont {J.}~\bibnamefont {Skilling}},\ ed.,\
  \href@noop {} {\emph {\bibinfo {title} {Maximum entropy and {Bayesian}
  methods}}}\ (\bibinfo  {publisher} {Springer},\ \bibinfo {address}
  {Cambridge, England},\ \bibinfo {year} {1988})\ \bibinfo {note} {p.
  455-466}\BibitemShut {NoStop}%
\bibitem [{\citenamefont {{De Raedt}}\ and\ \citenamefont {{de
  Vries}}(1989)}]{RAED89}%
  \BibitemOpen
  \bibfield  {author} {\bibinfo {author} {\bibfnamefont {H.}~\bibnamefont {{De
  Raedt}}}\ and\ \bibinfo {author} {\bibfnamefont {P.}~\bibnamefont {{de
  Vries}}},\ }\bibfield  {title} {\enquote {\bibinfo {title} {Simulation of two
  and three-dimensional disordered systems: {Lifshitz} tails and localization
  properties},}\ }\href@noop {} {\bibfield  {journal} {\bibinfo  {journal} {Z.
  Phys. B}\ }\textbf {\bibinfo {volume} {77}},\ \bibinfo {pages} {243}
  (\bibinfo {year} {1989})}\BibitemShut {NoStop}%
\bibitem [{\citenamefont {de~Vries}\ and\ \citenamefont {{De
  Raedt}}(1993)}]{RAED93}%
  \BibitemOpen
  \bibfield  {author} {\bibinfo {author} {\bibfnamefont {P.}~\bibnamefont
  {de~Vries}}\ and\ \bibinfo {author} {\bibfnamefont {H.}~\bibnamefont {{De
  Raedt}}},\ }\bibfield  {title} {\enquote {\bibinfo {title} {Solution of the
  time-dependent {Schr\"odinger} equation for two-dimensional spin-1/2
  {Heisenberg} systems},}\ }\href@noop {} {\bibfield  {journal} {\bibinfo
  {journal} {Phys. Rev. B}\ }\textbf {\bibinfo {volume} {47}},\ \bibinfo
  {pages} {7929 -- 7937} (\bibinfo {year} {1993})}\BibitemShut {NoStop}%
\bibitem [{\citenamefont {Fukamachi}\ and\ \citenamefont
  {Nishimori}(1994)}]{FUKA94}%
  \BibitemOpen
  \bibfield  {author} {\bibinfo {author} {\bibfnamefont {K.}~\bibnamefont
  {Fukamachi}}\ and\ \bibinfo {author} {\bibfnamefont {H.}~\bibnamefont
  {Nishimori}},\ }\bibfield  {title} {\enquote {\bibinfo {title} {Specific heat
  of the quantum {Heisenberg} antiferromagnet on the {kagom\'e} lattice},}\
  }\href@noop {} {\bibfield  {journal} {\bibinfo  {journal} {Phys. Rev. B}\
  }\textbf {\bibinfo {volume} {49}},\ \bibinfo {pages} {651 -- 654} (\bibinfo
  {year} {1994})}\BibitemShut {NoStop}%
\bibitem [{\citenamefont {Drabold}\ and\ \citenamefont
  {Sankey}(1993)}]{DRAB93}%
  \BibitemOpen
  \bibfield  {author} {\bibinfo {author} {\bibfnamefont {D.A.}\ \bibnamefont
  {Drabold}}\ and\ \bibinfo {author} {\bibfnamefont {O.F.}\ \bibnamefont
  {Sankey}},\ }\bibfield  {title} {\enquote {\bibinfo {title} {Maximum entropy
  approach for linear scaling in the electronic structure problem},}\ }\href
  {\doibase 10.1103/PhysRevLett.70.3631} {\bibfield  {journal} {\bibinfo
  {journal} {Phys. Rev. Lett.}\ }\textbf {\bibinfo {volume} {70}},\ \bibinfo
  {pages} {3631--3634} (\bibinfo {year} {1993})}\BibitemShut {NoStop}%
\bibitem [{\citenamefont {Silver}\ and\ \citenamefont
  {R\"oder}(1994)}]{SILV94}%
  \BibitemOpen
  \bibfield  {author} {\bibinfo {author} {\bibfnamefont {R.~N.}\ \bibnamefont
  {Silver}}\ and\ \bibinfo {author} {\bibfnamefont {H.}~\bibnamefont
  {R\"oder}},\ }\bibfield  {title} {\enquote {\bibinfo {title} {Densities of
  states of mega-dimensional {Hamiltonian} matrices},}\ }\href {\doibase
  10.1142/S0129183194000842} {\bibfield  {journal} {\bibinfo  {journal} {Int.
  J. Mod. Phys. C}\ }\textbf {\bibinfo {volume} {05}},\ \bibinfo {pages}
  {735--753} (\bibinfo {year} {1994})}\BibitemShut {NoStop}%
\bibitem [{\citenamefont {Silver}\ and\ \citenamefont
  {R\"oder}(1997)}]{SILV97}%
  \BibitemOpen
  \bibfield  {author} {\bibinfo {author} {\bibfnamefont {R.~N.}\ \bibnamefont
  {Silver}}\ and\ \bibinfo {author} {\bibfnamefont {H.}~\bibnamefont
  {R\"oder}},\ }\bibfield  {title} {\enquote {\bibinfo {title} {Calculation of
  densities of states and spectral functions by {Chebyshev} recursion and
  maximum entropy},}\ }\href {\doibase 10.1103/PhysRevE.56.4822} {\bibfield
  {journal} {\bibinfo  {journal} {Phys. Rev. E}\ }\textbf {\bibinfo {volume}
  {56}},\ \bibinfo {pages} {4822--4829} (\bibinfo {year} {1997})}\BibitemShut
  {NoStop}%
\bibitem [{\citenamefont {Iitaka}\ \emph {et~al.}(1997)\citenamefont {Iitaka},
  \citenamefont {Nomura}, \citenamefont {Hirayama}, \citenamefont {Zhao},
  \citenamefont {Aoyagi},\ and\ \citenamefont {Sugano}}]{IITA97}%
  \BibitemOpen
  \bibfield  {author} {\bibinfo {author} {\bibfnamefont {T.}~\bibnamefont
  {Iitaka}}, \bibinfo {author} {\bibfnamefont {S.}~\bibnamefont {Nomura}},
  \bibinfo {author} {\bibfnamefont {H.}~\bibnamefont {Hirayama}}, \bibinfo
  {author} {\bibfnamefont {X.}~\bibnamefont {Zhao}}, \bibinfo {author}
  {\bibfnamefont {Y.}~\bibnamefont {Aoyagi}}, \ and\ \bibinfo {author}
  {\bibfnamefont {T.}~\bibnamefont {Sugano}},\ }\bibfield  {title} {\enquote
  {\bibinfo {title} {Calculating the linear response functions of
  noninteracting electrons with a time-dependent {Schr\"odinger} equation},}\
  }\href@noop {} {\bibfield  {journal} {\bibinfo  {journal} {Phys. Rev. E}\
  }\textbf {\bibinfo {volume} {56}},\ \bibinfo {pages} {1222--1229} (\bibinfo
  {year} {1997})}\BibitemShut {NoStop}%
\bibitem [{\citenamefont {Nomura}\ \emph {et~al.}(1997)\citenamefont {Nomura},
  \citenamefont {Iitaka}, \citenamefont {Zhao}, \citenamefont {Sugano},\ and\
  \citenamefont {Aoyagi}}]{NOMU97b}%
  \BibitemOpen
  \bibfield  {author} {\bibinfo {author} {\bibfnamefont {S.}~\bibnamefont
  {Nomura}}, \bibinfo {author} {\bibfnamefont {T.}~\bibnamefont {Iitaka}},
  \bibinfo {author} {\bibfnamefont {X.}~\bibnamefont {Zhao}}, \bibinfo {author}
  {\bibfnamefont {T.}~\bibnamefont {Sugano}}, \ and\ \bibinfo {author}
  {\bibfnamefont {Y.}~\bibnamefont {Aoyagi}},\ }\bibfield  {title} {\enquote
  {\bibinfo {title} {Linear scaling calculation for optical-absorption spectra
  of large hydrogenated silicon nanocrystallites},}\ }\href@noop {} {\bibfield
  {journal} {\bibinfo  {journal} {Phys. Rev. B}\ }\textbf {\bibinfo {volume}
  {56}},\ \bibinfo {pages} {R4348--R4350} (\bibinfo {year} {1997})}\BibitemShut
  {NoStop}%
\bibitem [{\citenamefont {Nomura}\ \emph {et~al.}(1999)\citenamefont {Nomura},
  \citenamefont {Iitaka}, \citenamefont {Zhao}, \citenamefont {Sugano},\ and\
  \citenamefont {Aoyagi}}]{IITA99}%
  \BibitemOpen
  \bibfield  {author} {\bibinfo {author} {\bibfnamefont {S.}~\bibnamefont
  {Nomura}}, \bibinfo {author} {\bibfnamefont {T.}~\bibnamefont {Iitaka}},
  \bibinfo {author} {\bibfnamefont {X.}~\bibnamefont {Zhao}}, \bibinfo {author}
  {\bibfnamefont {T.}~\bibnamefont {Sugano}}, \ and\ \bibinfo {author}
  {\bibfnamefont {Y.}~\bibnamefont {Aoyagi}},\ }\bibfield  {title} {\enquote
  {\bibinfo {title} {Quantum-size effect in model nanocrystalline/amorphous
  mixed-phase silicon structures},}\ }\href@noop {} {\bibfield  {journal}
  {\bibinfo  {journal} {Phys. Rev. B}\ }\textbf {\bibinfo {volume} {59}},\
  \bibinfo {pages} {10309} (\bibinfo {year} {1999})}\BibitemShut {NoStop}%
\bibitem [{\citenamefont {Nishida}\ \emph {et~al.}(2020)\citenamefont
  {Nishida}, \citenamefont {Fujiuchi}, \citenamefont {Sugimoto},\ and\
  \citenamefont {Ohta}}]{NISH20}%
  \BibitemOpen
  \bibfield  {author} {\bibinfo {author} {\bibfnamefont {H.}~\bibnamefont
  {Nishida}}, \bibinfo {author} {\bibfnamefont {R.}~\bibnamefont {Fujiuchi}},
  \bibinfo {author} {\bibfnamefont {K.}~\bibnamefont {Sugimoto}}, \ and\
  \bibinfo {author} {\bibfnamefont {Y.}~\bibnamefont {Ohta}},\ }\bibfield
  {title} {\enquote {\bibinfo {title} {Typicality-based variational cluster
  approach to thermodynamic properties of the {Hubbard} model},}\ }\href@noop
  {} {\bibfield  {journal} {\bibinfo  {journal} {J. Phys. Soc. Jpn.}\ }\textbf
  {\bibinfo {volume} {89}},\ \bibinfo {pages} {023702} (\bibinfo {year}
  {2020})}\BibitemShut {NoStop}%
\bibitem [{\citenamefont {{Hams}}\ and\ \citenamefont {{De
  Raedt}}(2000)}]{HAMS00}%
  \BibitemOpen
  \bibfield  {author} {\bibinfo {author} {\bibfnamefont {A.}~\bibnamefont
  {{Hams}}}\ and\ \bibinfo {author} {\bibfnamefont {H.}~\bibnamefont {{De
  Raedt}}},\ }\bibfield  {title} {\enquote {\bibinfo {title} {{Fast algorithm
  for finding the eigenvalue distribution of very large matrices}},}\
  }\href@noop {} {\bibfield  {journal} {\bibinfo  {journal} {Phys. Rev. E}\
  }\textbf {\bibinfo {volume} {62}},\ \bibinfo {pages} {4365 -- 4377} (\bibinfo
  {year} {2000})}\BibitemShut {NoStop}%
\bibitem [{\citenamefont {Breuer}\ and\ \citenamefont
  {Petruccione}(2002)}]{BREU02}%
  \BibitemOpen
  \bibfield  {author} {\bibinfo {author} {\bibfnamefont {H.-P.}\ \bibnamefont
  {Breuer}}\ and\ \bibinfo {author} {\bibfnamefont {F.}~\bibnamefont
  {Petruccione}},\ }\href@noop {} {\emph {\bibinfo {title} {{The Theory of Open
  Quantum Systems}}}}\ (\bibinfo  {publisher} {Oxford University Press},\
  \bibinfo {address} {Oxford},\ \bibinfo {year} {2002})\BibitemShut {NoStop}%
\bibitem [{\citenamefont {Yuan}\ \emph {et~al.}(2006)\citenamefont {Yuan},
  \citenamefont {Katsnelson},\ and\ \citenamefont {{De Raedt}}}]{YUAN06}%
  \BibitemOpen
  \bibfield  {author} {\bibinfo {author} {\bibfnamefont {S.}~\bibnamefont
  {Yuan}}, \bibinfo {author} {\bibfnamefont {M.I.}\ \bibnamefont {Katsnelson}},
  \ and\ \bibinfo {author} {\bibfnamefont {H.}~\bibnamefont {{De Raedt}}},\
  }\bibfield  {title} {\enquote {\bibinfo {title} {Giant enhancement of quantum
  decoherence by frustrated environments},}\ }\href@noop {} {\bibfield
  {journal} {\bibinfo  {journal} {JETP Lett.}\ }\textbf {\bibinfo {volume}
  {84}},\ \bibinfo {pages} {99} (\bibinfo {year} {2006})}\BibitemShut {NoStop}%
\bibitem [{\citenamefont {Yuan}\ \emph {et~al.}(2008)\citenamefont {Yuan},
  \citenamefont {Katsnelson},\ and\ \citenamefont {{De Raedt}}}]{YUAN08}%
  \BibitemOpen
  \bibfield  {author} {\bibinfo {author} {\bibfnamefont {S.}~\bibnamefont
  {Yuan}}, \bibinfo {author} {\bibfnamefont {M.I.}\ \bibnamefont {Katsnelson}},
  \ and\ \bibinfo {author} {\bibfnamefont {H.}~\bibnamefont {{De Raedt}}},\
  }\bibfield  {title} {\enquote {\bibinfo {title} {Decoherence by a spin
  thermal bath: Role of spin-spin interactions and initial state of the
  bath},}\ }\href@noop {} {\bibfield  {journal} {\bibinfo  {journal} {Phys.
  Rev. B}\ }\textbf {\bibinfo {volume} {77}},\ \bibinfo {pages} {184301}
  (\bibinfo {year} {2008})}\BibitemShut {NoStop}%
\bibitem [{\citenamefont {Yuan}\ \emph {et~al.}(2009)\citenamefont {Yuan},
  \citenamefont {Katsnelson},\ and\ \citenamefont {{De Raedt}}}]{YUAN09}%
  \BibitemOpen
  \bibfield  {author} {\bibinfo {author} {\bibfnamefont {S.}~\bibnamefont
  {Yuan}}, \bibinfo {author} {\bibfnamefont {M.I.}\ \bibnamefont {Katsnelson}},
  \ and\ \bibinfo {author} {\bibfnamefont {H.}~\bibnamefont {{De Raedt}}},\
  }\bibfield  {title} {\enquote {\bibinfo {title} {Origin of the canonical
  ensemble: Thermalization with decoherence},}\ }\href@noop {} {\bibfield
  {journal} {\bibinfo  {journal} {J. Phys. Soc. Jpn.}\ }\textbf {\bibinfo
  {volume} {78}},\ \bibinfo {pages} {094003} (\bibinfo {year}
  {2009})}\BibitemShut {NoStop}%
\bibitem [{\citenamefont {Yuan}(2011)}]{YUAN11}%
  \BibitemOpen
  \bibfield  {author} {\bibinfo {author} {\bibfnamefont {S.}~\bibnamefont
  {Yuan}},\ }\bibfield  {title} {\enquote {\bibinfo {title} {Decoherence and
  thermalization of quantum spin systems},}\ }\href@noop {} {\bibfield
  {journal} {\bibinfo  {journal} {J. Comput. Theor. Nanoscience}\ }\textbf
  {\bibinfo {volume} {8}},\ \bibinfo {pages} {889 -- 911} (\bibinfo {year}
  {2011})}\BibitemShut {NoStop}%
\bibitem [{\citenamefont {{Jin}}\ \emph {et~al.}(2010)\citenamefont {{Jin}},
  \citenamefont {{De Raedt}}, \citenamefont {{Yuan}}, \citenamefont
  {{Katsnelson}}, \citenamefont {{Miyashita}},\ and\ \citenamefont
  {{Michielsen}}}]{JIN10x}%
  \BibitemOpen
  \bibfield  {author} {\bibinfo {author} {\bibfnamefont {F.}~\bibnamefont
  {{Jin}}}, \bibinfo {author} {\bibfnamefont {H.}~\bibnamefont {{De Raedt}}},
  \bibinfo {author} {\bibfnamefont {S.}~\bibnamefont {{Yuan}}}, \bibinfo
  {author} {\bibfnamefont {M.~I.}\ \bibnamefont {{Katsnelson}}}, \bibinfo
  {author} {\bibfnamefont {S.}~\bibnamefont {{Miyashita}}}, \ and\ \bibinfo
  {author} {\bibfnamefont {K.}~\bibnamefont {{Michielsen}}},\ }\bibfield
  {title} {\enquote {\bibinfo {title} {Approach to equilibrium in nano-scale
  systems at finite temperature},}\ }\href@noop {} {\bibfield  {journal}
  {\bibinfo  {journal} {J. Phys. Soc. Jpn.}\ }\textbf {\bibinfo {volume}
  {79}},\ \bibinfo {pages} {124005} (\bibinfo {year} {2010})}\BibitemShut
  {NoStop}%
\bibitem [{\citenamefont {{De Raedt}}\ \emph {et~al.}(2012)\citenamefont {{De
  Raedt}}, \citenamefont {Barbara}, \citenamefont {Miyashita}, \citenamefont
  {Michielsen}, \citenamefont {Bertaina},\ and\ \citenamefont
  {Gambarelli}}]{RAED12z}%
  \BibitemOpen
  \bibfield  {author} {\bibinfo {author} {\bibfnamefont {H.}~\bibnamefont {{De
  Raedt}}}, \bibinfo {author} {\bibfnamefont {B.}~\bibnamefont {Barbara}},
  \bibinfo {author} {\bibfnamefont {S.}~\bibnamefont {Miyashita}}, \bibinfo
  {author} {\bibfnamefont {K.}~\bibnamefont {Michielsen}}, \bibinfo {author}
  {\bibfnamefont {S.}~\bibnamefont {Bertaina}}, \ and\ \bibinfo {author}
  {\bibfnamefont {S.}~\bibnamefont {Gambarelli}},\ }\bibfield  {title}
  {\enquote {\bibinfo {title} {Quantum simulations and experiments on {Rabi}
  oscillations of spin qubits: Intrinsic vs extrinsic damping},}\ }\href@noop
  {} {\bibfield  {journal} {\bibinfo  {journal} {Phys. Rev. B}\ }\textbf
  {\bibinfo {volume} {85}},\ \bibinfo {pages} {014408} (\bibinfo {year}
  {2012})}\BibitemShut {NoStop}%
\bibitem [{\citenamefont {Novotny}\ \emph {et~al.}(2016)\citenamefont
  {Novotny}, \citenamefont {Jin}, \citenamefont {Yuan}, \citenamefont
  {Miyashita}, \citenamefont {{De Raedt}},\ and\ \citenamefont
  {Michielsen}}]{NOVO16}%
  \BibitemOpen
  \bibfield  {author} {\bibinfo {author} {\bibfnamefont {M.~A.}\ \bibnamefont
  {Novotny}}, \bibinfo {author} {\bibfnamefont {F.}~\bibnamefont {Jin}},
  \bibinfo {author} {\bibfnamefont {S.}~\bibnamefont {Yuan}}, \bibinfo {author}
  {\bibfnamefont {S.}~\bibnamefont {Miyashita}}, \bibinfo {author}
  {\bibfnamefont {H.}~\bibnamefont {{De Raedt}}}, \ and\ \bibinfo {author}
  {\bibfnamefont {K.}~\bibnamefont {Michielsen}},\ }\bibfield  {title}
  {\enquote {\bibinfo {title} {Quantum decoherence and thermalization at finite
  temperatures within the canonical-thermal-state ensemble},}\ }\href@noop {}
  {\bibfield  {journal} {\bibinfo  {journal} {Phys. Rev. A}\ }\textbf {\bibinfo
  {volume} {93}},\ \bibinfo {pages} {032110} (\bibinfo {year}
  {2016})}\BibitemShut {NoStop}%
\bibitem [{\citenamefont {Zhao}\ \emph {et~al.}(2016)\citenamefont {Zhao},
  \citenamefont {{De Raedt}}, \citenamefont {Miyashita}, \citenamefont {Jin},\
  and\ \citenamefont {Michielsen}}]{ZHAO16}%
  \BibitemOpen
  \bibfield  {author} {\bibinfo {author} {\bibfnamefont {P.}~\bibnamefont
  {Zhao}}, \bibinfo {author} {\bibfnamefont {H.}~\bibnamefont {{De Raedt}}},
  \bibinfo {author} {\bibfnamefont {S.}~\bibnamefont {Miyashita}}, \bibinfo
  {author} {\bibfnamefont {F.}~\bibnamefont {Jin}}, \ and\ \bibinfo {author}
  {\bibfnamefont {K.}~\bibnamefont {Michielsen}},\ }\bibfield  {title}
  {\enquote {\bibinfo {title} {{Dynamics of open quantum spin systems: An
  assessment of the quantum master equation approach}},}\ }\href@noop {}
  {\bibfield  {journal} {\bibinfo  {journal} {Phys. Rev. E}\ }\textbf {\bibinfo
  {volume} {94}},\ \bibinfo {pages} {022126} (\bibinfo {year}
  {2016})}\BibitemShut {NoStop}%
\bibitem [{\citenamefont {{De Raedt}}\ \emph {et~al.}(2017)\citenamefont {{De
  Raedt}}, \citenamefont {Jin}, \citenamefont {Katsnelson},\ and\ \citenamefont
  {Michielsen}}]{RAED17b}%
  \BibitemOpen
  \bibfield  {author} {\bibinfo {author} {\bibfnamefont {H.}~\bibnamefont {{De
  Raedt}}}, \bibinfo {author} {\bibfnamefont {F.}~\bibnamefont {Jin}}, \bibinfo
  {author} {\bibfnamefont {M.I.}\ \bibnamefont {Katsnelson}}, \ and\ \bibinfo
  {author} {\bibfnamefont {K.}~\bibnamefont {Michielsen}},\ }\bibfield  {title}
  {\enquote {\bibinfo {title} {Relaxation, thermalization, and {Markovian}
  dynamics of two spins coupled to a spin bath},}\ }\href@noop {} {\bibfield
  {journal} {\bibinfo  {journal} {Phys. Rev. E}\ }\textbf {\bibinfo {volume}
  {96}},\ \bibinfo {pages} {053306} (\bibinfo {year} {2017})}\BibitemShut
  {NoStop}%
\bibitem [{\citenamefont {Gelman}\ and\ \citenamefont
  {Kosloff}(2003)}]{GELM03}%
  \BibitemOpen
  \bibfield  {author} {\bibinfo {author} {\bibfnamefont {D.}~\bibnamefont
  {Gelman}}\ and\ \bibinfo {author} {\bibfnamefont {R.}~\bibnamefont
  {Kosloff}},\ }\bibfield  {title} {\enquote {\bibinfo {title} {Simulating
  dissipative phenomena with a random thermal phase wavefunctions, high
  temperature application of the surrogate {Hamiltonian} approach},}\
  }\href@noop {} {\bibfield  {journal} {\bibinfo  {journal} {Chem. Phys.
  Lett.}\ }\textbf {\bibinfo {volume} {381}},\ \bibinfo {pages} {129--138}
  (\bibinfo {year} {2003})}\BibitemShut {NoStop}%
\bibitem [{\citenamefont {Gelman}\ \emph {et~al.}(2004)\citenamefont {Gelman},
  \citenamefont {Koch},\ and\ \citenamefont {Kosloff}}]{GELM04}%
  \BibitemOpen
  \bibfield  {author} {\bibinfo {author} {\bibfnamefont {D.}~\bibnamefont
  {Gelman}}, \bibinfo {author} {\bibfnamefont {C.P.}\ \bibnamefont {Koch}}, \
  and\ \bibinfo {author} {\bibfnamefont {R.}~\bibnamefont {Kosloff}},\
  }\bibfield  {title} {\enquote {\bibinfo {title} {Dissipative quantum dynamics
  with the surrogate {Hamiltonian} approach. {A} comparison between spin and
  harmonic baths},}\ }\href@noop {} {\bibfield  {journal} {\bibinfo  {journal}
  {J. Chem. Phys.}\ }\textbf {\bibinfo {volume} {121}},\ \bibinfo {pages}
  {661--671} (\bibinfo {year} {2004})}\BibitemShut {NoStop}%
\bibitem [{\citenamefont {Katz}\ \emph {et~al.}(2008)\citenamefont {Katz},
  \citenamefont {Gelman}, \citenamefont {Ratner},\ and\ \citenamefont
  {Kosloff}}]{KATZ08}%
  \BibitemOpen
  \bibfield  {author} {\bibinfo {author} {\bibfnamefont {G.}~\bibnamefont
  {Katz}}, \bibinfo {author} {\bibfnamefont {D.}~\bibnamefont {Gelman}},
  \bibinfo {author} {\bibfnamefont {M.A.}\ \bibnamefont {Ratner}}, \ and\
  \bibinfo {author} {\bibfnamefont {R.}~\bibnamefont {Kosloff}},\ }\bibfield
  {title} {\enquote {\bibinfo {title} {Stochastic surrogate {Hamiltonian}},}\
  }\href@noop {} {\bibfield  {journal} {\bibinfo  {journal} {J. Chem.Phys.}\
  }\textbf {\bibinfo {volume} {129}},\ \bibinfo {pages} {034108} (\bibinfo
  {year} {2008})}\BibitemShut {NoStop}%
\bibitem [{\citenamefont {Monnai}\ and\ \citenamefont {Sugita}(2014)}]{MONN14}%
  \BibitemOpen
  \bibfield  {author} {\bibinfo {author} {\bibfnamefont {T.}~\bibnamefont
  {Monnai}}\ and\ \bibinfo {author} {\bibfnamefont {A.}~\bibnamefont
  {Sugita}},\ }\bibfield  {title} {\enquote {\bibinfo {title} {Typical pure
  states and nonequilibrium processes in quantum many-body systems},}\ }\href
  {\doibase 10.7566/JPSJ.83.094001} {\bibfield  {journal} {\bibinfo  {journal}
  {J. Phys. Soc. Jpn.}\ }\textbf {\bibinfo {volume} {83}},\ \bibinfo {pages}
  {094001} (\bibinfo {year} {2014})}\BibitemShut {NoStop}%
\bibitem [{\citenamefont {Steinigeweg}\ \emph
  {et~al.}(2016{\natexlab{b}})\citenamefont {Steinigeweg}, \citenamefont
  {Herbrych}, \citenamefont {Zotos},\ and\ \citenamefont {Brenig}}]{STEI16}%
  \BibitemOpen
  \bibfield  {author} {\bibinfo {author} {\bibfnamefont {R.}~\bibnamefont
  {Steinigeweg}}, \bibinfo {author} {\bibfnamefont {J.}~\bibnamefont
  {Herbrych}}, \bibinfo {author} {\bibfnamefont {X.}~\bibnamefont {Zotos}}, \
  and\ \bibinfo {author} {\bibfnamefont {W.}~\bibnamefont {Brenig}},\
  }\bibfield  {title} {\enquote {\bibinfo {title} {Heat conductivity of the
  {Heisenberg} spin-$1/2$ ladder: From weak to strong breaking of
  integrability},}\ }\href {\doibase 10.1103/PhysRevLett.116.017202} {\bibfield
   {journal} {\bibinfo  {journal} {Phys. Rev. Lett.}\ }\textbf {\bibinfo
  {volume} {116}},\ \bibinfo {pages} {017202} (\bibinfo {year}
  {2016}{\natexlab{b}})}\BibitemShut {NoStop}%
\bibitem [{\citenamefont {Steinigeweg}\ \emph
  {et~al.}(2017{\natexlab{a}})\citenamefont {Steinigeweg}, \citenamefont {Jin},
  \citenamefont {Schmidtke}, \citenamefont {{De Raedt}}, \citenamefont
  {Michielsen},\ and\ \citenamefont {Gemmer}}]{STEI17a}%
  \BibitemOpen
  \bibfield  {author} {\bibinfo {author} {\bibfnamefont {R.}~\bibnamefont
  {Steinigeweg}}, \bibinfo {author} {\bibfnamefont {F.}~\bibnamefont {Jin}},
  \bibinfo {author} {\bibfnamefont {D.}~\bibnamefont {Schmidtke}}, \bibinfo
  {author} {\bibfnamefont {H.}~\bibnamefont {{De Raedt}}}, \bibinfo {author}
  {\bibfnamefont {K.}~\bibnamefont {Michielsen}}, \ and\ \bibinfo {author}
  {\bibfnamefont {J.}~\bibnamefont {Gemmer}},\ }\bibfield  {title} {\enquote
  {\bibinfo {title} {Real-time broadening of nonequilibrium density profiles
  and the role of the specific initial-state realization},}\ }\href {\doibase
  10.1103/PhysRevB.95.035155} {\bibfield  {journal} {\bibinfo  {journal} {Phys.
  Rev. B}\ }\textbf {\bibinfo {volume} {95}},\ \bibinfo {pages} {035155}
  (\bibinfo {year} {2017}{\natexlab{a}})}\BibitemShut {NoStop}%
\bibitem [{\citenamefont {Yamaji}\ \emph {et~al.}(2018)\citenamefont {Yamaji},
  \citenamefont {Suzuki},\ and\ \citenamefont {Kawamura}}]{YAMA18}%
  \BibitemOpen
  \bibfield  {author} {\bibinfo {author} {\bibfnamefont {Y.}~\bibnamefont
  {Yamaji}}, \bibinfo {author} {\bibfnamefont {T.}~\bibnamefont {Suzuki}}, \
  and\ \bibinfo {author} {\bibfnamefont {M.}~\bibnamefont {Kawamura}},\
  }\bibfield  {title} {\enquote {\bibinfo {title} {Numerical algorithm for
  exact finite temperature spectra and its application to frustrated quantum
  spin systems},}\ }\href@noop {} {\bibfield  {journal} {\bibinfo  {journal}
  {arXiv:1802.02854}\ } (\bibinfo {year} {2018})}\BibitemShut {NoStop}%
\bibitem [{\citenamefont {Richter}\ \emph {et~al.}(2018)\citenamefont
  {Richter}, \citenamefont {Jin}, \citenamefont {{De Raedt}}, \citenamefont
  {Michielsen}, \citenamefont {Gemmer},\ and\ \citenamefont
  {Steinigeweg}}]{RICH18}%
  \BibitemOpen
  \bibfield  {author} {\bibinfo {author} {\bibfnamefont {J.}~\bibnamefont
  {Richter}}, \bibinfo {author} {\bibfnamefont {F.}~\bibnamefont {Jin}},
  \bibinfo {author} {\bibfnamefont {H.}~\bibnamefont {{De Raedt}}}, \bibinfo
  {author} {\bibfnamefont {K.}~\bibnamefont {Michielsen}}, \bibinfo {author}
  {\bibfnamefont {J.}~\bibnamefont {Gemmer}}, \ and\ \bibinfo {author}
  {\bibfnamefont {R.}~\bibnamefont {Steinigeweg}},\ }\bibfield  {title}
  {\enquote {\bibinfo {title} {Real-time dynamics of typical and untypical
  states in nonintegrable systems},}\ }\href {\doibase
  10.1103/PhysRevB.97.174430} {\bibfield  {journal} {\bibinfo  {journal} {Phys.
  Rev. B}\ }\textbf {\bibinfo {volume} {97}},\ \bibinfo {pages} {174430}
  (\bibinfo {year} {2018})}\BibitemShut {NoStop}%
\bibitem [{\citenamefont {Richter}\ \emph
  {et~al.}(2019{\natexlab{a}})\citenamefont {Richter}, \citenamefont {Jin},
  \citenamefont {Knipschild}, \citenamefont {Herbrych}, \citenamefont {{De
  Raedt}}, \citenamefont {Michielsen}, \citenamefont {Gemmer},\ and\
  \citenamefont {Steinigeweg}}]{RICH19}%
  \BibitemOpen
  \bibfield  {author} {\bibinfo {author} {\bibfnamefont {J.}~\bibnamefont
  {Richter}}, \bibinfo {author} {\bibfnamefont {F.}~\bibnamefont {Jin}},
  \bibinfo {author} {\bibfnamefont {L.}~\bibnamefont {Knipschild}}, \bibinfo
  {author} {\bibfnamefont {J.}~\bibnamefont {Herbrych}}, \bibinfo {author}
  {\bibfnamefont {H.}~\bibnamefont {{De Raedt}}}, \bibinfo {author}
  {\bibfnamefont {K.}~\bibnamefont {Michielsen}}, \bibinfo {author}
  {\bibfnamefont {J.}~\bibnamefont {Gemmer}}, \ and\ \bibinfo {author}
  {\bibfnamefont {R.}~\bibnamefont {Steinigeweg}},\ }\bibfield  {title}
  {\enquote {\bibinfo {title} {Magnetization and energy dynamics in spin
  ladders: Evidence of diffusion in time, frequency, position, and momentum},}\
  }\href {\doibase 10.1103/PhysRevB.99.144422} {\bibfield  {journal} {\bibinfo
  {journal} {Phys. Rev. B}\ }\textbf {\bibinfo {volume} {99}},\ \bibinfo
  {pages} {144422} (\bibinfo {year} {2019}{\natexlab{a}})}\BibitemShut
  {NoStop}%
\bibitem [{\citenamefont {Jin}\ \emph {et~al.}(2015{\natexlab{a}})\citenamefont
  {Jin}, \citenamefont {Steinigeweg}, \citenamefont {Heidrich-Meisner},
  \citenamefont {Michielsen},\ and\ \citenamefont {De~Raedt}}]{JIN15}%
  \BibitemOpen
  \bibfield  {author} {\bibinfo {author} {\bibfnamefont {F.}~\bibnamefont
  {Jin}}, \bibinfo {author} {\bibfnamefont {R.}~\bibnamefont {Steinigeweg}},
  \bibinfo {author} {\bibfnamefont {F.}~\bibnamefont {Heidrich-Meisner}},
  \bibinfo {author} {\bibfnamefont {K.}~\bibnamefont {Michielsen}}, \ and\
  \bibinfo {author} {\bibfnamefont {H.}~\bibnamefont {De~Raedt}},\ }\bibfield
  {title} {\enquote {\bibinfo {title} {Finite-temperature charge transport in
  the one-dimensional {Hubbard} model},}\ }\href {\doibase
  10.1103/PhysRevB.92.205103} {\bibfield  {journal} {\bibinfo  {journal} {Phys.
  Rev. B}\ }\textbf {\bibinfo {volume} {92}},\ \bibinfo {pages} {205103}
  (\bibinfo {year} {2015}{\natexlab{a}})}\BibitemShut {NoStop}%
\bibitem [{\citenamefont {Steinigeweg}\ \emph
  {et~al.}(2017{\natexlab{b}})\citenamefont {Steinigeweg}, \citenamefont {Jin},
  \citenamefont {{De Raedt}}, \citenamefont {Michielsen},\ and\ \citenamefont
  {Gemmer}}]{STEI17b}%
  \BibitemOpen
  \bibfield  {author} {\bibinfo {author} {\bibfnamefont {R.}~\bibnamefont
  {Steinigeweg}}, \bibinfo {author} {\bibfnamefont {F.}~\bibnamefont {Jin}},
  \bibinfo {author} {\bibfnamefont {H.}~\bibnamefont {{De Raedt}}}, \bibinfo
  {author} {\bibfnamefont {K.}~\bibnamefont {Michielsen}}, \ and\ \bibinfo
  {author} {\bibfnamefont {J.}~\bibnamefont {Gemmer}},\ }\bibfield  {title}
  {\enquote {\bibinfo {title} {Charge diffusion in the one-dimensional
  {Hubbard} model},}\ }\href {\doibase 10.1103/PhysRevE.96.020105} {\bibfield
  {journal} {\bibinfo  {journal} {Phys. Rev. E}\ }\textbf {\bibinfo {volume}
  {96}},\ \bibinfo {pages} {020105} (\bibinfo {year}
  {2017}{\natexlab{b}})}\BibitemShut {NoStop}%
\bibitem [{\citenamefont {Yuan}\ \emph
  {et~al.}(2010{\natexlab{a}})\citenamefont {Yuan}, \citenamefont {{De
  Raedt}},\ and\ \citenamefont {Katsnelson}}]{YUAN10d}%
  \BibitemOpen
  \bibfield  {author} {\bibinfo {author} {\bibfnamefont {S.}~\bibnamefont
  {Yuan}}, \bibinfo {author} {\bibfnamefont {H.}~\bibnamefont {{De Raedt}}}, \
  and\ \bibinfo {author} {\bibfnamefont {M.I.}\ \bibnamefont {Katsnelson}},\
  }\bibfield  {title} {\enquote {\bibinfo {title} {Modeling electronic
  structure and transport properties of graphene with resonant scattering
  centers},}\ }\href {\doibase 10.1103/PhysRevB.82.115448} {\bibfield
  {journal} {\bibinfo  {journal} {Phys. Rev. B}\ }\textbf {\bibinfo {volume}
  {82}},\ \bibinfo {pages} {115448} (\bibinfo {year}
  {2010}{\natexlab{a}})}\BibitemShut {NoStop}%
\bibitem [{\citenamefont {Yuan}\ \emph
  {et~al.}(2010{\natexlab{b}})\citenamefont {Yuan}, \citenamefont {{De
  Raedt}},\ and\ \citenamefont {Katsnelson}}]{YUAN10e}%
  \BibitemOpen
  \bibfield  {author} {\bibinfo {author} {\bibfnamefont {S.}~\bibnamefont
  {Yuan}}, \bibinfo {author} {\bibfnamefont {H.}~\bibnamefont {{De Raedt}}}, \
  and\ \bibinfo {author} {\bibfnamefont {M.I.}\ \bibnamefont {Katsnelson}},\
  }\bibfield  {title} {\enquote {\bibinfo {title} {Electronic transport in
  disordered bilayer and trilayer graphene},}\ }\href {\doibase
  10.1103/PhysRevB.82.235409} {\bibfield  {journal} {\bibinfo  {journal} {Phys.
  Rev. B}\ }\textbf {\bibinfo {volume} {82}},\ \bibinfo {pages} {235409}
  (\bibinfo {year} {2010}{\natexlab{b}})}\BibitemShut {NoStop}%
\bibitem [{\citenamefont {Yuan}\ \emph {et~al.}(2014)\citenamefont {Yuan},
  \citenamefont {Rold\'an}, \citenamefont {Katsnelson},\ and\ \citenamefont
  {Guinea}}]{YUAN14g}%
  \BibitemOpen
  \bibfield  {author} {\bibinfo {author} {\bibfnamefont {S.}~\bibnamefont
  {Yuan}}, \bibinfo {author} {\bibfnamefont {R.}~\bibnamefont {Rold\'an}},
  \bibinfo {author} {\bibfnamefont {M.~I.}\ \bibnamefont {Katsnelson}}, \ and\
  \bibinfo {author} {\bibfnamefont {F.}~\bibnamefont {Guinea}},\ }\bibfield
  {title} {\enquote {\bibinfo {title} {Effect of point defects on the optical
  and transport properties of {${\mathrm{MoS}}_{2}$} and
  {${\mathrm{WS}}_{2}$}},}\ }\href {\doibase 10.1103/PhysRevB.90.041402}
  {\bibfield  {journal} {\bibinfo  {journal} {Phys. Rev. B}\ }\textbf {\bibinfo
  {volume} {90}},\ \bibinfo {pages} {041402} (\bibinfo {year}
  {2014})}\BibitemShut {NoStop}%
\bibitem [{\citenamefont {Jin}\ \emph {et~al.}(2015{\natexlab{b}})\citenamefont
  {Jin}, \citenamefont {Rold\'an}, \citenamefont {Katsnelson},\ and\
  \citenamefont {Yuan}}]{YUAN15}%
  \BibitemOpen
  \bibfield  {author} {\bibinfo {author} {\bibfnamefont {F.}~\bibnamefont
  {Jin}}, \bibinfo {author} {\bibfnamefont {R.}~\bibnamefont {Rold\'an}},
  \bibinfo {author} {\bibfnamefont {M.I.}\ \bibnamefont {Katsnelson}}, \ and\
  \bibinfo {author} {\bibfnamefont {S.}~\bibnamefont {Yuan}},\ }\bibfield
  {title} {\enquote {\bibinfo {title} {Screening and plasmons in pure and
  disordered single- and bilayer black phosphorus},}\ }\href {\doibase
  10.1103/PhysRevB.92.115440} {\bibfield  {journal} {\bibinfo  {journal} {Phys.
  Rev. B}\ }\textbf {\bibinfo {volume} {92}},\ \bibinfo {pages} {115440}
  (\bibinfo {year} {2015}{\natexlab{b}})}\BibitemShut {NoStop}%
\bibitem [{\citenamefont {{De Raedt}}(1987)}]{RAED87}%
  \BibitemOpen
  \bibfield  {author} {\bibinfo {author} {\bibfnamefont {H.}~\bibnamefont {{De
  Raedt}}},\ }\bibfield  {title} {\enquote {\bibinfo {title} {Product formula
  algorithms for solving the time dependent {Schr{\" o}dinger} equation},}\
  }\href {\doibase DOI: 10.1016/0167-7977(87)90002-5} {\bibfield  {journal}
  {\bibinfo  {journal} {Comp. Phys. Rep.}\ }\textbf {\bibinfo {volume} {7}},\
  \bibinfo {pages} {1 -- 72} (\bibinfo {year} {1987})}\BibitemShut {NoStop}%
\bibitem [{\citenamefont {{De Raedt}}\ and\ \citenamefont
  {Michielsen}(2006)}]{RAED06}%
  \BibitemOpen
  \bibfield  {author} {\bibinfo {author} {\bibfnamefont {H.}~\bibnamefont {{De
  Raedt}}}\ and\ \bibinfo {author} {\bibfnamefont {K.}~\bibnamefont
  {Michielsen}},\ }\bibfield  {title} {\enquote {\bibinfo {title}
  {{Computational Methods for Simulating Quantum Computers}},}\ }in\ \href@noop
  {} {\emph {\bibinfo {booktitle} {Handbook of Theoretical and Computational
  Nanotechnology}}},\ \bibinfo {editor} {edited by\ \bibinfo {editor}
  {\bibfnamefont {M.}~\bibnamefont {Rieth}}\ and\ \bibinfo {editor}
  {\bibfnamefont {W.}~\bibnamefont {Schommers}}}\ (\bibinfo  {publisher}
  {American Scientific Publishers},\ \bibinfo {address} {Los Angeles},\
  \bibinfo {year} {2006})\ pp.\ \bibinfo {pages} {2 -- 48}\BibitemShut
  {NoStop}%
\bibitem [{\citenamefont {{Tal-Ezer}}\ and\ \citenamefont
  {Kosloff}(1984)}]{TALE84}%
  \BibitemOpen
  \bibfield  {author} {\bibinfo {author} {\bibfnamefont {H.}~\bibnamefont
  {{Tal-Ezer}}}\ and\ \bibinfo {author} {\bibfnamefont {R.}~\bibnamefont
  {Kosloff}},\ }\bibfield  {title} {\enquote {\bibinfo {title} {An accurate and
  efficient scheme for propagating the time dependent {Schr\"odinger}
  equation},}\ }\href@noop {} {\bibfield  {journal} {\bibinfo  {journal} {J.
  Chem. Phys.}\ }\textbf {\bibinfo {volume} {81}},\ \bibinfo {pages}
  {3967--3971} (\bibinfo {year} {1984})}\BibitemShut {NoStop}%
\bibitem [{\citenamefont {Dobrovitski}\ and\ \citenamefont {{De
  Raedt}}(2003)}]{DOBR03}%
  \BibitemOpen
  \bibfield  {author} {\bibinfo {author} {\bibfnamefont {V.~V.}\ \bibnamefont
  {Dobrovitski}}\ and\ \bibinfo {author} {\bibfnamefont {H.}~\bibnamefont {{De
  Raedt}}},\ }\bibfield  {title} {\enquote {\bibinfo {title} {Efficient scheme
  for numerical simulations of the spin-bath decoherence},}\ }\href@noop {}
  {\bibfield  {journal} {\bibinfo  {journal} {Phys. Rev. E}\ }\textbf {\bibinfo
  {volume} {67}},\ \bibinfo {pages} {056702} (\bibinfo {year}
  {2003})}\BibitemShut {NoStop}%
\bibitem [{\citenamefont {Wei\ss{}e}\ \emph {et~al.}(2006)\citenamefont
  {Wei\ss{}e}, \citenamefont {Wellein}, \citenamefont {Alvermann},\ and\
  \citenamefont {Fehske}}]{WEIS06}%
  \BibitemOpen
  \bibfield  {author} {\bibinfo {author} {\bibfnamefont {A.}~\bibnamefont
  {Wei\ss{}e}}, \bibinfo {author} {\bibfnamefont {G.}~\bibnamefont {Wellein}},
  \bibinfo {author} {\bibfnamefont {A.}~\bibnamefont {Alvermann}}, \ and\
  \bibinfo {author} {\bibfnamefont {H.}~\bibnamefont {Fehske}},\ }\bibfield
  {title} {\enquote {\bibinfo {title} {The kernel polynomial method},}\ }\href
  {\doibase 10.1103/RevModPhys.78.275} {\bibfield  {journal} {\bibinfo
  {journal} {Rev. Mod. Phys.}\ }\textbf {\bibinfo {volume} {78}},\ \bibinfo
  {pages} {275--306} (\bibinfo {year} {2006})}\BibitemShut {NoStop}%
\bibitem [{\citenamefont {Prelov\v{s}ek}(2017)}]{PREL17}%
  \BibitemOpen
  \bibfield  {author} {\bibinfo {author} {\bibfnamefont {P.}~\bibnamefont
  {Prelov\v{s}ek}},\ }\bibfield  {title} {\enquote {\bibinfo {title} {The
  finite temperature {Lanczos} method and its applications},}\ }in\ \href@noop
  {} {\emph {\bibinfo {booktitle} {The Physics of Correlated Insulators,
  Metals, and Superconductors Modeling and Simulation}}},\ \bibinfo {editor}
  {edited by\ \bibinfo {editor} {\bibfnamefont {E.}~\bibnamefont {Pavarini}},
  \bibinfo {editor} {\bibfnamefont {E.}~\bibnamefont {Koch}}, \bibinfo {editor}
  {\bibfnamefont {R.}~\bibnamefont {Scalettar}}, \ and\ \bibinfo {editor}
  {\bibfnamefont {R.}~\bibnamefont {Martin}}}\ (\bibinfo  {publisher}
  {Forschungszentrum J\"{u}lich},\ \bibinfo {address} {J\"{u}lich},\ \bibinfo
  {year} {2017})\BibitemShut {NoStop}%
\bibitem [{\citenamefont {Schnack}\ \emph {et~al.}(2020)\citenamefont
  {Schnack}, \citenamefont {Richter},\ and\ \citenamefont
  {Steinigeweg}}]{SCHN20}%
  \BibitemOpen
  \bibfield  {author} {\bibinfo {author} {\bibfnamefont {J.}~\bibnamefont
  {Schnack}}, \bibinfo {author} {\bibfnamefont {J.}~\bibnamefont {Richter}}, \
  and\ \bibinfo {author} {\bibfnamefont {R.}~\bibnamefont {Steinigeweg}},\
  }\bibfield  {title} {\enquote {\bibinfo {title} {Accuracy of the
  finite-temperature {Lanczos} method compared to simple typicality-based
  estimates},}\ }\href {\doibase 10.1103/PhysRevResearch.2.013186} {\bibfield
  {journal} {\bibinfo  {journal} {Phys. Rev. Research}\ }\textbf {\bibinfo
  {volume} {2}},\ \bibinfo {pages} {013186} (\bibinfo {year}
  {2020})}\BibitemShut {NoStop}%
\bibitem [{\citenamefont {Elsayed}\ and\ \citenamefont {Fine}(2013)}]{TARE13}%
  \BibitemOpen
  \bibfield  {author} {\bibinfo {author} {\bibfnamefont {T.A.}\ \bibnamefont
  {Elsayed}}\ and\ \bibinfo {author} {\bibfnamefont {B.V.}\ \bibnamefont
  {Fine}},\ }\bibfield  {title} {\enquote {\bibinfo {title} {Regression
  relation for pure quantum states and its implications for efficient
  computing},}\ }\href {\doibase 10.1103/PhysRevLett.110.070404} {\bibfield
  {journal} {\bibinfo  {journal} {Phys. Rev. Lett.}\ }\textbf {\bibinfo
  {volume} {110}},\ \bibinfo {pages} {070404} (\bibinfo {year}
  {2013})}\BibitemShut {NoStop}%
\bibitem [{\citenamefont {{Google AI Quantum}}\ and\ \citenamefont
  {collaborators}(2019)}]{GOOG19}%
  \BibitemOpen
  \bibfield  {author} {\bibinfo {author} {\bibnamefont {{Google AI Quantum}}}\
  and\ \bibinfo {author} {\bibnamefont {collaborators}},\ }\bibfield  {title}
  {\enquote {\bibinfo {title} {Quantum supremacy using a programmable
  superconducting processor},}\ }\href@noop {} {\bibfield  {journal} {\bibinfo
  {journal} {Nature}\ }\textbf {\bibinfo {volume} {574}},\ \bibinfo {pages}
  {505--510} (\bibinfo {year} {2019})}\BibitemShut {NoStop}%
\bibitem [{\citenamefont {Nielsen}(2002)}]{nielsen2002gatefidelity}%
  \BibitemOpen
  \bibfield  {author} {\bibinfo {author} {\bibfnamefont {M.A.}\ \bibnamefont
  {Nielsen}},\ }\bibfield  {title} {\enquote {\bibinfo {title} {A simple
  formula for the average gate fidelity of a quantum dynamical operation},}\
  }\href {\doibase 10.1016/S0375-9601(02)01272-0} {\bibfield  {journal}
  {\bibinfo  {journal} {Phys. Lett. A}\ }\textbf {\bibinfo {volume} {303}},\
  \bibinfo {pages} {249 -- 252} (\bibinfo {year} {2002})}\BibitemShut {NoStop}%
\bibitem [{\citenamefont {Gilchrist}\ \emph {et~al.}(2005)\citenamefont
  {Gilchrist}, \citenamefont {Langford},\ and\ \citenamefont
  {Nielsen}}]{Gilchrist2005fidelities}%
  \BibitemOpen
  \bibfield  {author} {\bibinfo {author} {\bibfnamefont {A.}~\bibnamefont
  {Gilchrist}}, \bibinfo {author} {\bibfnamefont {N.K.}\ \bibnamefont
  {Langford}}, \ and\ \bibinfo {author} {\bibfnamefont {M.A.}\ \bibnamefont
  {Nielsen}},\ }\bibfield  {title} {\enquote {\bibinfo {title} {Distance
  measures to compare real and ideal quantum processes},}\ }\href {\doibase
  10.1103/physreva.71.062310} {\bibfield  {journal} {\bibinfo  {journal} {Phys.
  Rev. A}\ }\textbf {\bibinfo {volume} {71}},\ \bibinfo {pages} {062310}
  (\bibinfo {year} {2005})}\BibitemShut {NoStop}%
\bibitem [{\citenamefont {Iitaka}\ and\ \citenamefont
  {Ebisuzaki}(2004)}]{IITA04}%
  \BibitemOpen
  \bibfield  {author} {\bibinfo {author} {\bibfnamefont {T.}~\bibnamefont
  {Iitaka}}\ and\ \bibinfo {author} {\bibfnamefont {T.}~\bibnamefont
  {Ebisuzaki}},\ }\bibfield  {title} {\enquote {\bibinfo {title} {Random phase
  vector for calculating the trace of a large matrix},}\ }\href@noop {}
  {\bibfield  {journal} {\bibinfo  {journal} {Phys. Rev. E}\ }\textbf {\bibinfo
  {volume} {69}},\ \bibinfo {pages} {057701} (\bibinfo {year}
  {2004})}\BibitemShut {NoStop}%
\bibitem [{\citenamefont {Grimmet}\ and\ \citenamefont
  {Stirzaker}(2001)}]{GRIM01}%
  \BibitemOpen
  \bibfield  {author} {\bibinfo {author} {\bibfnamefont {G.~R.}\ \bibnamefont
  {Grimmet}}\ and\ \bibinfo {author} {\bibfnamefont {D.~R.}\ \bibnamefont
  {Stirzaker}},\ }\href@noop {} {\emph {\bibinfo {title} {Probability and
  Random Processes}}}\ (\bibinfo  {publisher} {Clarendon Press},\ \bibinfo
  {address} {Oxford},\ \bibinfo {year} {2001})\BibitemShut {NoStop}%
\bibitem [{Jug(2020)}]{JugitRST}%
  \BibitemOpen
  \href@noop {} {}\bibinfo {howpublished}
  {\url{https://jugit.fz-juelich.de/qip/random-state-technology}} (\bibinfo
  {year} {2020})\BibitemShut {NoStop}%
\bibitem [{\citenamefont {Gershgorin}(1931)}]{GERS31}%
  \BibitemOpen
  \bibfield  {author} {\bibinfo {author} {\bibfnamefont {S.}~\bibnamefont
  {Gershgorin}},\ }\bibfield  {title} {\enquote {\bibinfo {title} {{\"U}ber die
  {Abgrenzung} der {Eigenwerte} einer {Matrix}},}\ }\href@noop {} {\bibfield
  {journal} {\bibinfo  {journal} {Bulletin de l'Acad\'emie des Sciences de
  l'URSS. Classe des sciences math\'ematiques et na}\ }\textbf {\bibinfo
  {volume} {6}},\ \bibinfo {pages} {749--754} (\bibinfo {year}
  {1931})}\BibitemShut {NoStop}%
\bibitem [{\citenamefont {Leforestier}\ \emph {et~al.}(1991)\citenamefont
  {Leforestier}, \citenamefont {Bisseling}, \citenamefont {Cerjan},
  \citenamefont {Feit}, \citenamefont {Friesner}, \citenamefont {Guldberg},
  \citenamefont {Hammerich}, \citenamefont {Jolicard}, \citenamefont
  {Karrlein}, \citenamefont {Meyer}, \citenamefont {Lipkin}, \citenamefont
  {Roncero},\ and\ \citenamefont {Kosloff}}]{LEFO91}%
  \BibitemOpen
  \bibfield  {author} {\bibinfo {author} {\bibfnamefont {C.}~\bibnamefont
  {Leforestier}}, \bibinfo {author} {\bibfnamefont {R.~H.}\ \bibnamefont
  {Bisseling}}, \bibinfo {author} {\bibfnamefont {C.}~\bibnamefont {Cerjan}},
  \bibinfo {author} {\bibfnamefont {M.~D.}\ \bibnamefont {Feit}}, \bibinfo
  {author} {\bibfnamefont {R.}~\bibnamefont {Friesner}}, \bibinfo {author}
  {\bibfnamefont {A.}~\bibnamefont {Guldberg}}, \bibinfo {author}
  {\bibfnamefont {A.}~\bibnamefont {Hammerich}}, \bibinfo {author}
  {\bibfnamefont {G.}~\bibnamefont {Jolicard}}, \bibinfo {author}
  {\bibfnamefont {W.}~\bibnamefont {Karrlein}}, \bibinfo {author}
  {\bibfnamefont {H.-D.}\ \bibnamefont {Meyer}}, \bibinfo {author}
  {\bibfnamefont {N.}~\bibnamefont {Lipkin}}, \bibinfo {author} {\bibfnamefont
  {O.}~\bibnamefont {Roncero}}, \ and\ \bibinfo {author} {\bibfnamefont
  {R.}~\bibnamefont {Kosloff}},\ }\bibfield  {title} {\enquote {\bibinfo
  {title} {A comparison of different propagation schemes for the time-dependent
  {Schr\"odinger} equation},}\ }\href@noop {} {\bibfield  {journal} {\bibinfo
  {journal} {J. Comput. Phys.}\ }\textbf {\bibinfo {volume} {94}},\ \bibinfo
  {pages} {59--80} (\bibinfo {year} {1991})}\BibitemShut {NoStop}%
\bibitem [{\citenamefont {Wilkinson}(1965)}]{WILK65}%
  \BibitemOpen
  \bibfield  {author} {\bibinfo {author} {\bibfnamefont {J.H.}\ \bibnamefont
  {Wilkinson}},\ }\href@noop {} {\emph {\bibinfo {title} {The Algebraic
  Eigenvalue Problem}}}\ (\bibinfo  {publisher} {Clarendon Press},\ \bibinfo
  {address} {Oxford},\ \bibinfo {year} {1965})\BibitemShut {NoStop}%
\bibitem [{\citenamefont {Elstner}\ and\ \citenamefont {Young}(1994)}]{ELST94}%
  \BibitemOpen
  \bibfield  {author} {\bibinfo {author} {\bibfnamefont {N.}~\bibnamefont
  {Elstner}}\ and\ \bibinfo {author} {\bibfnamefont {A.~P.}\ \bibnamefont
  {Young}},\ }\bibfield  {title} {\enquote {\bibinfo {title} {Spin-1/2
  {Heisenberg} antiferromagnet on the kagome lattice: High-temperature
  expansion and exact-diagonalization studies},}\ }\href {\doibase
  10.1103/PhysRevB.50.6871} {\bibfield  {journal} {\bibinfo  {journal} {Phys.
  Rev. B}\ }\textbf {\bibinfo {volume} {50}},\ \bibinfo {pages} {6871--6876}
  (\bibinfo {year} {1994})}\BibitemShut {NoStop}%
\bibitem [{\citenamefont {Schnack}\ \emph {et~al.}(2018)\citenamefont
  {Schnack}, \citenamefont {Schulenburg},\ and\ \citenamefont
  {Richter}}]{SCHN18}%
  \BibitemOpen
  \bibfield  {author} {\bibinfo {author} {\bibfnamefont {J.}~\bibnamefont
  {Schnack}}, \bibinfo {author} {\bibfnamefont {J.}~\bibnamefont
  {Schulenburg}}, \ and\ \bibinfo {author} {\bibfnamefont {J.}~\bibnamefont
  {Richter}},\ }\bibfield  {title} {\enquote {\bibinfo {title} {Magnetism of
  the $n=42$ kagome lattice antiferromagnet},}\ }\href {\doibase
  10.1103/PhysRevB.98.094423} {\bibfield  {journal} {\bibinfo  {journal} {Phys.
  Rev. B}\ }\textbf {\bibinfo {volume} {98}},\ \bibinfo {pages} {094423}
  (\bibinfo {year} {2018})}\BibitemShut {NoStop}%
\bibitem [{\citenamefont {Prelov\v{s}ek}\ and\ \citenamefont
  {Kokalj}(2018)}]{PREL18}%
  \BibitemOpen
  \bibfield  {author} {\bibinfo {author} {\bibfnamefont {P.}~\bibnamefont
  {Prelov\v{s}ek}}\ and\ \bibinfo {author} {\bibfnamefont {J.}~\bibnamefont
  {Kokalj}},\ }\bibfield  {title} {\enquote {\bibinfo {title}
  {Finite-temperature properties of the extended {Heisenberg} model on a
  triangular lattice},}\ }\href {\doibase 10.1103/PhysRevB.98.035107}
  {\bibfield  {journal} {\bibinfo  {journal} {Phys. Rev. B}\ }\textbf {\bibinfo
  {volume} {98}},\ \bibinfo {pages} {035107} (\bibinfo {year}
  {2018})}\BibitemShut {NoStop}%
\bibitem [{\citenamefont {Nakamura}\ and\ \citenamefont
  {Miyashita}(1995)}]{NAKA95}%
  \BibitemOpen
  \bibfield  {author} {\bibinfo {author} {\bibfnamefont {T.}~\bibnamefont
  {Nakamura}}\ and\ \bibinfo {author} {\bibfnamefont {S.}~\bibnamefont
  {Miyashita}},\ }\bibfield  {title} {\enquote {\bibinfo {title} {Thermodynamic
  properties of the quantum {Heisenberg} antiferromagnet on the kagome
  lattice},}\ }\href {\doibase 10.1103/PhysRevB.52.9174} {\bibfield  {journal}
  {\bibinfo  {journal} {Phys. Rev. B}\ }\textbf {\bibinfo {volume} {52}},\
  \bibinfo {pages} {9174--9177} (\bibinfo {year} {1995})}\BibitemShut {NoStop}%
\bibitem [{\citenamefont {Shimokawa}\ and\ \citenamefont
  {Kawamura}(2016)}]{SHIM16}%
  \BibitemOpen
  \bibfield  {author} {\bibinfo {author} {\bibfnamefont {T.}~\bibnamefont
  {Shimokawa}}\ and\ \bibinfo {author} {\bibfnamefont {H.}~\bibnamefont
  {Kawamura}},\ }\bibfield  {title} {\enquote {\bibinfo {title}
  {Finite-temperature crossover phenomenon in the $s = 1/2$ antiferromagnetic
  {Heisenberg} model on the kagome lattice},}\ }\href {\doibase
  10.7566/JPSJ.85.113702} {\bibfield  {journal} {\bibinfo  {journal} {J. Phys.
  Soc. Jpn.}\ }\textbf {\bibinfo {volume} {85}},\ \bibinfo {pages} {113702}
  (\bibinfo {year} {2016})}\BibitemShut {NoStop}%
\bibitem [{\citenamefont {Cullum}\ and\ \citenamefont
  {Willoughby}(2002)}]{CULL02}%
  \BibitemOpen
  \bibfield  {author} {\bibinfo {author} {\bibfnamefont {J.K.}\ \bibnamefont
  {Cullum}}\ and\ \bibinfo {author} {\bibfnamefont {R.A.}\ \bibnamefont
  {Willoughby}},\ }\href {\doibase 10.1137/1.9780898719192} {\emph {\bibinfo
  {title} {Lanczos Algorithms for Large Symmetric Eigenvalue Computations}}}\
  (\bibinfo  {publisher} {Society for Industrial and Applied Mathematics},\
  \bibinfo {year} {2002})\ \Eprint
  {http://arxiv.org/abs/https://epubs.siam.org/doi/pdf/10.1137/1.9780898719192}
  {https://epubs.siam.org/doi/pdf/10.1137/1.9780898719192} \BibitemShut
  {NoStop}%
\bibitem [{\citenamefont {Sakurai}\ and\ \citenamefont
  {Sugiura}(2003)}]{SAKU03}%
  \BibitemOpen
  \bibfield  {author} {\bibinfo {author} {\bibfnamefont {T.}~\bibnamefont
  {Sakurai}}\ and\ \bibinfo {author} {\bibfnamefont {H.}~\bibnamefont
  {Sugiura}},\ }\bibfield  {title} {\enquote {\bibinfo {title} {A projection
  method for generalized eigenvalue problems using numerical integration},}\
  }\href {\doibase https://doi.org/10.1016/S0377-0427(03)00565-X} {\bibfield
  {journal} {\bibinfo  {journal} {Journal of Computational and Applied
  Mathematics}\ }\textbf {\bibinfo {volume} {159}},\ \bibinfo {pages} {119 --
  128} (\bibinfo {year} {2003})},\ \bibinfo {note} {6th Japan-China Joint
  Seminar on Numerical Mathematics; In Search for the Frontier of Computational
  and Applied Mathematics toward the 21st Century}\BibitemShut {NoStop}%
\bibitem [{\citenamefont {Morita}\ and\ \citenamefont
  {Tohyama}(2020)}]{MORI20}%
  \BibitemOpen
  \bibfield  {author} {\bibinfo {author} {\bibfnamefont {K.}~\bibnamefont
  {Morita}}\ and\ \bibinfo {author} {\bibfnamefont {T.}~\bibnamefont
  {Tohyama}},\ }\bibfield  {title} {\enquote {\bibinfo {title}
  {Finite-temperature properties of the {Kitaev-Heisenberg} models on kagome
  and triangular lattices studied by improved finite-temperature {Lanczos}
  methods},}\ }\href {\doibase 10.1103/PhysRevResearch.2.013205} {\bibfield
  {journal} {\bibinfo  {journal} {Phys. Rev. Research}\ }\textbf {\bibinfo
  {volume} {2}},\ \bibinfo {pages} {013205} (\bibinfo {year}
  {2020})}\BibitemShut {NoStop}%
\bibitem [{\citenamefont {Steinigeweg}\ \emph
  {et~al.}(2014{\natexlab{b}})\citenamefont {Steinigeweg}, \citenamefont
  {Gemmer},\ and\ \citenamefont {Brenig}}]{STEI14b}%
  \BibitemOpen
  \bibfield  {author} {\bibinfo {author} {\bibfnamefont {R.}~\bibnamefont
  {Steinigeweg}}, \bibinfo {author} {\bibfnamefont {J.}~\bibnamefont {Gemmer}},
  \ and\ \bibinfo {author} {\bibfnamefont {W.}~\bibnamefont {Brenig}},\
  }\bibfield  {title} {\enquote {\bibinfo {title} {Spin-current
  autocorrelations from single pure-state propagation},}\ }\href@noop {}
  {\bibfield  {journal} {\bibinfo  {journal} {Phys. Rev. Lett.}\ }\textbf
  {\bibinfo {volume} {112}},\ \bibinfo {pages} {120601} (\bibinfo {year}
  {2014}{\natexlab{b}})}\BibitemShut {NoStop}%
\bibitem [{\citenamefont {Karrasch}\ \emph {et~al.}(2012)\citenamefont
  {Karrasch}, \citenamefont {Bardarson},\ and\ \citenamefont {Moore}}]{KARR12}%
  \BibitemOpen
  \bibfield  {author} {\bibinfo {author} {\bibfnamefont {C.}~\bibnamefont
  {Karrasch}}, \bibinfo {author} {\bibfnamefont {J.~H.}\ \bibnamefont
  {Bardarson}}, \ and\ \bibinfo {author} {\bibfnamefont {J.~E.}\ \bibnamefont
  {Moore}},\ }\bibfield  {title} {\enquote {\bibinfo {title}
  {Finite-temperature dynamical density matrix renormalization group and the
  {Drude} weight of spin-$1/2$ chains},}\ }\href {\doibase
  10.1103/PhysRevLett.108.227206} {\bibfield  {journal} {\bibinfo  {journal}
  {Phys. Rev. Lett.}\ }\textbf {\bibinfo {volume} {108}},\ \bibinfo {pages}
  {227206} (\bibinfo {year} {2012})}\BibitemShut {NoStop}%
\bibitem [{\citenamefont {Karrasch}\ \emph {et~al.}(2013)\citenamefont
  {Karrasch}, \citenamefont {Hauschild}, \citenamefont {Langer},\ and\
  \citenamefont {Heidrich-Meisner}}]{KARR13}%
  \BibitemOpen
  \bibfield  {author} {\bibinfo {author} {\bibfnamefont {C.}~\bibnamefont
  {Karrasch}}, \bibinfo {author} {\bibfnamefont {J.}~\bibnamefont {Hauschild}},
  \bibinfo {author} {\bibfnamefont {S.}~\bibnamefont {Langer}}, \ and\ \bibinfo
  {author} {\bibfnamefont {F.}~\bibnamefont {Heidrich-Meisner}},\ }\bibfield
  {title} {\enquote {\bibinfo {title} {{Drude} weight of the spin-1/2 {XXZ}
  chain: Density matrix renormalization group versus exact diagonalization},}\
  }\href {\doibase 10.1103/PhysRevB.87.245128} {\bibfield  {journal} {\bibinfo
  {journal} {Phys. Rev. B}\ }\textbf {\bibinfo {volume} {87}},\ \bibinfo
  {pages} {245128} (\bibinfo {year} {2013})}\BibitemShut {NoStop}%
\bibitem [{\citenamefont {Richter}\ and\ \citenamefont
  {Steinigeweg}(2019)}]{RICH19b}%
  \BibitemOpen
  \bibfield  {author} {\bibinfo {author} {\bibfnamefont {J.}~\bibnamefont
  {Richter}}\ and\ \bibinfo {author} {\bibfnamefont {R.}~\bibnamefont
  {Steinigeweg}},\ }\bibfield  {title} {\enquote {\bibinfo {title} {Combining
  dynamical quantum typicality and numerical linked cluster expansions},}\
  }\href {\doibase 10.1103/PhysRevB.99.094419} {\bibfield  {journal} {\bibinfo
  {journal} {Phys. Rev. B}\ }\textbf {\bibinfo {volume} {99}},\ \bibinfo
  {pages} {094419} (\bibinfo {year} {2019})}\BibitemShut {NoStop}%
\bibitem [{\citenamefont {Richter}\ \emph
  {et~al.}(2019{\natexlab{b}})\citenamefont {Richter}, \citenamefont {Jin},
  \citenamefont {Knipschild}, \citenamefont {{De Raedt}}, \citenamefont
  {Michielsen}, \citenamefont {Gemmer},\ and\ \citenamefont
  {Steinigeweg}}]{RICH19c}%
  \BibitemOpen
  \bibfield  {author} {\bibinfo {author} {\bibfnamefont {J.}~\bibnamefont
  {Richter}}, \bibinfo {author} {\bibfnamefont {F.}~\bibnamefont {Jin}},
  \bibinfo {author} {\bibfnamefont {L.}~\bibnamefont {Knipschild}}, \bibinfo
  {author} {\bibfnamefont {H.}~\bibnamefont {{De Raedt}}}, \bibinfo {author}
  {\bibfnamefont {K.}~\bibnamefont {Michielsen}}, \bibinfo {author}
  {\bibfnamefont {J.}~\bibnamefont {Gemmer}}, \ and\ \bibinfo {author}
  {\bibfnamefont {R.}~\bibnamefont {Steinigeweg}},\ }\bibfield  {title}
  {\enquote {\bibinfo {title} {Exponential damping induced by random and
  realistic perturbations},}\ }\href@noop {} {\bibfield  {journal} {\bibinfo
  {journal} {arXiv:1906.09268}\ } (\bibinfo {year}
  {2019}{\natexlab{b}})}\BibitemShut {NoStop}%
\bibitem [{\citenamefont {Karrasch}\ \emph {et~al.}(2014)\citenamefont
  {Karrasch}, \citenamefont {Moore},\ and\ \citenamefont
  {Heidrich-Meisner}}]{KARR14}%
  \BibitemOpen
  \bibfield  {author} {\bibinfo {author} {\bibfnamefont {C.}~\bibnamefont
  {Karrasch}}, \bibinfo {author} {\bibfnamefont {J.~E.}\ \bibnamefont {Moore}},
  \ and\ \bibinfo {author} {\bibfnamefont {F.}~\bibnamefont
  {Heidrich-Meisner}},\ }\bibfield  {title} {\enquote {\bibinfo {title}
  {Real-time and real-space spin and energy dynamics in one-dimensional
  spin-$\frac{1}{2}$ systems induced by local quantum quenches at finite
  temperatures},}\ }\href {\doibase 10.1103/PhysRevB.89.075139} {\bibfield
  {journal} {\bibinfo  {journal} {Phys. Rev. B}\ }\textbf {\bibinfo {volume}
  {89}},\ \bibinfo {pages} {075139} (\bibinfo {year} {2014})}\BibitemShut
  {NoStop}%
\bibitem [{\citenamefont {Kubo}(1957)}]{KUBO57}%
  \BibitemOpen
  \bibfield  {author} {\bibinfo {author} {\bibfnamefont {R.}~\bibnamefont
  {Kubo}},\ }\bibfield  {title} {\enquote {\bibinfo {title}
  {Statistical-mechanical theory of irreversible processes. {I.}}}\ }\href@noop
  {} {\bibfield  {journal} {\bibinfo  {journal} {J. Phys. Soc. Jpn.}\ }\textbf
  {\bibinfo {volume} {12}},\ \bibinfo {pages} {570--586} (\bibinfo {year}
  {1957})}\BibitemShut {NoStop}%
\bibitem [{\citenamefont {Miyashita}\ \emph {et~al.}(1999)\citenamefont
  {Miyashita}, \citenamefont {Yoshino},\ and\ \citenamefont
  {Ogasahara}}]{MIYA99}%
  \BibitemOpen
  \bibfield  {author} {\bibinfo {author} {\bibfnamefont {S.}~\bibnamefont
  {Miyashita}}, \bibinfo {author} {\bibfnamefont {T.}~\bibnamefont {Yoshino}},
  \ and\ \bibinfo {author} {\bibfnamefont {A.}~\bibnamefont {Ogasahara}},\
  }\bibfield  {title} {\enquote {\bibinfo {title} {Direct calculation of
  dynamical susceptibility in strongly fluctuating quantum spin systems},}\
  }\href {\doibase 10.1143/JPSJ.68.655} {\bibfield  {journal} {\bibinfo
  {journal} {J. Phys. Soc. Jpn.}\ }\textbf {\bibinfo {volume} {68}},\ \bibinfo
  {pages} {655--661} (\bibinfo {year} {1999})}\BibitemShut {NoStop}%
\bibitem [{\citenamefont {{El Shawish}}\ \emph {et~al.}(2010)\citenamefont {{El
  Shawish}}, \citenamefont {C\'epas},\ and\ \citenamefont
  {Miyashita}}]{ELSH10}%
  \BibitemOpen
  \bibfield  {author} {\bibinfo {author} {\bibfnamefont {S.}~\bibnamefont {{El
  Shawish}}}, \bibinfo {author} {\bibfnamefont {O.}~\bibnamefont {C\'epas}}, \
  and\ \bibinfo {author} {\bibfnamefont {S.}~\bibnamefont {Miyashita}},\
  }\bibfield  {title} {\enquote {\bibinfo {title} {Electron spin resonance in
  $s=\frac{1}{2}$ antiferromagnets at high temperature},}\ }\href {\doibase
  10.1103/PhysRevB.81.224421} {\bibfield  {journal} {\bibinfo  {journal} {Phys.
  Rev. B}\ }\textbf {\bibinfo {volume} {81}},\ \bibinfo {pages} {224421}
  (\bibinfo {year} {2010})}\BibitemShut {NoStop}%
\bibitem [{\citenamefont {Oshikawa}\ and\ \citenamefont
  {Affleck}(1999)}]{OSHI99}%
  \BibitemOpen
  \bibfield  {author} {\bibinfo {author} {\bibfnamefont {M.}~\bibnamefont
  {Oshikawa}}\ and\ \bibinfo {author} {\bibfnamefont {I.}~\bibnamefont
  {Affleck}},\ }\bibfield  {title} {\enquote {\bibinfo {title} {Low-temperature
  electron spin resonance theory for half-integer spin antiferromagnetic
  chains},}\ }\href {\doibase 10.1103/PhysRevLett.82.5136} {\bibfield
  {journal} {\bibinfo  {journal} {Phys. Rev. Lett.}\ }\textbf {\bibinfo
  {volume} {82}},\ \bibinfo {pages} {5136--5139} (\bibinfo {year}
  {1999})}\BibitemShut {NoStop}%
\bibitem [{\citenamefont {Ikeuchi}\ \emph {et~al.}(2015)\citenamefont
  {Ikeuchi}, \citenamefont {De~Raedt}, \citenamefont {Bertaina},\ and\
  \citenamefont {Miyashita}}]{IKEU15}%
  \BibitemOpen
  \bibfield  {author} {\bibinfo {author} {\bibfnamefont {H.}~\bibnamefont
  {Ikeuchi}}, \bibinfo {author} {\bibfnamefont {H.}~\bibnamefont {De~Raedt}},
  \bibinfo {author} {\bibfnamefont {S.}~\bibnamefont {Bertaina}}, \ and\
  \bibinfo {author} {\bibfnamefont {S.}~\bibnamefont {Miyashita}},\ }\bibfield
  {title} {\enquote {\bibinfo {title} {Computation of {ESR} spectra from the
  time evolution of the magnetization: Comparison of autocorrelation and
  {Wiener-Khinchin}-relation-based methods},}\ }\href {\doibase
  10.1103/PhysRevB.92.214431} {\bibfield  {journal} {\bibinfo  {journal} {Phys.
  Rev. B}\ }\textbf {\bibinfo {volume} {92}},\ \bibinfo {pages} {214431}
  (\bibinfo {year} {2015})}\BibitemShut {NoStop}%
\bibitem [{\citenamefont {Ikeuchi}\ \emph {et~al.}(2017)\citenamefont
  {Ikeuchi}, \citenamefont {De~Raedt}, \citenamefont {Bertaina},\ and\
  \citenamefont {Miyashita}}]{IKEU17}%
  \BibitemOpen
  \bibfield  {author} {\bibinfo {author} {\bibfnamefont {H.}~\bibnamefont
  {Ikeuchi}}, \bibinfo {author} {\bibfnamefont {H.}~\bibnamefont {De~Raedt}},
  \bibinfo {author} {\bibfnamefont {S.}~\bibnamefont {Bertaina}}, \ and\
  \bibinfo {author} {\bibfnamefont {S.}~\bibnamefont {Miyashita}},\ }\bibfield
  {title} {\enquote {\bibinfo {title} {Size and temperature dependence of the
  line shape of {ESR} spectra of the $\mathit{XXZ}$ antiferromagnetic chain},}\
  }\href {\doibase 10.1103/PhysRevB.95.024402} {\bibfield  {journal} {\bibinfo
  {journal} {Phys. Rev. B}\ }\textbf {\bibinfo {volume} {95}},\ \bibinfo
  {pages} {024402} (\bibinfo {year} {2017})}\BibitemShut {NoStop}%
\bibitem [{\citenamefont {Preskill}(2018)}]{PRES18}%
  \BibitemOpen
  \bibfield  {author} {\bibinfo {author} {\bibfnamefont {J.}~\bibnamefont
  {Preskill}},\ }\bibfield  {title} {\enquote {\bibinfo {title} {Quantum
  computing in the {NISQ} era and beyond},}\ }\href {\doibase
  10.22331/q-2018-08-06-79} {\bibfield  {journal} {\bibinfo  {journal}
  {Quantum}\ }\textbf {\bibinfo {volume} {2}},\ \bibinfo {pages} {79} (\bibinfo
  {year} {2018})}\BibitemShut {NoStop}%
\bibitem [{\citenamefont {Emerson}\ \emph {et~al.}(2005)\citenamefont
  {Emerson}, \citenamefont {Alicki},\ and\ \citenamefont
  {{\.Zyczkowski}}}]{emerson2005randomunitaryoperatorstwirling}%
  \BibitemOpen
  \bibfield  {author} {\bibinfo {author} {\bibfnamefont {J.}~\bibnamefont
  {Emerson}}, \bibinfo {author} {\bibfnamefont {R.}~\bibnamefont {Alicki}}, \
  and\ \bibinfo {author} {\bibfnamefont {K.}~\bibnamefont {{\.Zyczkowski}}},\
  }\bibfield  {title} {\enquote {\bibinfo {title} {Scalable noise estimation
  with random unitary operators},}\ }\href
  {http://stacks.iop.org/1464-4266/7/i=10/a=021} {\bibfield  {journal}
  {\bibinfo  {journal} {J. Opt. B: Quantum Semiclassical Opt.}\ }\textbf
  {\bibinfo {volume} {7}},\ \bibinfo {pages} {S347} (\bibinfo {year}
  {2005})}\BibitemShut {NoStop}%
\bibitem [{\citenamefont {Knill}\ \emph {et~al.}(2008)\citenamefont {Knill},
  \citenamefont {Leibfried}, \citenamefont {Reichle}, \citenamefont {Britton},
  \citenamefont {Blakestad}, \citenamefont {Jost}, \citenamefont {Langer},
  \citenamefont {Ozeri}, \citenamefont {Seidelin},\ and\ \citenamefont
  {Wineland}}]{Knill2008randomizedbenchmarking}%
  \BibitemOpen
  \bibfield  {author} {\bibinfo {author} {\bibfnamefont {E.}~\bibnamefont
  {Knill}}, \bibinfo {author} {\bibfnamefont {D.}~\bibnamefont {Leibfried}},
  \bibinfo {author} {\bibfnamefont {R.}~\bibnamefont {Reichle}}, \bibinfo
  {author} {\bibfnamefont {J.}~\bibnamefont {Britton}}, \bibinfo {author}
  {\bibfnamefont {R.~B.}\ \bibnamefont {Blakestad}}, \bibinfo {author}
  {\bibfnamefont {J.~D.}\ \bibnamefont {Jost}}, \bibinfo {author}
  {\bibfnamefont {C.}~\bibnamefont {Langer}}, \bibinfo {author} {\bibfnamefont
  {R.}~\bibnamefont {Ozeri}}, \bibinfo {author} {\bibfnamefont
  {S.}~\bibnamefont {Seidelin}}, \ and\ \bibinfo {author} {\bibfnamefont
  {D.~J.}\ \bibnamefont {Wineland}},\ }\bibfield  {title} {\enquote {\bibinfo
  {title} {Randomized benchmarking of quantum gates},}\ }\href {\doibase
  10.1103/physreva.77.012307} {\bibfield  {journal} {\bibinfo  {journal} {Phys.
  Rev. A}\ }\textbf {\bibinfo {volume} {77}},\ \bibinfo {pages} {012307}
  (\bibinfo {year} {2008})}\BibitemShut {NoStop}%
\bibitem [{\citenamefont {Jaynes}(2003)}]{JAYN03}%
  \BibitemOpen
  \bibfield  {author} {\bibinfo {author} {\bibfnamefont {E.~T.}\ \bibnamefont
  {Jaynes}},\ }\href@noop {} {\emph {\bibinfo {title} {Probability Theory: The
  Logic of Science}}}\ (\bibinfo  {publisher} {Cambridge University Press},\
  \bibinfo {address} {Cambridge},\ \bibinfo {year} {2003})\BibitemShut
  {NoStop}%
\bibitem [{\citenamefont {Nielsen}\ and\ \citenamefont
  {Chuang}(2000)}]{NIEL00}%
  \BibitemOpen
  \bibfield  {author} {\bibinfo {author} {\bibfnamefont {M.}~\bibnamefont
  {Nielsen}}\ and\ \bibinfo {author} {\bibfnamefont {I.}~\bibnamefont
  {Chuang}},\ }\href@noop {} {\emph {\bibinfo {title} {Quantum Computation and
  Quantum Information}}}\ (\bibinfo  {publisher} {Cambridge University Press},\
  \bibinfo {address} {Cambridge},\ \bibinfo {year} {2000})\BibitemShut
  {NoStop}%
\bibitem [{\citenamefont {Michielsen}\ \emph {et~al.}(2017)\citenamefont
  {Michielsen}, \citenamefont {Nocon}, \citenamefont {Willsch}, \citenamefont
  {Jin}, \citenamefont {{Th. Lippert}},\ and\ \citenamefont {{De
  Raedt}}}]{MICH17b}%
  \BibitemOpen
  \bibfield  {author} {\bibinfo {author} {\bibfnamefont {K.}~\bibnamefont
  {Michielsen}}, \bibinfo {author} {\bibfnamefont {M.}~\bibnamefont {Nocon}},
  \bibinfo {author} {\bibfnamefont {D.}~\bibnamefont {Willsch}}, \bibinfo
  {author} {\bibfnamefont {F.}~\bibnamefont {Jin}}, \bibinfo {author}
  {\bibnamefont {{Th. Lippert}}}, \ and\ \bibinfo {author} {\bibfnamefont
  {H.}~\bibnamefont {{De Raedt}}},\ }\bibfield  {title} {\enquote {\bibinfo
  {title} {Benchmarking gate-based quantum computers},}\ }\href@noop {}
  {\bibfield  {journal} {\bibinfo  {journal} {Comp. Phys. Comm.}\ }\textbf
  {\bibinfo {volume} {220}},\ \bibinfo {pages} {44 -- 55} (\bibinfo {year}
  {2017})}\BibitemShut {NoStop}%
\bibitem [{\citenamefont {Porter}\ and\ \citenamefont {Thomas}(1956)}]{PORT56}%
  \BibitemOpen
  \bibfield  {author} {\bibinfo {author} {\bibfnamefont {C.E.}\ \bibnamefont
  {Porter}}\ and\ \bibinfo {author} {\bibfnamefont {R.G.}\ \bibnamefont
  {Thomas}},\ }\bibfield  {title} {\enquote {\bibinfo {title} {Fluctuations of
  nuclear reaction widths},}\ }\href@noop {} {\bibfield  {journal} {\bibinfo
  {journal} {Phys. Rev.}\ }\textbf {\bibinfo {volume} {104}},\ \bibinfo {pages}
  {483 -- 491} (\bibinfo {year} {1956})}\BibitemShut {NoStop}%
\bibitem [{\citenamefont {Zlokapa}\ \emph {et~al.}(2020)\citenamefont
  {Zlokapa}, \citenamefont {Boixo},\ and\ \citenamefont
  {Lidar}}]{Zlokapa2005BoundariesQS}%
  \BibitemOpen
  \bibfield  {author} {\bibinfo {author} {\bibfnamefont {A.}~\bibnamefont
  {Zlokapa}}, \bibinfo {author} {\bibfnamefont {S.}~\bibnamefont {Boixo}}, \
  and\ \bibinfo {author} {\bibfnamefont {D.}~\bibnamefont {Lidar}},\ }\bibfield
   {title} {\enquote {\bibinfo {title} {Boundaries of quantum supremacy via
  random circuit sampling},}\ }\href@noop {} {\bibfield  {journal} {\bibinfo
  {journal} {arXiv:2005.02464}\ } (\bibinfo {year} {2020})}\BibitemShut
  {NoStop}%
\bibitem [{\citenamefont {{De Raedt}}\ \emph {et~al.}(2019)\citenamefont {{De
  Raedt}}, \citenamefont {Jin}, \citenamefont {Willsch}, \citenamefont
  {Willsch}, \citenamefont {Yoshioka}, \citenamefont {Ito}, \citenamefont
  {Yuan},\ and\ \citenamefont {Michielsen}}]{RAED19a}%
  \BibitemOpen
  \bibfield  {author} {\bibinfo {author} {\bibfnamefont {H.}~\bibnamefont {{De
  Raedt}}}, \bibinfo {author} {\bibfnamefont {F.}~\bibnamefont {Jin}}, \bibinfo
  {author} {\bibfnamefont {D.}~\bibnamefont {Willsch}}, \bibinfo {author}
  {\bibfnamefont {M.}~\bibnamefont {Willsch}}, \bibinfo {author} {\bibfnamefont
  {N.}~\bibnamefont {Yoshioka}}, \bibinfo {author} {\bibfnamefont
  {N.}~\bibnamefont {Ito}}, \bibinfo {author} {\bibfnamefont {S.}~\bibnamefont
  {Yuan}}, \ and\ \bibinfo {author} {\bibfnamefont {K.}~\bibnamefont
  {Michielsen}},\ }\bibfield  {title} {\enquote {\bibinfo {title} {Massively
  parallel quantum computer simulator, eleven years later},}\ }\href@noop {}
  {\bibfield  {journal} {\bibinfo  {journal} {Comp. Phys. Comm.}\ }\textbf
  {\bibinfo {volume} {237}},\ \bibinfo {pages} {47 -- 61} (\bibinfo {year}
  {2019})}\BibitemShut {NoStop}%
\bibitem [{\citenamefont {{De Raedt}}\ \emph {et~al.}(2007)\citenamefont {{De
  Raedt}}, \citenamefont {Michielsen}, \citenamefont {{De Raedt}},
  \citenamefont {Trieu}, \citenamefont {Arnold}, \citenamefont {Richter},
  \citenamefont {{Th. Lippert}}, \citenamefont {Watanabe},\ and\ \citenamefont
  {Ito}}]{RAED07x}%
  \BibitemOpen
  \bibfield  {author} {\bibinfo {author} {\bibfnamefont {K.}~\bibnamefont {{De
  Raedt}}}, \bibinfo {author} {\bibfnamefont {K.}~\bibnamefont {Michielsen}},
  \bibinfo {author} {\bibfnamefont {H.}~\bibnamefont {{De Raedt}}}, \bibinfo
  {author} {\bibfnamefont {B.}~\bibnamefont {Trieu}}, \bibinfo {author}
  {\bibfnamefont {G.}~\bibnamefont {Arnold}}, \bibinfo {author} {\bibfnamefont
  {M.}~\bibnamefont {Richter}}, \bibinfo {author} {\bibnamefont {{Th.
  Lippert}}}, \bibinfo {author} {\bibfnamefont {H.}~\bibnamefont {Watanabe}}, \
  and\ \bibinfo {author} {\bibfnamefont {N.}~\bibnamefont {Ito}},\ }\bibfield
  {title} {\enquote {\bibinfo {title} {Massively parallel quantum computer
  simulator},}\ }\href@noop {} {\bibfield  {journal} {\bibinfo  {journal}
  {Comp. Phys. Comm.}\ }\textbf {\bibinfo {volume} {176}},\ \bibinfo {pages}
  {121 -- 136} (\bibinfo {year} {2007})}\BibitemShut {NoStop}%
\bibitem [{\citenamefont {Willsch}\ \emph {et~al.}(2020)\citenamefont
  {Willsch}, \citenamefont {Lagemann}, \citenamefont {Willsch}, \citenamefont
  {Jin}, \citenamefont {{De Raedt}},\ and\ \citenamefont
  {Michielsen}}]{JUQCSNIC}%
  \BibitemOpen
  \bibfield  {author} {\bibinfo {author} {\bibfnamefont {D.}~\bibnamefont
  {Willsch}}, \bibinfo {author} {\bibfnamefont {H.}~\bibnamefont {Lagemann}},
  \bibinfo {author} {\bibfnamefont {M.}~\bibnamefont {Willsch}}, \bibinfo
  {author} {\bibfnamefont {F.}~\bibnamefont {Jin}}, \bibinfo {author}
  {\bibfnamefont {H.}~\bibnamefont {{De Raedt}}}, \ and\ \bibinfo {author}
  {\bibfnamefont {K.}~\bibnamefont {Michielsen}},\ }\bibfield  {title}
  {\enquote {\bibinfo {title} {{B}enchmarking {S}upercomputers with the
  {J}\"ulich {U}niversal {Q}uantum {C}omputer {S}imulator},}\ }in\ \href
  {https://juser.fz-juelich.de/record/874419} {\emph {\bibinfo {booktitle}
  {{NIC} {S}ymposium 2020}}},\ \bibinfo {series} {Publication Series of the
  John von Neumann Institute for Computing (NIC) NIC Series}, Vol.~\bibinfo
  {volume} {50},\ \bibinfo {editor} {edited by\ \bibinfo {editor}
  {\bibfnamefont {M.}~\bibnamefont {M\"uller}}, \bibinfo {editor}
  {\bibfnamefont {K.}~\bibnamefont {Binder}}, \ and\ \bibinfo {editor}
  {\bibfnamefont {A.}~\bibnamefont {Trautmann}}}\ (\bibinfo  {publisher}
  {Forschungszentrum J\"ulich GmbH Zentralbibliothek, Verlag},\ \bibinfo
  {address} {J\"ulich},\ \bibinfo {year} {2020})\ pp.\ \bibinfo {pages}
  {255--264}\BibitemShut {NoStop}%
\bibitem [{\citenamefont {Zhou}\ \emph {et~al.}(2020)\citenamefont {Zhou},
  \citenamefont {Stoudenmire},\ and\ \citenamefont {Waintal}}]{ZHOU20}%
  \BibitemOpen
  \bibfield  {author} {\bibinfo {author} {\bibfnamefont {Y.}~\bibnamefont
  {Zhou}}, \bibinfo {author} {\bibfnamefont {E.~M.}\ \bibnamefont
  {Stoudenmire}}, \ and\ \bibinfo {author} {\bibfnamefont {X.}~\bibnamefont
  {Waintal}},\ }\bibfield  {title} {\enquote {\bibinfo {title} {What limits the
  simulation of quantum computers?}}\ }\href@noop {} {\bibfield  {journal}
  {\bibinfo  {journal} {arXiv:quant-ph/2002.07730}\ } (\bibinfo {year}
  {2020})}\BibitemShut {NoStop}%
\bibitem [{\citenamefont {Jozsa}(1994)}]{Jozsa1994fidelity}%
  \BibitemOpen
  \bibfield  {author} {\bibinfo {author} {\bibfnamefont {R.}~\bibnamefont
  {Jozsa}},\ }\bibfield  {title} {\enquote {\bibinfo {title} {Fidelity for
  mixed quantum states},}\ }\href {\doibase 10.1080/09500349414552171}
  {\bibfield  {journal} {\bibinfo  {journal} {J. Mod. Opt.}\ }\textbf {\bibinfo
  {volume} {41}},\ \bibinfo {pages} {2315--2323} (\bibinfo {year}
  {1994})}\BibitemShut {NoStop}%
\bibitem [{\citenamefont {Schumacher}(1996)}]{SCHU96}%
  \BibitemOpen
  \bibfield  {author} {\bibinfo {author} {\bibfnamefont {Benjamin}\
  \bibnamefont {Schumacher}},\ }\bibfield  {title} {\enquote {\bibinfo {title}
  {Sending entanglement through noisy quantum channels},}\ }\href@noop {}
  {\bibfield  {journal} {\bibinfo  {journal} {Phys. Rev. A}\ }\textbf {\bibinfo
  {volume} {54}},\ \bibinfo {pages} {2614--2628} (\bibinfo {year}
  {1996})}\BibitemShut {NoStop}%
\bibitem [{\citenamefont {Horodecki}\ \emph {et~al.}(1999)\citenamefont
  {Horodecki}, \citenamefont {Horodecki},\ and\ \citenamefont
  {Horodecki}}]{horodecki1999fidelity}%
  \BibitemOpen
  \bibfield  {author} {\bibinfo {author} {\bibfnamefont {M.}~\bibnamefont
  {Horodecki}}, \bibinfo {author} {\bibfnamefont {P.}~\bibnamefont
  {Horodecki}}, \ and\ \bibinfo {author} {\bibfnamefont {R.}~\bibnamefont
  {Horodecki}},\ }\bibfield  {title} {\enquote {\bibinfo {title} {General
  teleportation channel, singlet fraction, and quasidistillation},}\ }\href
  {\doibase 10.1103/PhysRevA.60.1888} {\bibfield  {journal} {\bibinfo
  {journal} {Phys. Rev. A}\ }\textbf {\bibinfo {volume} {60}},\ \bibinfo
  {pages} {1888--1898} (\bibinfo {year} {1999})}\BibitemShut {NoStop}%
\bibitem [{\citenamefont {{J\"{u}lich Supercomputing Centre}}(2018)}]{JURECA}%
  \BibitemOpen
  \bibfield  {author} {\bibinfo {author} {\bibnamefont {{J\"{u}lich
  Supercomputing Centre}}},\ }\bibfield  {title} {\enquote {\bibinfo {title}
  {{JURECA: Modular supercomputer at J\"{u}lich Supercomputing Centre}},}\
  }\href {\doibase 10.17815/jlsrf-4-121-1} {\bibfield  {journal} {\bibinfo
  {journal} {J. of Large-Scale Res. Facil.}\ }\textbf {\bibinfo {volume} {4}},\
  \bibinfo {pages} {A132} (\bibinfo {year} {2018})}\BibitemShut {NoStop}%
\bibitem [{\citenamefont {Ullah}(1964)}]{ULLA64}%
  \BibitemOpen
  \bibfield  {author} {\bibinfo {author} {\bibfnamefont {N.}~\bibnamefont
  {Ullah}},\ }\bibfield  {title} {\enquote {\bibinfo {title} {Invariance
  hypothesis and higher correlations of {Hamiltonian} matrix elements},}\
  }\href {\doibase https://doi.org/10.1016/0029-5582(64)90522-X} {\bibfield
  {journal} {\bibinfo  {journal} {Nuclear Physics}\ }\textbf {\bibinfo {volume}
  {58}},\ \bibinfo {pages} {65 -- 71} (\bibinfo {year} {1964})}\BibitemShut
  {NoStop}%
\bibitem [{\citenamefont {Muller}(1959)}]{MULL59}%
  \BibitemOpen
  \bibfield  {author} {\bibinfo {author} {\bibfnamefont {M.~E.}\ \bibnamefont
  {Muller}},\ }\bibfield  {title} {\enquote {\bibinfo {title} {{A note on a
  method for generating points uniformly on N-dimensional spheres}},}\
  }\href@noop {} {\bibfield  {journal} {\bibinfo  {journal} {Comm. Assoc.
  Comput. Mach.}\ }\textbf {\bibinfo {volume} {2}},\ \bibinfo {pages} {19 --
  20} (\bibinfo {year} {1959})}\BibitemShut {NoStop}%
\bibitem [{\citenamefont {Miller}(1964)}]{MILL64}%
  \BibitemOpen
  \bibfield  {author} {\bibinfo {author} {\bibfnamefont {K.S.}\ \bibnamefont
  {Miller}},\ }\href@noop {} {\emph {\bibinfo {title} {Multidimensional
  Gaussian Distributions}}}\ (\bibinfo  {publisher} {John Wiley and Sons,
  Inc.},\ \bibinfo {address} {New York},\ \bibinfo {year} {1964})\BibitemShut
  {NoStop}%
\bibitem [{\citenamefont {Bengtsson}\ and\ \citenamefont
  {{\.Zyczkowski}}(2006)}]{BENG06}%
  \BibitemOpen
  \bibfield  {author} {\bibinfo {author} {\bibfnamefont {I.}~\bibnamefont
  {Bengtsson}}\ and\ \bibinfo {author} {\bibfnamefont {K.}~\bibnamefont
  {{\.Zyczkowski}}},\ }\href@noop {} {\emph {\bibinfo {title} {Geometry of
  quantum states}}}\ (\bibinfo  {publisher} {Cambridge University Press},\
  \bibinfo {address} {Cambridge},\ \bibinfo {year} {2006})\BibitemShut
  {NoStop}%
\end{thebibliography}%
\end{document}